\documentclass[%
reprint,
showpacs,
amsmath,
amssymb,
aps,
prb,
floatfix,
]{revtex4-1}

\usepackage{graphicx}
\usepackage{dcolumn}
\usepackage{multirow}
\usepackage{bm}
\usepackage{color}
\definecolor{lightgray}{rgb}{0.7,0.7,0.7}
\usepackage{colortbl}
\usepackage{hyperref}

\newcommand{\braket}[2]{\langle #1 \rvert #2\rangle}

\newcommand{\br}{\mathbf{r}}

\newcommand{\bk}{\mathbf{k}}
\newcommand{\bq}{\mathbf{q}}

\newcommand{\qatk}{QuantumATK}
\newcommand{\nanolab}{NanoLab}
\newcommand{\cpp}{C++}
\newcommand{\python}{Python}

\newcommand{\atkpython}{ATK-Python}

\newcommand{\atklcao}{ATK-LCAO}
\newcommand{\atkpw}{ATK-PlaneWave}
\newcommand{\atkff}{ATK-ForceField}

\RequirePackage[normalem]{ulem} 
\RequirePackage{color}\definecolor{RED}{rgb}{1,0,0}\definecolor{BLUE}{rgb}{0,0,1} 

\begin{document}


\title{QuantumATK: An integrated platform of electronic and atomic-scale modelling tools}

\author{S\o{}ren Smidstrup$^1$}
\author{Troels Markussen$^1$}
\author{Pieter Vancraeyveld$^1$}
\author{Jess Wellendorff$^1$}
\author{Julian Schneider$^1$}
\author{Tue Gunst$^{1,2}$}
\author{Brecht Verstichel$^1$}
\author{Daniele Stradi$^1$}
\author{Petr A. Khomyakov$^1$}
\author{Ulrik G. Vej-Hansen$^1$}
\author{Maeng-Eun Lee$^1$}
\author{Samuel T. Chill$^1$}
\author{Filip Rasmussen$^1$}
\author{Gabriele Penazzi$^1$}
\author{Fabiano Corsetti$^1$}
\author{Ari Ojanper{\"a}$^1$}
\author{Kristian Jensen$^1$}
\author{Mattias L. N. Palsgaard$^{1,2}$}
\author{Umberto Martinez$^1$}
\author{Anders Blom$^1$}
\author{Mads Brandbyge$^2$}
\author{Kurt Stokbro$^1$}

\affiliation{$^1$ Synopsys Denmark, Fruebjergvej 3, Postbox 4, DK-2100 Copenhagen, Denmark}

\affiliation{$^2$ DTU Physics, Center for Nanostructured Graphene (CNG), Technical University of Denmark, DK-2800 Kgs.\ Lyngby, Denmark.}

\begin{abstract}
\qatk\ is an integrated set of atomic-scale modelling tools developed since 2003
by professional software engineers in collaboration with academic researchers.
While different aspects and individual modules of the platform have been previously presented,
the purpose of this paper is to give a general overview of the platform.
The \qatk\ simulation engines enable electronic-structure calculations
using density functional theory or tight-binding model Hamiltonians,
and also offers bonded or reactive empirical force fields in many different parametrizations.
Density functional theory is implemented using either a plane-wave basis
or expansion of electronic states in a linear combination of atomic orbitals.
The platform includes a long list of advanced modules,
including Green's-function methods for electron transport simulations and surface calculations,
first-principles electron-phonon and electron-photon couplings,
simulation of atomic-scale heat transport, ion dynamics,
spintronics, optical properties of materials, static polarization, and more.
Seamless integration of the different simulation engines into a common platform
allows for easy combination of different simulation methods into complex workflows.
Besides giving a general overview and presenting a number of implementation details
not previously published, we also present four different application examples.
These are calculations of the phonon-limited mobility of Cu, Ag and Au,
electron transport in a gated 2D device,
multi-model simulation of lithium ion drift through a battery cathode
in an external electric field,
and electronic-structure calculations of the composition-dependent band gap of SiGe alloys.
\end{abstract}

\maketitle

\tableofcontents

\section{Introduction}
%
Atomic-scale modelling is increasingly important for industrial and academic
research and development in a wide range of technology areas,
including semiconductors,\cite{shankar2008density, zographos2017multiscale}
batteries,\cite{shi2015multi}
catalysis,\cite{norskov2008nature}
renewable energy,\cite{islam2010recent}
advanced materials,\cite{Saal2013}
next-generation pharmaceuticals,\cite{trau2001novel}
and many others.
Surveys indicate that the return on investment of atomic-scale modelling
is typically around 5:1.\cite{goldbeck2012economic}
With development of increasingly advanced simulation algorithms and more powerful computers,
we expect that the economic benefits of atomic-scale modelling will only increase.

The current main application of atomic-scale modelling
is in early-stage research into new materials and technology designs,
see Refs.~\onlinecite{Yamamoto2014,Bernholc2019} for examples.
The early research stage often has a very large design space,
and experimental trial and error is a linear process
that will explore only a small part of this space.
Atomic-scale simulations make it possible to guide experimental investigations
towards the most promising part of the technology design space.
Such insights are typically achieved by simulating the underlying atomic-scale
processes behind failed or successful experiments,
to understand the physical or (bio-)chemical origins.
Such insight can often rule out or focus research to
certain designs or material systems.\cite{goldbeck2012economic}
Recently, materials screening has also shown great promise.
In this approach, atomic-scale calculations
are used to obtain important properties of a large pool of materials,
and the most promising candidates are then selected for experimental verification
and/or further theoretical refinement.\cite{Greeley2006,Saal2013,Armiento2011}

The scientific field of atomic-scale modelling covers everything from
near-exact quantum chemical calculations
to approximate simulations using empirical force fields.
Quantum chemical methods (based on wave-function theory)
attempt to fully solve the many-body Schr\"o{}dinger equation
for all electrons in the system,
and can provide remarkably accurate descriptions of molecules.\cite{bartlett2007coupled}
However, the computational cost is high:
in practice, one is usually limited to calculations
involving far below 100 atoms in total.
Such methods are currently not generally useful for industrial research
into advanced materials and next-generation electronic devices.

On the contrary,
force-field (FF) methods are empirical but computationally efficient:
all inter-atomic interactions are described by analytic functions
with pre-adjusted parameters.
It is thereby possible in practice to simulate systems with millions of atoms.
Unfortunately, this often also hampers the applicability of a force field
for system types not included when fitting the FF parameters.

As an attractive intermediate methodology,
density functional theory (DFT)\cite{Hohenberg1964,Kohn1965,Parr1994,Kohn1996}
provides an approximate but computationally tractable
solution to the electronic many-body problem.
This allows for good predictive power with respect to experiments
with minimal use of empirical parameters
at a reduced computational cost.
Standard DFT simulations
may routinely be applied
to systems containing more than one thousand atoms,
and DFT is today the preferred framework for industrial applications
of \textit{ab initio} electronic-structure theory.

Semi-empirical (SE) electronic-structure methods
based on tight-binding (TB) model Hamiltonians
are more approximate,
but have a long tradition in semiconductor research.\cite{vogl1983semi}
Whereas DFT ultimately aims to approximate the true many-body electronic Hamiltonian
in an efficient but parameter-free fashion,
a TB model relies on parameters that are
adjusted to very accurately describe the properties of a number of reference systems.
This leads to highly specialized electronic-structure models
that typically reduce the computational expense by an order of magnitude
compared to DFT methods.
Such SE methods may be convenient for large-scale electronic-structure calculations,
for example in simulations of electron transport in semiconductor devices.
\begin{figure}
\centering
\includegraphics[width=\columnwidth]{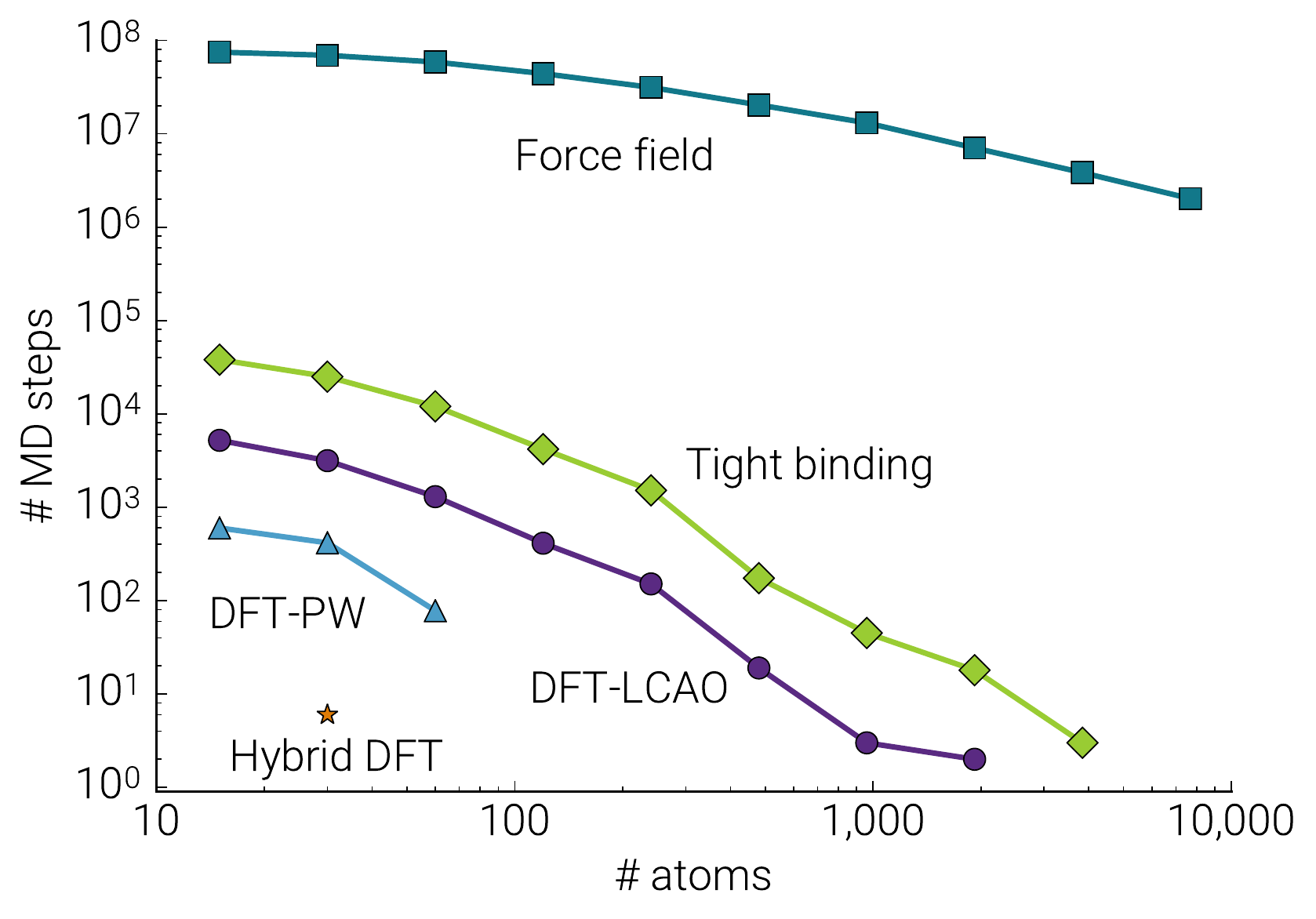}
\caption{Comparison of the simulation methods available in \qatk,
showing the total number of molecular dynamics steps performed
in 24 hours (\# MD steps) against system size
(\# atoms) for amorphous Al$_2$O$_3$ with constant density.
Each step includes evaluation of the total energy and atomic forces.
The simulations were run on a 16-core central processing unit (CPU) of the type
Intel\textsuperscript{\textregistered} Xeon\textsuperscript{\textregistered} E5-2670.
The FF simulations (Section~\ref{sec:ff})
were performed using threading only,
whereas full MPI parallelization was used for the TB
(Section~\ref{sec:tb}) and DFT (Section~\ref{sec:dft}) simulations.
For the latter, we have considered either semi-local exchange-correlation
functionals using linear-combination-of-atomic-orbitals (DFT-LCAO) and plane-wave (DFT-PW) basis sets,
or a hybrid exchange-correlation functional using a PW basis set (Hybrid DFT).
Further details of the calculations are given in Appendix~\ref{appendix}.}
\label{fig:ComparingModels}
\end{figure}

The \qatk\ platform offers simulation engines covering the entire range
of atomic-scale simulation methods relevant to the semiconductor industry
and materials science in general.
This includes force fields, SE methods, and several flavors of DFT.
These are summarized in Table~\ref{tab:engines},
including examples of other platforms that offer similar methodology.

To give a bird's-eye view of the computational cost of the different
atomic-scale simulation methods mentioned above,
we compare in Fig.~\ref{fig:ComparingModels} the computational speed of the methods
when simulating increasingly larger structures of amorphous Al$_2$O$_3$.
The measure of speed is here the number of molecular dynamics steps
that are feasible within 24 hours when run in parallel on 16 computing cores.
Although the parallel computing techniques used may differ between some of the methods,
we find that Fig.~\ref{fig:ComparingModels} gives a good overview
of the scaling between the different methods.
\begin{table*}
  \caption{Simulation engines in the \qatk\ platform, with examples
    of other simulation platforms using the same underlying methodology.
    LCAO and PW means linear combination of atomic orbitals and plane wave, respectively.}
  \label{tab:engines}
  \centering
  \begin{tabular}{l|lcl}
    Engine & Description & First release & Related platforms \\\hline
    \atklcao & Pseudopotential DFT using LCAO basis\cite{Smidstrup2017} & 2003 & SIESTA,\cite{Soler2002}
    OpenMX\cite{ozaki2003variationally} \\
    \atkpw & Pseudopotential DFT using PW basis & 2016 & VASP,\cite{VASPKresse}
    Quantum ESPRESSO\cite{giannozzi2009quantum}\\
    ATK-SE & Semi-empirical TB methods\cite{stokbro2010semiempirical}  & 2010 & DFTB+,\cite{Aradi2007} NEMO,\cite{klimeck2002development} OMEN\cite{klimeck2010atomistic} \\
    \atkff & All types of empirical force fields\cite{ATKForceField} & 2014 & LAMMPS,\cite{LAMMPS}
    GULP\cite{GULPGale2003}
  \end{tabular}
\end{table*}

It is important to realize that the simulation methods
listed in Table~\ref{tab:engines} should ideally complement each other:
for successful use of atomic-scale modelling,
it is essential to have easy access to all the methods,
in order to use them in combination.
The vast majority of atomic-scale simulation tools
are developed by academic groups,
and most of them focus on a single method.
Using the tool typically requires a large effort for compilation,
installation, learning the input/output syntax, etc.
The tool is often not fully compatible with any other tool,
so learning an additional tool within a new modelling class
requires yet another large effort.
Even within one modelling class, for example DFT,
a single simulation tool may not have all the required functionality
for a given application,
so several different tools within each modelling class
may be needed to solve a given problem,
and a significant effort must be invested to master each of them.
As a commercially developed platform,
QuantumATK aims to circumvent these issues.

Academic development of atomic-scale simulation platforms,
often made available through open-source licenses,
is essential for further technical progress of the field.
However, the importance of commercial platforms in progressing the industrial
uptake of the technology is often underestimated.
Commercial software relies on payment from end users.
This results in a strong focus on satisfying end-user requirements
in terms of usability, functionality, efficiency, reliability, and support.
The revenue enables the commercial software provider to establish
a stable team of developers and thereby provide a software
solution that will be maintained, extended, and supported for decades.

The ambition of the \qatk\ platform is to provide a state-of-the-art
and easy-to-use integrated toolbox with all important atomic-scale modelling
methodologies for a growing number of application areas. The methods are made available through
a modern graphical user interface (GUI) and a
Python scripting-based frontend for expert users.
Our current focus is semiconductor devices, polymers, glasses, catalysis,
batteries, and materials science in general.
In this context,
semiconductor devices is a broad area,
ranging from silicon-based electronic logics and memory elements,\cite{Thirunavukkarasu2017,Dong2018a}
to solar cells composed of novel materials\cite{crovetto_interface_2017}
and next-generation electronic devices based on spintronic phenomona.\cite{Sankaran2016}
One key strength of a unified framework
for a large selection of simulation engines and modelling tools
is within multiphysics and multiscale problems. Such problems often arise in physical modelling of semiconductor devices, and the \qatk\ platform is widely used for coupling technology computer-aided design (TCAD) tools with atomic-scale detail, for instance to provide first-principles simulations of defect migration paths and subsequently the temperature-dependent diffusion constant for continuum-level simulation of semiconductor processes.\cite{zographos2017multiscale}
Furthermore, QuantumATK provides
a highly flexible and efficient framework
for coupling advanced electrostatic setups
with state-of-the-art transport simulations
including electron-phonon coupling and light-matter interaction. This has  enabled predictions of gate-induced phonon scattering
in graphene gate stacks,\cite{gunst_flexural-phonon_2017}
atomistic description of ferroelectricity driven by edge-absorbed
polar molecules in gated graphene,\cite{caridad_graphene-edge_2018}
and new 2D material science such as prediction
of the room-temperature photocurrent in emerging layered Janus materials
with a large dipole across the plane.\cite{palsgaard_stacked_2018}
The flexibility of the \qatk\ framework supports the imagination of researchers,
and at the same time enables solutions to both real-world
and cutting-edge semiconductor device and material science problems.

The purpose of this paper is to give a general overview of the \qatk\ platform
with appropriate references to more thorough descriptions
of several aspects of the platform.
We also provide application examples that illustrate how the different
simulation engines can complement each other.
The paper is organized as follows:
In Section~\ref{sec:overview} we give a general overview of the \qatk\ platform,
while Section~\ref{sec:configurations} introduces the types of system
geometries handled by the platform.
The next three sections (\ref{sec:dft}--\ref{sec:ff})
describe the DFT, SE, and FF simulation engines, respectively.
We then introduce a number of simulation modules that work with the different engines.
These modules include ion dynamics (\ref{sec:optimization}),
phonon properties (\ref{sec:phonons}),
polarization (\ref{sec:berryphase}),
magnetic anisotropy energy (\ref{sec:mae}),
and quantum transport (\ref{sec:negf}).
We next describe the parallel computing strategies of the different engines,
and present parallel scaling plots in Section~\ref{sec:parallel}.
We then in Section~\ref{sec:nanolab} describe the scripting
and GUI simulation environment in the \qatk\ platform.
This is followed by four application examples in Section~\ref{sec:applications},
and the paper is summarized in Section~\ref{sec:Summary}.

\section{Overview}
\label{sec:overview}
%
The core of \qatk\ is implemented in \cpp\ modules with \python\ bindings,
such that all \cpp\ modules are accessible from \atkpython,
a customized version of \python\ built into the software.
The combination of a \cpp\ backend and a \python-based frontend
offers both high computational performance and a powerful but user-friendly
scripting platform for setting up, running, and analyzing atomic-scale simulations.
All simulation engines listed in Table~\ref{tab:engines}
are invoked using \atkpython\ scripting.
More details are given in Section~\ref{sec:atkpython}.
\qatk\ also relies on a number of open-source packages,
including high-performance numerical solvers.

All computationally demanding simulation modules may be run in parallel on
many processors at once,
using message passing between processes and/or shared-memory threading,
and often in a multi-level approach.
More details are given in Section~\ref{sec:parallel}.

The full \qatk\ package is installed on Windows or Linux/Unix operating systems
using a binary installer obtained from the Synopsys SolvNet website,
\url{https://solvnet.synopsys.com}.
All required external software libraries
are pre-compiled and shipped with the installer.
Licensing is handled using the Synopsys Common Licensing (SCL) system.

\section{Atomistic Configurations}
%
\label{sec:configurations}
\begin{figure*}
\begin{center}
    \centering
    \includegraphics[width=\textwidth]{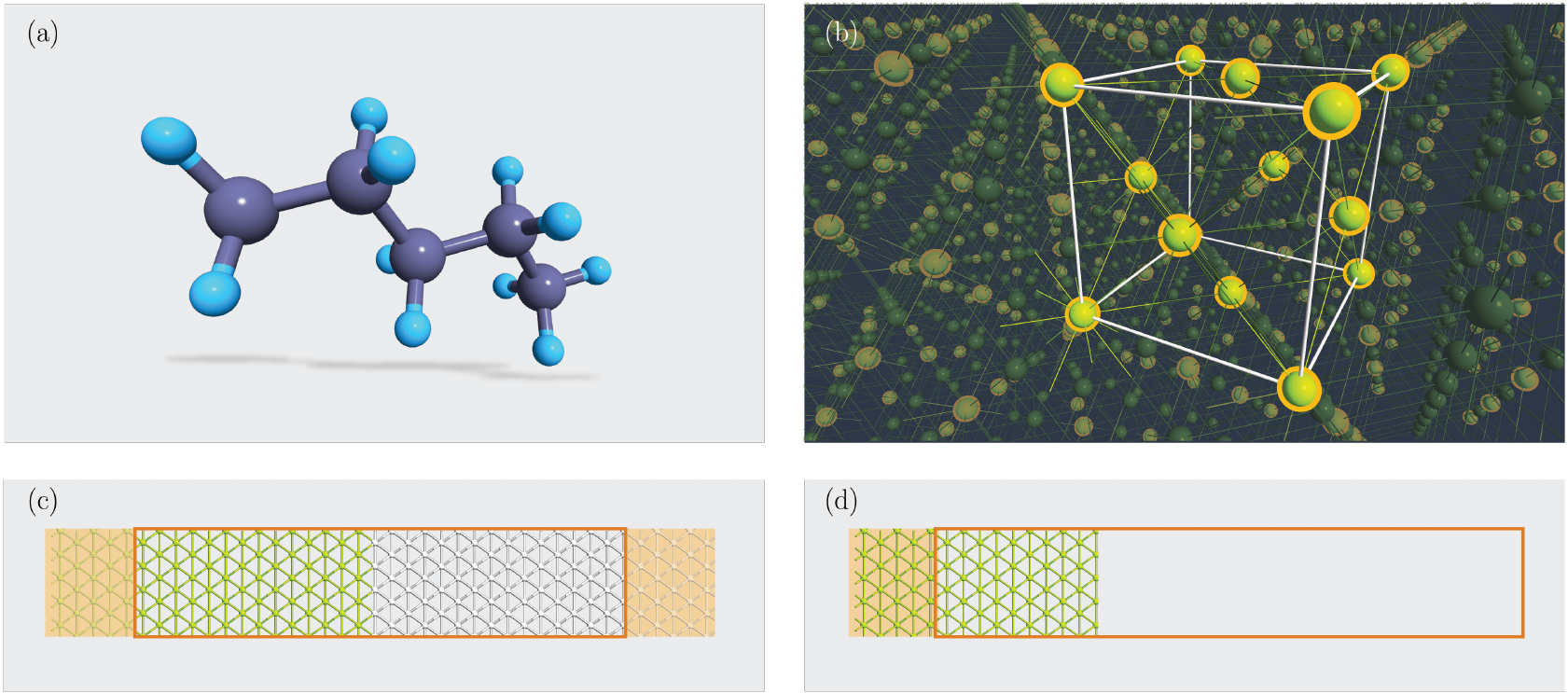}
    \caption{Supported atomistic configurations in QuantumATK.
    (a) \textit{Molecule configuration} of a pentane molecule.
    (b) \textit{Bulk configuration} of a gold crystal.
    (c) \textit{Device configuration} of a gold-silver interface.
    The structure consists of a left electrode (transparant yellow), central region (orange box), and a right electrode (transparent yellow).
    Both electrodes are semi-infinite in the left-right transport direction.
    The device is in this case periodic in the two directions perpendicular to the transport direction,
    but would be nonperiodic in one or both perpendicular directions in case of a nanosheet or nanotube device, respectively.
    (d) \textit{Surface configuration} of a gold surface. The structure consists of a left electrode (transparant yellow) and a central region (orange box).
    We note that an electric field can be applied to the surface by choice of boundary condition on the right-hand face of the central region.
    }
    \label{fig:configurations}
\end{center}
\end{figure*}
The real-space physical system to be simulated is defined as an \atkpython\
\textit{configuration object},
including lattice vectors, element types and positions, etc.
\qatk\ currently offers four main types of such configurations:
molecule, bulk, device, and surface.
Examples of these are given in Fig.~\ref{fig:configurations}.

The simplest configuration is the \textit{molecule configuration} shown in Fig.~\ref{fig:configurations}(a).
It is used for isolated (non-periodic) systems,
and is defined by a list of elements and their positions
in Cartesian coordinates.

The \textit{bulk configuration}, shown in Fig.~\ref{fig:configurations}(b),  defines an atomic-scale system
that repeats itself in one or more directions,
for example a fully periodic crystal (periodic in 3D),
a 2D nanosheet (or a slab), or a 1D nanowire.
The  bulk system is defined by the Bravais lattice
and the position of the atomic elements inside the primitive cell.

The two-probe \textit{device configuration} is used for quantum transport simulations.
As shown in Fig.~\ref{fig:configurations}(c),
the device consists of a central region connected to two semi-infinite bulk electrodes.
The central region,
where scattering of electrons travelling from one electrode to the other
may take place, can be periodic in zero (1D wire), one (2D sheet),
or two (3D bulk) directions,
but is bounded by the electrodes along the third dimension.
The device configuration is used to simulate electron and/or phonon transport
via the non-equilibrium Green's function (NEGF) method.\cite{Brandbyge2002}

Finally, for physically correct simulations of a surface,
\qatk\ provides the one-probe \textit{surface configuration}.
This is basically a device configuration with only one electrode,
as illustrated in Fig.~\ref{fig:configurations}(d).
By construction,
the surface configuration realistically describes the electronic structure
of a semi-infinite crystal beyond the approximate slab model.\cite{Smidstrup2017}

The remainder of this paper is devoted to describing the computational methods
available for calculating the properties of such configurations using \qatk.

\section{DFT Simulation Engines}
\label{sec:dft}
%
Density functional theory is implemented
in the Kohn--Sham (KS) formulation\cite{Hohenberg1964,Kohn1965,Parr1994,Kohn1996}
within the framework of the linear combination of atomic orbitals (LCAO)
and plane-wave (PW) basis set approaches,
combined with the pseudopotential method.
The electronic system is seen as a non-interacting electron gas of density $n$ in the effective potential $V^\text{eff}[n]$,
\begin{equation}
\label{eq:dft2}
V^\text{eff}[n] = V^\text{H}[n] + V^\text{xc}[n] + V^\text{ext}[n] ,
\end{equation}
where $V^\text{H}$ is the Hartree potential describing the classical electrostatic interaction between the electrons,
$V^\text{xc}$ is the exchange-correlation (XC) potential, which in practise needs to be approximated,
and $V^\text{ext}$ is the sum of the electrostatic potential energy of the electrons in the external potential of ions
and other electrostatic field sources.
The total external potential is in \qatk\ given by
\begin{equation}
\label{eq:dft3}
    V^\text{ext} = \sum_{a} V_{a}^\text{pseudo} + V^\text{gate} ,
\end{equation}
where $V_{a}^\text{pseudo}$ includes the local ($V^\text{loc}_{a}$) and nonlocal ($V^\text{nl}_{a}$) contributions to the pseudopotential of the $a$-th atom.
The term $V^\text{gate}$ is a potential that may originate from other external sources of electrostatic fields, for example metallic gates.

The KS Hamiltonian consists of the single-electron kinetic energy
and the effective potential,
\begin{equation}
\label{eq:dft1}
\hat{H}^\text{KS} = -\frac{\hbar^2}{2m} \nabla^2 + V^\text{eff} ,
\end{equation}
and the single-electron energies ($\epsilon_{\alpha}$) and wave functions ($\psi_{\alpha}$)
are solutions to eigenvalue problem
\begin{equation}
\hat{H}^\text{KS} \psi_{\alpha} = \epsilon_{\alpha}\, \psi_{\alpha} .
\end{equation}

The electronic ground state is found by iteratively minimizing
the KS total-energy density functional, $E[n]$, with respect to the electron density,
\begin{equation}
\label{eq:dft4}
E[n] = T + E^\text{H}[n] + E^\text{xc}[n] + E^\text{ext}[n] ,
\end{equation}
where $T$ is the kinetic energy.
The forces (acting on the atoms) and stress tensor of the electronic system
may then be computed as derivatives of the ground-state total energy
with respect to the atomic coordinates and the strain tensor, respectively.
%
\subsection{LCAO Representation}
\label{sec:dft_lcao}
%
The DFT-LCAO method uses a LCAO numerical representation of the KS equations,
closely resembling the SIESTA formalism.\cite{Soler2002}
This allows for a localized matrix representation of the KS Hamiltonian in \eqref{eq:dft1}, and therefore an efficient implementation of
KS-DFT for molecules, bulk materials, interface structures, and nanoscaled devices.

In the DFT-LCAO method, the single-electron KS eigenfunctions, $\psi_{\alpha}$,
are expanded in a set of finite-range atomic-like basis functions $\phi_i$,
\begin{equation}
\label{lcao1}
\psi_{\alpha}({\bf r}) = \sum_i c_{\alpha i} \phi_i({\bf r}) .
\end{equation}
The KS equation can then be represented as a matrix equation for determining
the expansion coefficients $c_{\alpha i}$,
\begin{equation}
\label{lcao2}
\sum_{j} H_{ij}^\text{KS} c_{\alpha j} = \varepsilon_\alpha \sum_{j } S_{ij} c_{\alpha j} ,
\end{equation}
where the Hamiltonian matrix
$H_{ij}^\text{KS} = \langle \phi_i | \hat{H}^\text{KS} | \phi_j \rangle$
and overlap matrix $S_{ij} = \langle \phi_i | \phi_j \rangle$
are given by integrals with respect to the electron coordinates.
Two-center integrals are computed using 1D radial integration
schemes employing a Fourier transform technique,
while multiple-center integrals are computed on a real-space grid.\cite{Soler2002}

For molecules and bulk systems, diagonalization of the Hamiltonian matrix yields the density matrix $D_{ij}$,
\begin{equation}
\label{lcao3}
D_{ij} = \sum_{\alpha}c_{\alpha i}^* c_{\alpha j}
         f \left(\frac{\varepsilon_\alpha-\varepsilon_\text{F}}{k_\text{B} T}\right) ,
\end{equation}
where $f$ is the Fermi--Dirac distribution of electrons over energy states,
$\varepsilon_\text{F}$ the Fermi energy,
$T$ the electron temperature,
and $k_\text{B}$ the Boltzmann constant.
For device and surface configurations,
the density matrix is calculated using the NEGF method,
as described in Section~\ref{sec:negf}.

The electron density is computed from the density matrix,
\begin{equation}
\label{lcao4}
n({\bf r}) = \sum_{ij} D_{ij} \phi_i({\bf r}) \phi_j({\bf r}),
\end{equation}
and is represented on a regular real-space grid,
which is the same grid as used for the effective potential in \eqref{eq:dft2}.

%
\subsection{PW Representation}
%
A PW representation of the KS equations was recently implemented in \qatk.
It is complimentary to the LCAO representation discussed above.
The ATK-PlaneWave engine is intended mainly for simulating bulk configuratins with periodic boundary conditions. The KS eigenfunctions are expanded in terms of PW basis functions,
\begin{equation}
    \psi_\alpha({\bf r}) = \sum_{|{\bf g}| < g_\text{max}} c_{\alpha, {\bf g}} \mathrm{e}^{\mathrm{i} {\bf g} \cdot {\bf r}},
\end{equation}
where $\alpha$ denotes both the wave vector $\bk$ and the band index $n$, and ${\bf g}$ are reciprocal lattice vectors. The upper threshold for the reciprocal lattice-vector lengths included in the PW expansion ($g_\text{max}$) is determined by a kinetic-energy (wave-function) cutoff energy $E_\text{cut}$,
\begin{equation}
\frac{\hbar^2 g_\text{max}^2}{2m} < E_\text{cut}.
\end{equation}

The DFT-PW method has its distinct advantages and disadvantages compared to the DFT-LCAO approach.
In particular, the PW expansion is computationally efficient for relatively small bulk systems,
and the obtained physical quantities can be systematically converged with respect to the PW basis-set size by increasing $E_\text{cut}$.
However, the PW representation is computationally inefficient for low-dimensional systems with large vacuum regions. It is also incompatible with the DFT-NEGF methodology for electron transport calculations in nanoscaled devices, unlike the LCAO representation, which is ideally suited for dealing with open boundary conditions, and is also more efficient for large systems.

The ATK-PlaneWave engine was implemented on the same infrastructure as used by the ATK-LCAO engine,
though a number of routines were modified to reach state-of-the-art PW efficiency. For example, we have adopted
iterative algorithms for solving the KS equations,\cite{Davidson1975} and fast Fourier transform (FFT) techniques for applying the Hamiltonian operator and evaluating the electron density.\cite{Payne1992,WendeMarsmanSteinke2016}

\begin{figure}
  \centering
  \includegraphics[width=\columnwidth]{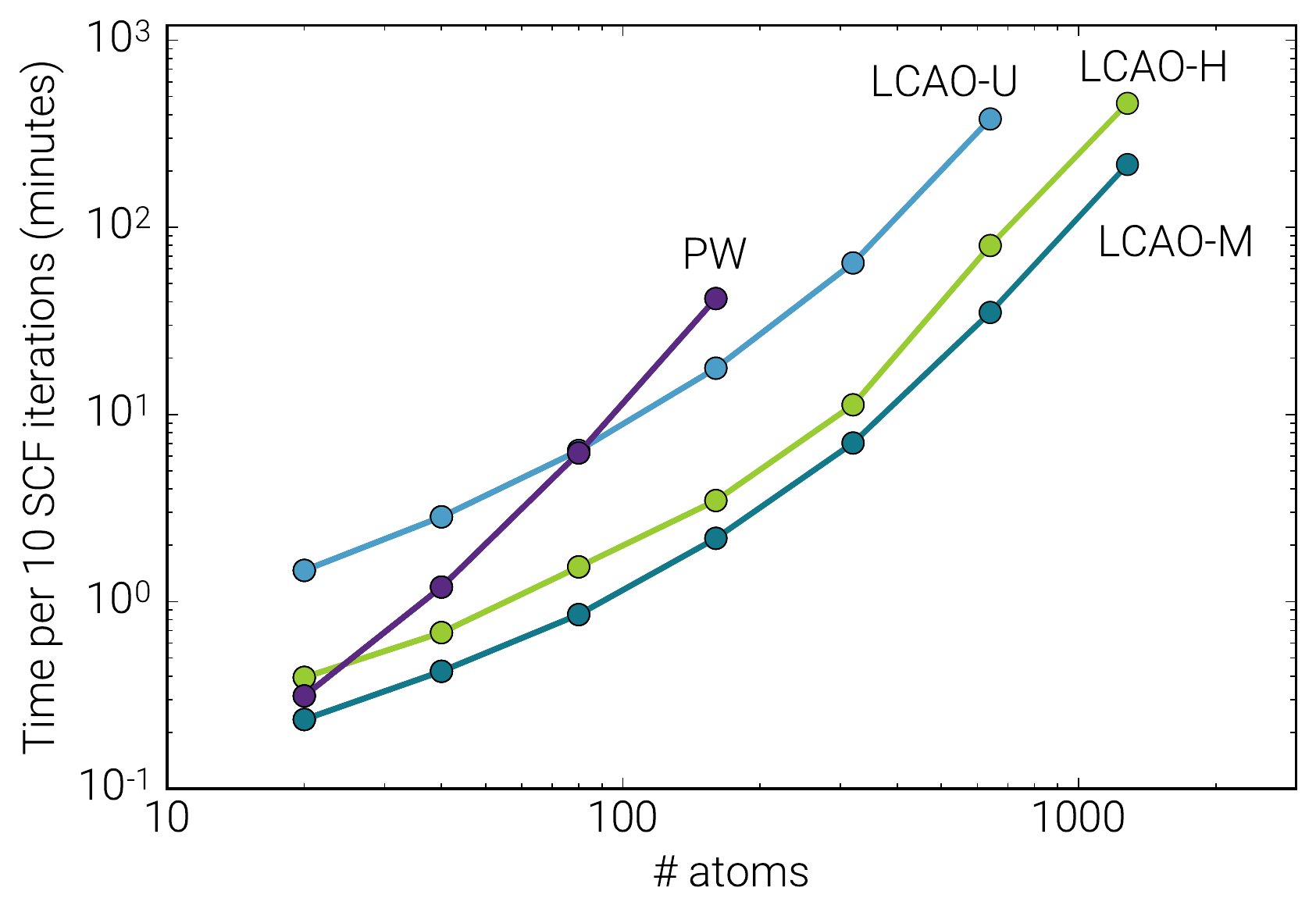}
  \caption{Time per 10 selfconsistent field (SCF) iterations for different sized gold melts at 900~K.
  For each system, we use a single $\bk$-point and the simulation runs on a 16-core CPU.
  The timings of the DFT-PW method are compared to those of the DFT-LCAO method using the Ultra (LCAO-U),
  High (LCAO-H), and Medium (LCAO-M) basis sets.}
  \label{fig:gold_melts}
\end{figure}

In Fig.~\ref{fig:gold_melts} we compare the CPU times of DFT-PW vs.\ DFT-LCAO calculations for different LCAO basis sets.
The figure shows the CPU time for the different methods as function of the system size.
The PW approach is computationally efficient for smaller systems, while the LCAO approach can be more than an order of magnitude faster for systems with more than 100 atoms.

\subsection{Pseudopotentials and LCAO Basis Sets}
%
\label{sec:lcaobasis}
\qatk\ uses pseudopotentials (PPs) to avoid explicit DFT calculations of core electrons,
and currently supports both scalar-relativistic and fully relativistic normconserving PPs.\cite{dal2005spin}
Projector augmented-wave (PAW) potentials\cite{blochl1994projector}
is currently available for the ATK-PlaneWave simulation engine only.

The \qatk\ platform is shipped with built-in databases of well-tested PPs,
covering all elements up to $Z=83$ (Bi), excluding lanthanides.
The current default PPs are those of the published
SG15\cite{Schlipf2015} and PseudoDojo\cite{van2018pseudodojo} sets.
These are two modern normconserving PP types
with multiple projectors for each angular momentum, to ensure high accuracy.
Both sets contain scalar-relativistic and fully relativistic
PPs for each element.
The fully relativistic PPs are generated by solving the Dirac equation for the atom,
which naturally includes spin-orbit coupling,
and then mapping the solution onto the scalar-relativistic formalism.\cite{Theurich2001,dal2005spin}

For each PP, we have generated an optimized LCAO basis set,
consisting of orbitals $\phi_{nlm}$,
\begin{equation}
\label{eq:basis1}
\phi_{nlm}({\bf r}) = \chi_{nl}(r) Y_{lm}(\hat{\br}),
\end{equation}
where $Y_{lm}$ are spherical harmonics, and $\chi_{nl}$ are radial functions
with compact support, being exactly zero outside a confinement radius.
The basis orbitals are obtained by solving the radial Schr\"{o}dinger equation
for the atom in a confinement potential.\cite{Soler2002}
For the shape of the confinement potential, we follow Ref.~\onlinecite{Blum2009}.

To construct high-accuracy LCAO basis sets for the SG15 and PseudoDojo PPs,
we have adopted a large set of pseudo-atomic orbitals that are
similar to the ``tight tier 2'' basis sets used in the FHI-aims package.\cite{Blum2009}
These basis sets typically have 5 orbitals per PP valence electron,
and a range of 5~\AA\ for all orbitals,
and include angular momenta up to $l=5$.
From this large set, we have constructed three different series of reduced DFT-LCAO basis sets implemented in QuantumATK:
\begin{enumerate}
\item \textit{Ultra}: Generated by reducing the range of the original pseudo-atomic orbitals,
requiring that the overlap of each contracted orbital
with the corresponding original orbital can change by no more than 0.1\%.
Also denoted ``LCAO-U''.
\item \textit{High}: Generated by reducing the number of basis orbitals in the Ultra basis set,
requiring that the DFT-obtained total energy of suitably chosen test systems
change by no more than 1~meV/atom. 
Also denoted ``LCAO-H''.
\item \textit{Medium}: Generated by further reduction of the number of orbitals in the High basis set,
requiring that the subsequent change of the DFT-obtained total energies do not exceed 4~meV/atom. Also denoted ``LCAO-M''.
The number of pseudo-atomic orbitals in a Medium basis set is typically comparable to that of a double-zeta polarized (DZP) basis set.
\end{enumerate}
\begin{table}
\caption{\label{tab:delta} Summary of \qatk\ $\Delta$-tests for elemental solids,\cite{lejaeghere2016reproducibility}
and RMS errors of the lattice constant ($a$) and bulk modulus ($B$) of rock-salt and
perovskite test sets,\cite{Garrity2014} using SG15 and PseudoDojo PPs,
and the \atkpw\ and \atklcao\ engines,
the latter with different basis sets.
All errors are relative to all-electron calculations.}
\begin{ruledtabular}
\begin{tabular}{lcccc}
& Medium & High & Ultra & PW \\
\hline \cline{1-5} \\ [-2ex]
\multicolumn{5}{c}{Elemental solids: Delta tests}\\
\hline \cline{1-5} \\ [-2ex]
SG15 $\Delta$ (meV)       & 3.45 & 1.88 & 2.03 & 1.32 \\ [-0.25ex]
PseudoDojo $\Delta$ (meV) & 4.53 & 1.52 & 1.40 & 1.04 \\ [-0.25ex]
\hline \cline{1-5} \\ [-2ex]
\multicolumn{5}{c}{Rock salts: RMS of $a$ and $B$}\\
\hline \cline{1-5} \\ [-2ex]
SG15 (\%)       & 0.40 & 0.24 & 0.23 & 0.16 \\
PseudoDojo (\%) & 0.50 & 0.18 & 0.15 & 0.09 \\
\hline \cline{1-5} \\ [-2ex]
\multicolumn{5}{c}{Perovskites: RMS of $a$ and $B$ }\\
\hline \cline{1-5} \\ [-2ex]
SG15 (\%)       & 0.36 & 0.24 & 0.18 & 0.13 \\
PseudoDojo (\%) & 0.35 & 0.21 & 0.13 & 0.06
\end{tabular}
\end{ruledtabular}
\end{table}

To validate the PPs and basis-sets,
we have used the $\Delta$-test\cite{DeltaTest, lejaeghere2016reproducibility} to check the accuracy of the equation of state for elemental, rock-salt, and perovskite solids against all-electron reference calculations, as shown in Table~\ref{tab:delta}.
For each bulk crystal, the equation of state was calculated at fixed internal ion coordinates, and the equilibrium lattice constant and bulk modulus were computed.
In Table~\ref{tab:delta}, the $\Delta$-value is defined as the root-mean-square (RMS) energy difference between the equations of state obtained with QuantumATK and the all-electron reference, averaged over all crystals in a purely elemental benchmark set.

Table~\ref{tab:delta} suggests a general trend that the PseudoDojo PPs are slightly more accurate than the SG15 ones. Since the PseudoDojo PPs are in general also softer, requiring a lower real-space density mesh cutoff energy, these are the default PPs in QuantumATK.

Table~\ref{tab:delta} also shows that the accuracy of the DFT-LCAO calculations done with High or Ultra basis sets
is rather close to that of the PW calculations. The Medium basis sets give on average a larger deviation from the PW results.
However, we also find that LCAO-M provides sufficient accuracy for many applications,
and it is therefore the default \atklcao\ basis set in \qatk.
We note that in typical applications, using Medium instead of the High (Ultra) basis sets decreases the computational cost
by a factor of 2--4 (10--20), as seen in Fig.~\ref{fig:gold_melts}.

More details on the construction and validation of the LCAO basis sets
can be found in Ref.~\onlinecite{Smidstrup2017}.
%
%
\subsection{Exchange-Correlation Methods}
\label{sec:xc}
%
The XC functional in \eqref{eq:dft4} is the only formal approximation in KS-DFT, since the exact functional is unknown.\cite{Kohn1965,Parr1994,Kohn1996}
\qatk\ supports a large range of approximate XC functionals, including the local density approximation (LDA),
generalized gradient approximations (GGAs), and meta-GGA functionals, all supplied through the Libxc library.\cite{marques2012libxc}
The \atkpw\ engine also allows for calculations using the HSE06 screened hybrid functional.\cite{heyd2003hybrid,Heyd2005,Krukau2006}
The \atklcao\ and \atkpw\ engines both support van der Waals dispersion methods
using the two-body and three-body dispersion corrections by Grimme.\cite{grimme2006semiempirical}
Both DFT engines support different spin variants for each XC functional:
spin-unpolarized and spin-polarized (both collinear and noncollinear).
Spin-polarized noncollinear calculations may include spin-orbit interaction through the use of fully relativistic PPs.

\subsubsection{Semilocal functionals}
During the past 20 years, the semilocal (GGA) XC approximations have been widely used,
owing to a good balance between accuracy and efficiency for DFT calculations.
QuantumATK implements many of the popular GGAs,
including the general-purpose PBE,\cite{Perdew1996} the PBEsol (designed for solids),\cite{perdew2008restoring}
and the revPBE/RPBE functionals (designed for chemistry applications).\cite{hammer1999improved}
Recently, the meta-GGA SCAN functional\cite{scan2015} was also included in QuantumATK,
often providing improved accuracy of DFT calculations as compared to PBE.

\subsubsection{Hybrid functionals}
Hybrid XC approximations mix local and/or semilocal functionals with some amount of exact exchange
in order to provide higher accuracy for electronic-structure calculations.\cite{heyd2003hybrid,Paier2006}
However, the computational cost is usually much higher than for semilocal approximations.
New methodological developments based on the adaptively compressed exchange operator (ACE) method\cite{lin2016}
allow reducing the computational burden of hybrid functionals.
The ACE algorithm was recently implemented in QuantumATK for HSE06 calculations,
which gives a systematically good description of the band gap of most semiconductors and insulators,
see Table~\ref{tab:bandgaps}.

\subsubsection{Semiempirical methods}
Using hybrid functionals is computationally demanding for simulating large systems,
often even prohibitive.
\qatk\ offers a number of semiempirical XC methods
that allow for computationally efficient simulations
while giving rather accurate semiconductor band gaps.
These include the DFT-1/2 method,\cite{Ferreira2008,Ferreira2011} the TB09 XC potential,\cite{TB09}
and the pseudopotential projector-shift approach of Ref.~\onlinecite{Smidstrup2017}.

The selfconsistent DFT-1/2 methods, including LDA-1/2 and GGA-1/2,
do contain empirical parameters.
In \qatk, these parameters are chosen by fitting the calculated band gaps
to measured ones for bulk crystals. Table~\ref{tab:bandgaps} suggests that
the DFT-1/2 method, as implemented in \qatk, allows for significantly improved
band gaps at almost no extra computational cost.
We note that a recent study has shown certain limitations of the DFT-1/2 method,
in particular for anti-ferromagnetic transition metal oxides.\cite{doumont2019limitations}
Furthermore, this method does not provide reliable force and stress calculations.
It is also important to note that not all species in the system necessarily require the DFT-1/2 correction.
In general, it is advisable to apply this correction to the anionic species only,
keeping the cationic species as normal.\cite{Ferreira2008,Ferreira2011}

The Tran--Blaha meta-GGA XC functional (TB09)\cite{TB09}
introduces a parameter, $c$, which can be calculated selfconsistently according to
an empirical formula given in Ref.~\onlinecite{TB09}.
Table~\ref{tab:bandgaps} includes band gaps computed using this approach.
The $c$-parameter may also be adjusted to obtain a particular band gap for a given material,
and \qatk\ allows for setting different TB09 $c$-parameters on different regions
in the simulation cell.
This may be useful for studying electronic effects at interfaces between
dissimilar materials, for example in oxide-semiconductor junctions,
where the appropriate (and material-dependent) $c$-parameter
may be significantly different in the oxide and in the semiconductor.

\qatk\ also offers a pseudopotential projector-shift (PPS) method,
that introduces empirical shifts of the nonlocal projectors in the PPs,
in spirit of the empirical PPs proposed by Zunger and co-workers.\cite{Wang1995}
The PPS method is usually combined with ordinary PBE calculations.\cite{Smidstrup2017}
The two main advantages of this PPS-PBE approach are that
(1) for each semiconductor, the projector shifts can be fitted such that the DFT-predicted fundamental band gap
and lattice parameters are both fairly accurate compared to measured ones,
and (2) the PPS method does yield first-principles forces and stress,
and therefore can be used for geometry optimization,
unlike the DFT-1/2 and TB09 methods.
Table~\ref{tab:ppspbe} shows that the PPS-PBE predicted equilibrium lattice parameters are only slightly overestimated, and
the PPS-PBE band gaps are fairly close to experiments. We note that the PPS-PBE parameters are currently
available in QuantumATK for the elements silicon and germanium only.
\begin{table}
\caption{\label{tab:bandgaps}
Fundamental band gaps (in units of eV)
for a range of semiconductors and simple oxides,
calculated using different XC methods,
and compared to experimental values.
The ATK-LCAO simulation engine was used for PBE, TB09,
and PBE-1/2 calculations,
while the ATK-PlaneWave engine was used for simulations using HSE06.
PseudoDojo PPs were used,
combined with Ultra basis sets for DFT-LCAO,
except for TB09 calculations, which were done using FHI-DZP.
Default cutoff energies were used,
and a $\bk$-point grid density of 7~\AA.
For bulk silicon, this corresponds to a $15 \times 15 \times 15$ $\bk$-point grid.
Experimental band gaps are from Ref.~\onlinecite{Heyd2005}
unless otherwise noted.
The bottom row lists the RMS deviation between theory and experiments.
}
\begin{ruledtabular}
\begin{tabular}{lddddd}
Material & \multicolumn{1}{c}{Experiment} & \multicolumn{1}{c}{PBE} & \multicolumn{1}{c}{TB09} & \multicolumn{1}{c}{PBE-1/2} & \multicolumn{1}{c}{HSE06} \\
[0.5ex] \hline \cline{1-6} \\ [-2ex]
    C     &  5.48  &  4.19  &  5.11  &  5.59  &  5.33  \\
   Si     &  1.17  &  0.57  &  1.20  &  1.16  &  1.17  \\
   Ge     &  0.74  &  0.00  &  1.11  &  0.81  &  0.55\footnote{Direct band gap ($\Gamma \rightarrow \Gamma$), different in size from the 0.72~eV
reported in Ref.~\onlinecite{Schimka2011}, but similar to the 0.56~eV reported in Ref.~\onlinecite{Heyd2005}, both using theoretical lattice constants rather than experimental ones.} \\
  SiC     &  2.42  &  1.36  &  2.31  &  2.66  &  2.27  \\
   BP     &  2.40  &  1.24  &  1.79  &  1.63  &  2.01  \\
  BAs     &  1.46  &  1.25  &  1.94  &  1.58  &  2.05  \\
  AlN     &  6.13  &  4.16  &  6.97  &  5.83  &  5.54  \\
  AlP     &  2.51  &  1.55  &  2.36  &  2.46  &  2.30  \\
 AlAs     &  2.23  &  1.45  &  2.45  &  2.38  &  2.27  \\
 AlSb     &  1.68  &  1.22  &  1.82  &  1.92  &  1.76  \\
  GaN     &  3.50  &  1.89  &  4.10  &  3.27  &  2.87  \\
  GaP     &  2.35  &  1.59  &  2.38  &  2.22  &  2.26  \\
 GaAs     &  1.52  &  0.63  &  1.81  &  1.23  &  1.11  \\
 GaSb     &  0.73  &  0.11  &  0.76  &  0.52  &  0.64  \\
  InN     &  0.69  &  0.00  &  1.74  &  1.20  &  0.49  \\
  InP     &  1.42  &  0.69  &  2.17  &  1.30  &  1.26  \\
 InAs     &  0.41  &  0.00  &  1.08  &  0.51  &  0.23  \\
 InSb     &  0.23  &  0.00  &  0.49  &  0.32  &  0.27  \\
 TiO$_2$  &  3.0\footnote{Ref.~\onlinecite{Landmann2012}.}        &  1.91  &  3.11  &  3.00  &  3.37  \\
 SiO$_2$  &  8.9\footnote{Ref.~\onlinecite{Bersch2008}.}          &  6.07  &  11.31 &  8.16  &  7.83  \\
 ZrO$_2$  &  5.5\footnotemark[3]          &  3.65  &  4.96  &  5.26  &  5.16  \\
 HfO$_2$  &  5.7\footnotemark[3]          &  4.17  &  5.54  &  5.87  &  5.76  \\
  ZnO     &  3.44\footnote{Ref.~\onlinecite{CRCsemiconductors}.}  &  0.95  &  3.24  &  2.78  &  2.47  \\
  MgO     &  7.22  &  4.79  &  8.51  &  6.75  &  6.49  \\
[0.5ex] \hline \cline{1-6} \\ [-2ex]
RMS error &        &  1.34  &  0.71  &  0.33  &  0.43
\end{tabular}
\end{ruledtabular}
\end{table}
\begin{table}
\caption{\label{tab:ppspbe}
Silicon and germanium equilibrium lattice constants and fundamental band gaps,
both calculated using the PPS-PBE XC method,
and compared to experiments at 300~K.
The SG15-High combination of PPs and LCAO basis sets was used,
and a $15 \times 15 \times 15$ $\bk$-point grid.
The lattice constants were determined by minimizing the first-principles
stress on the primitive unit cells,
using a maximum stress criterion of 0.1~GPa (0.6~meV/\AA$^3$).
}
\begin{ruledtabular}
\begin{tabular}{llcc}
Material & Property & PPS-PBE & Experiment \\
[0.5ex] \hline \cline{1-4} \\ [-2ex]
\multirow{2}{*}{Silicon}   & Lattice constant & 5.439~\AA & 5.431~\AA \\
                           & Band gap         & 1.14~eV   & 1.12~eV   \\
\hline
\multirow{2}{*}{Germanium} & Lattice constant & 5.736~\AA & 5.658~\AA \\
                           & Band gap         & 0.65~eV   & 0.67~eV
\end{tabular}
\end{ruledtabular}
\end{table}
\subsubsection{DFT+U methods}
\qatk\ supports the mean-field Hubbard-U correction
by Dudarev \textit{et al.}\cite{Dudarev1998} and Cococcioni \textit{et al.},\cite{Cococcioni2005}
denoted DFT+U, LDA+U, GGA+U, or XC+U.
This method aims to include the strong on-site Coulomb interaction of localized electrons
(often localized $d$ and $f$ electrons),
which are not correctly described by LDA or GGA.
A Hubbard-like term is added to the XC functional,
\begin{equation}
    E_{U} = \frac{1}{2} \sum_{l} U_{l} (n_{l} - n_{l}^2) ,
\end{equation}
where $n_{l}$ is the projection onto an atomic shell $l$,
and $U_{l}$ is the Hubbard U for that shell.
The energy term $E_{U}$ is zero for a fully occupied or unoccupied shell,
but positive for a fractionally occupied shell.
This favors localization of electrons in the shell $l$, typically increasing the band gap of semiconductors.
%
%
%
\subsection{Boundary Conditions and Poisson Solvers}
\label{sec:poisson}
%
\begin{figure}
\begin{center}
\includegraphics[width=\columnwidth]{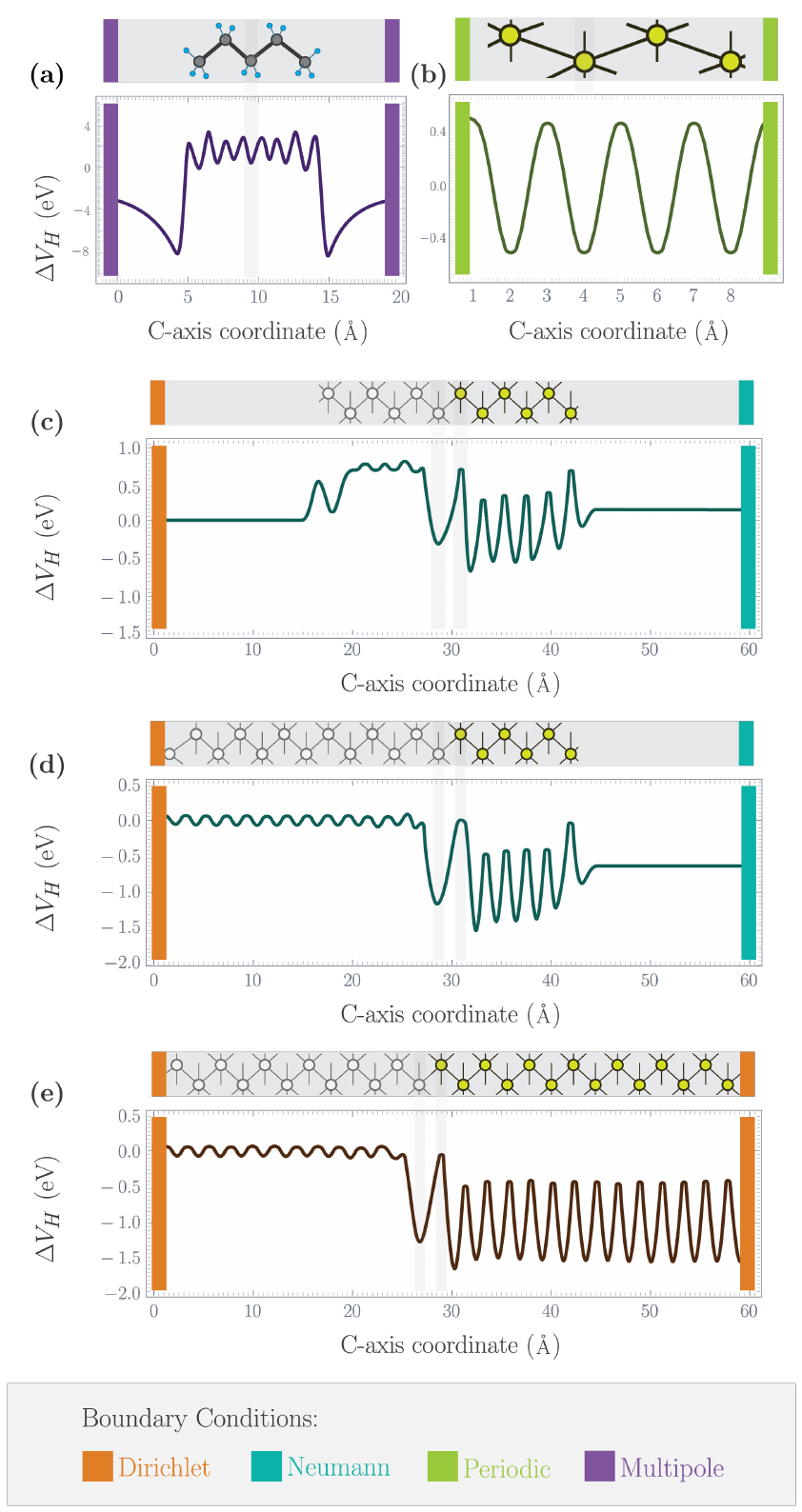}
\caption{\qatk\ supports many different BCs.
(a) Multipole BCs for a charged molecule in all directions,
(b) 3D periodic BCs for a bulk configuration,
(c) mixed Dirichlet and Neumann BCs for a slab model,
(d) Dirichlet and Neumann BCs are also the natural choice for a surface configuration,
(e) Dirichlet BCs at the interfaces between the semi-infinite electrodes and the central region in a device configuration.
Note that periodic BCs are imposed in the directions perpendicular to the C-axis in (b)-(e).
}
\label{fig:bcs}
\end{center}
\end{figure}
As already mentioned in Section~\ref{sec:dft_lcao},
the electron density, $n(\br)$ in \eqref{lcao4}, and the effective potential, $V^\text{eff}(\br)$ in \eqref{eq:dft1}, are in QuantumATK represented on a real-space regular grid. The corresponding Hartree potential $V^\text{H}(\bf r)$ is then calculated by solving the Poisson equation on this grid with appropriate boundary conditions (BCs) imposed on the six facets of the simulation cell,
\begin{equation}
\label{eq:poisson_general}
\nabla^2 V^{\text{H}}({\bf r}) = -\frac{e^2}{4 \pi \epsilon_0} n({\bf r}),
\end{equation}
where $e$ is the elementary charge, and $\epsilon_0$ is the vacuum permittivity.

In QuantumATK, one may also specify metallic or dielectric continuum regions in combination with a microscopic, atomistic structure,
as demonstrated for a 2D device in Fig.~\ref{fig:2Ddevice_structure} in Section~\ref{sec:2DFET}. This
affects the solution of the Poisson equation \eqref{eq:poisson_general} in the following way.
For a metallic region denoted $\Omega$, the electrostatic potential is fixed to a constant potential value ($V_0$) within this region, i.e.,
the Poisson equation is solved with the constraint
\begin{equation}
V^{\text{H}}({\bf r}) = V_0, \, {\bf r} \in \Omega.
\end{equation}

For a dielectric region denoted $\Upsilon$, the right-hand side of the Poisson equation will be modified
as follows:
\begin{align}
\label{eq:poisson_dielectric}
\nabla^2 V^{\text{H}}(\bf r) &= -\frac{e^2}{4 \pi \epsilon_0} n({\bf r}), \, {\bf r} \not\in \Upsilon, \nonumber\\
\nabla^2 V^{\text{H}}(\bf r) &= -\frac{e^2}{4 \pi \epsilon_r \epsilon_0} n({\bf r}), \, {\bf r} \in \Upsilon,
\end{align}
where $\epsilon_r$ is the relative dielectric constant, which can be specified as an external
parameter in QuantumATK calculations.

\subsubsection{Boundary conditions}
QuantumATK implements four basic types of BCs; multipole, periodic, Dirichlet and Neumann BCs.
It is also possible to impose mixed BCs on the six facets of the simulation cell to simulate a large variety of
physical systems at different levels of approximation.

A {\it multipole} BC means that the Hartree potential at the
boundary is determined by calculating the monopole, dipole and quadrupole moments of the charge distribution inside
the simulation cell, and that these moments are used to extrapolate the value of the potential at the boundary.
A {\it Dirichlet} BC means that the potential has been fixed
to a certain potential $V_0({\bf r})$ at the boundary, such that, for a facet $S$ of the simulation cell,
\begin{equation}
V^\text{H}({\bf r}) = V_0 ({\bf r}), \, {\bf r} \in S.
\end{equation}
A {\it Neumann} BC means that the normal derivative of the
potential on a facet has been fixed to a given function $V_{0}^{\prime}({\bf r})$,
\begin{equation}
\frac{\partial{V^\text{H}({\bf r})}}{\partial {\bf n}} = {\bf{n}} \cdot \nabla V^\text{H} ({\bf r}) = V_{0}^{\prime}({\bf r}), \, {\bf r} \in S,
\end{equation}
where ${\bf n}$ denotes the normal vector of the facet. Next, we briefly describe applications of
the different BCs.

\begin{itemize}
\item {\it Multipole BCs} are used for molecule configurations,
ensuring the correct asymptotic behavior of the Hartree potential,
even for charged systems (ions or charged molecules), as shown in Fig.~\ref{fig:bcs}(a).

\item {\it Periodic BCs} is the natural choice along all directions for fully periodic bulk materials, as shown in Fig.~\ref{fig:bcs}(b). Periodic BCs are also often used to model heterostructures or interfaces, as well as surfaces using a slab model.

\item {\it Dirichlet-Neumann BCs for a slab model}. In slab calculations, it can be more advantageous to impose mixed BCs, such as Neumann (fixed potential gradient) and Dirichlet (fixed potential) on the left- and right-hand side of the slab, respectively, combined with periodic BCs in the in-plane-directions, as shown in Fig.~\ref{fig:bcs}(c). These mixed BCs provide a physically sound alternative to the often-used dipole correction for slab calculations.\cite{neugebauer1992adsorbate}

\item {\it Dirichlet-Neumann BCs for a surface configuration}. For accurate surface simulations, the surface configuration may be used,
in combination with mixed BCs: Neumann in the right-hand-side vacuum region, Dirichlet at the left electrode, and periodic BCs in the in-plane directions, see Fig.~\ref{fig:bcs}(d). In this case, one can account, e.g., for the charge transfer from the near-surface region to the semi-infinite electrode, which acts as an electron reservoir.\cite{Smidstrup2017}

\item {\it Dirichlet BCs for a device configuration}. Two-probe device simulations are in QuantumATK done using Dirichlet BCs at the left and right boundaries to the electrodes. Periodic BCs may then be applied in the directions perpendicular to the electron transport direction, as shown in Fig.~\ref{fig:bcs}(e). For complex devices, one may need to apply a more mixed set of BCs, as discussed in the following.

\item {\it General mixed BCs}. {QuantumATK also allows for combining Neumann, Dirichlet and periodic BCs. This can be used to, e.g., model a 2D device in a field-effect transistor setup, such as that in Fig.~\ref{fig:2Ddevice_structure}.}
\end{itemize}

We note that for systems with periodic or Neumann BCs in all directions,
the Hartree potential can only be determined up to an additive constant. In this case,
in order to obtain a uniquely defined solution, we require the average
of the Hartree potential to be zero when solving the Poisson equation.

\subsubsection{Poisson solvers}
To handle such different BCs, the QuantumATK simulation engines
use Poisson solvers based on either FFT methods
or real-space finite-difference (FD) methods.
The FD methods are implemented using a multigrid solver,\cite{holst1993multigrid}
a parallel conjugate-gradient-based solver,\cite{concus1976generalized}
and the MUMPS direct solver.\cite{mumps}
The real-space methods also allow for specifying spatial regions
with specific dielectric constants or values of the electrostatic potential,
as mentioned above.

For systems with 2D or 3D periodic BCs, and no dielectric regions or metallic gates,
the Poisson equation \eqref{eq:poisson_general} is most efficiently solved using the FFT solvers.
For a bulk configuration with 3D periodic directions,
we use a 3D-FFT method, see Fig.~\ref{fig:bcs}(b).
In the case of only 2 periodic directions, for example in slab models, surface configurations, and device configurations,
we use a 2D-FFT method combined with a 1D finite-difference method,
see Figs.~\ref{fig:bcs}(c)-(d).\cite{ozaki2010efficient}


\section{Semi-Empirical Models}
\label{sec:tb}
%
As a computationally fast alternative to DFT,
the ATK-SE engine allows for semi-empirical TB-type simulations.\cite{stokbro2010semiempirical}
The TB models consist of a non-selfconsistent Hamiltonian
that can be extended with a selfconsistent correction
for charge fluctuations and spin polarization.
These corrections closely follow the density functional tight-binding (DFTB) approach.\cite{elstner1998self}
The main aspects of these TB models have been described in Ref.~\onlinecite{stokbro2010semiempirical}
and below we give only a brief description of the models.
\begin{table}
\caption{\label{tab:tbmodels} Classes of TB models currently supported by ATK-SE. The model types are either two-center Slater--Koster (SK) or based on environment-dependent parameters (Env). The model may be orthogonal ($H$) or non-orthogonal ($H,S$). Short-ranged models include nearest-neighbour interactions only (range up to a few \AA),
while the long-ranged H\"{u}ckel models have a typical range of 5--10~\AA.
As indicated in the right-hand column, not all models support calculation of total energies, forces, and stress,
but are used mainly for simulating the electronic structure of materials.}
\begin{ruledtabular}
\begin{tabular}{lcccc}
Model & Ref. & Type & Range & $E,F,\sigma$ \\
\hline \cline{1-5} \\ [-2ex]
H\"{u}ckel & \onlinecite{ammeter1978counterintuitive} & SK, ($H,S$) &  long & no \\
Empirical TB & \onlinecite{vogl1983semi} &  SK, ($H$) & short & no \\
DFTB & \onlinecite{elstner1998self} & SK, ($H,S$) & medium & yes \\
Purdue & \onlinecite{boykin2002diagonal} & Env, ($H$) & short & no\\
NRL & \onlinecite{bernstein2000energetic} & Env, ($H,S$) & long & yes\\
\end{tabular}
\end{ruledtabular}
\end{table}

Table~\ref{tab:tbmodels} summarizes the available models
for the non-selfconsistent part of the SE Hamiltonian, $H^0_{ij}$.
Most of the models are non-orthogonal, that is,
include a parametrization of the overlap matrix $S_{ij}$.
In most of the models,
the Hamiltonian matrix elements depend only on two centers,
parameterized in terms of Slater--Koster parameters.
These models include H\"{u}ckel models,\cite{ammeter1978counterintuitive,cerda2000accurate}
Slater--Koster orthogonal TB models,\cite{vogl1983semi,jancu1998empirical}
and DFTB models.\cite{elstner1998self}
The ATK-SE engine also supports models that take into account
the position of atoms around the two centers.
These currently include the environment-dependent TB models from Purdue\cite{boykin2002diagonal}
and those from the U.S.\ Naval Research Laboratory.\cite{bernstein2000energetic}

It is possible to add a selfconsistent correction
to the non-selfconsistent TB models.\cite{stokbro2010semiempirical}
The selfconsistent correction use the change
in the onsite Mulliken population of each orbital,
relative to a reference system,
to assign an orbital-dependent charge to each atom.
The charge on the orbital is represented by a Gaussian orbital,
and the width of the Gaussian, $\sigma_l$,
can be related to an onsite repulsion, $U_l$,
where $l$ is the angular momentum of the orbital.
The relation is given by\cite{stokbro2010semiempirical}
\begin{equation}
U_l= \frac{2 e^2}{ \sqrt{\pi}\sigma_l} .
\end{equation}

This onsite repulsion can be calculated from the charge-dependent onsite energies,\cite{elstner1998self}
\begin{equation}
U_l = \frac{d \varepsilon_l}{d n_l},
\end{equation}
where $\varepsilon_l$ is the orbital energy of the atom
and $n_l$ the charge in orbital $l$.
\qatk\ comes with a database of $U_l$ calculated using DFT all-electron simulations of the atom.
In practice, it is more reliable for each element to use a single averaged value,\cite{elstner1998self}
\begin{equation}
    U = \frac{1}{N} \sum_l  n_{l}\, U_{l} ,
\end{equation}
where the average is determined by the number of valence electrons of each orbital, $n_l$; $N=\sum_{l} n_{l}$.
The ATK-SE default is to use such a single value.

In the ATK-SE selfconsistent loop,
the Mulliken population is calculated for each orbital.
Based on the change in charge relative to the reference system,
a Gaussian charge is added at the orbital position.
We note that in the default case,
where an atom-averaged $U$ is used on each orbital,
only changes in the atomic charge will have an affect.
From the atom-centered charge we set up a real-space charge density
from which the Hartree potential $V(\br)$ is calculated
using the same methods as used for DFT,
see Section~\ref{sec:poisson}.
It is added to the TB Hamiltonian through
\begin{equation}
H_{ij} = H^0_{ij}+ \frac{1}{2}  (V({\bf r}_i) +V({\bf r}_j) ) S_{ij} ,
\end{equation}
where ${\bf r}_i$ is the position of orbital $i$.

The ATK-SE engine also supports spin polarization through the term\cite{kohler2007treatment}
\begin{equation}
H_{ij}^\sigma= \pm \frac{1}{2}  S_{ij} \left(dE_{l_i} + dE_{l_j} \right) ,
\end{equation}
where the sign depends on the spin.
The spin splitting of shell $l$, $dE_{l_i}$,
is calculated from the spin-dependent Mulliken populations $m_{l \uparrow}, m_{l\downarrow}$
of each shell at the local site ${\mu_l}$:
\begin{equation}
dE_{l} = \sum_{l' \in \mu_l} W_{l l'} \, (m_{l'\uparrow}- m_{l'\downarrow}).
\end{equation}
The shell-dependent spin-splitting strength $W_{l l'}$
is calculated from a spin-polarized atomic calculation,\cite{kohler2007treatment}
and ATK-SE provides a database with the parameters.

The main advantage of the SE models compared to DFT methods are their computational efficiency.
For large systems, the main computational cost of both DFT and TB simulations
is related to diagonalization of the Hamiltonian,
the speed of which depends strongly on the number of orbitals on each site and their range.
This makes TB Hamiltonians very attractive for large systems,
provided the SE parametrization is appropriate for the particular simulation.
Furthermore, orthogonal Hamiltonians have inherent performance advantages.
The Empirical and Purdue environment-dependent models
are the most popular TB models for electron transport calculations.
We also note that for many two-probe device systems,
it is mainly the band structure and quantum confinement
that determine electrical characteristics such as current-voltage curves.
TB model Hamiltonians can provide good results for such simple devices.
Finally, DFTB models are popular for total-energy calculations,
although we find in general that the accuracy should be cross-checked against DFT.

\section{Empirical Force Fields}
%
\label{sec:ff}
\atkff\ is a state-of-the-art FF simulation engine
that is fully integrated into the Python framework.
This has already been described in detail in Ref.~\onlinecite{ATKForceField},
and we therefore only summarize some of the main features.
\begin{table*}
  \caption{\label{tab:ff}Selected potential models included in \atkff.}
  \begin{ruledtabular}
  \begin{tabular}{lll} 
    Potential model                         & Special properties                                                & References\\
    \hline \cline{1-3} \\ [-2ex]
    \arrayrulecolor{lightgray}
    Stillinger--Weber (SW)                  & Three-body                                                        & \onlinecite{Stillinger1985}\\
    \hline \\ [-2ex]
    Embedded atom model (EAM)               & Many-body                                                         & \onlinecite{mishin2001structural}\\
    \hline \\ [-2ex]
    \multirow{2}{*}{Modified embedded atom model (MEAM)} & Many-body                                            & \multirow{2}{*}{\onlinecite{Baskes1997}}\\
                                            & Directional bonding & \\
    \hline \\ [-2ex]
    Tersoff/Brenner                         & Bond-order                                                        & \onlinecite{Tersoff1988,Brenner2002}\\
    \hline \\ [-2ex]
    \multirow{2}{*}{ReaxFF}                 & Bond-order                                                        & \multirow{2}{*}{\onlinecite{chenoweth2008reaxff}}\\
                                            & Dynamical charges & \\
    \hline \\ [-2ex]
    \multirow{3}{*}{COMB/COMB3}             & Bond-order                                                        & \multirow{3}{*}{\onlinecite{Yu2007}}\\
                                            & Dynamical charges & \\
                                            & Induced dipoles & \\
    \hline \\ [-2ex]
    Core-shell                              & Dynamical charge fluctuations                                     & \onlinecite{mitchell1993shell}\\
    \hline \\ [-2ex]
    Tangney--Scandolo (TS)                  & Induced dipoles                                                   & \onlinecite{tangney2002}\\
    \hline \\ [-2ex]
    \multirow{2}{*}{Aspherical ion model}   & Induced dipoles and quadrupoles                                   & \multirow{2}{*}{\onlinecite{Rowley1998}}\\
                                            & Dynamical ion distortion & \\
    \hline \\ [-2ex]
    Biomolecular and valence force fields   & Static bonds                                                      & \onlinecite{mackerell2004empirical, keating1966effect} \\
    \arrayrulecolor{black}
  \end{tabular}
  \end{ruledtabular}
\end{table*}

Table~\ref{tab:ff} lists the empirical potential models supported by \atkff,
which includes all major FF types.
The simulation engine also allows for combining models,
such that different FFs
can be assigned to different sub-systems.
The empirical potential for each sub-system,
and the interactions between them,
can be customized as desired, again using Python scripting.
\atkff\ currently includes more than 300 predefined literature parameter sets,
which can conveniently be invoked from the \nanolab\ GUI.
Additionally, it is also possible to specify custom FF parameters
via the Potential Editor tool in NanoLab or in a Python script,
or even use built-in Python optimization modules
to optimize the parameters against reference data.

Table~\ref{tab:md_timings} compares the computational speed
of \atkff\ molecular dynamics simulations
to that of the popular LAMMPS package.\cite{Plimpton1995}
For most of the FF potential types,
the two codes have similar performance.

\begin{table*}
\caption{\label{tab:md_timings}\atkff\ timings as compared to LAMMPS\@.\cite{Plimpton1995}
Absolute timings for molecular dynamics (MD) simulations,
in units of microseconds per atom per MD step, using one computing core for all potentials.
Potential abbreviations are defined in Table~\ref{tab:ff}.
In addition, LJ means Lennard--Jones.
More details of the benchmark systems can be found in Ref.~\onlinecite{ATKForceField}.}
\begin{ruledtabular}
\begin{tabular}{lccccccc}
       & LJ  & Tersoff & SW  & EAM & ReaxFF & COMB & TS\\
\hline \cline{1-8} \\ [-1.5ex]
\qatk  & 3.8 & 6.3     & 5.2 & 3.8 & 180    & 320  & 360\\
LAMMPS & 1.9 & 7.8     & 5.2 & 2.4 & 190    & 240  & N/A\\
\end{tabular}
\end{ruledtabular}
\end{table*}

\section{Ion Dynamics}
%
%
%
One very powerful feature of \qatk\ is that ion dynamics is executed
using common modules that are not specific to the chosen simulation engine.
This means that modules for calculating energy, forces, and stress
may be used with any of the supported engines,
including DFT, SE methods, and classical FFs.
Options for ion-dynamics simulations are defined using Python scripting,
which allows for easy customization, extension, and combination
of different simulation methods, without loss of performance.
In Section~\ref{sec:mdefield} we illustrate this
by combining the DFT and FF engines in a single molecular dynamics simulation.
Several methods related to ion dynamics in \qatk\ have been
described in detail in Ref.~\onlinecite{ATKForceField},
so here we only summarize the main features.
%
\subsection{Local Structural Optimization}
\label{sec:optimization}
%
The atomic positions in molecules and clusters are optimized by minimizing the
forces, while for periodic crystals, the unit-cell vectors can also be
included in the optimization, possibly under an external pressure that may be anisotropic.
The simultaneous optimization of positions and cell vectors is based
on Ref.~\onlinecite{Sheppard2012}, where the changes to the system are
described as a combined vector of atomic and strain coordinates.

The default method for optimization is the limited-memory
Broyden--Fletcher--Goldfarb--Shanno (L-BFGS)
quasi-Newton-type minimization algorithm,\cite{Liu1989}
but \qatk\ also implements the fast inertial relaxation engine (FIRE) method.\cite{Bitzek2006}
%
\subsection{Global Structural Optimization}
%
The previous section considered methods for \textit{local} geometry optimization,
which locate the closest local minimum-energy configuration.
However, often the goal is to find the \textit{globally} most stable configuration,
for example, the minimum-energy crystal structure.
\qatk\ therefore implements a genetic algorithm for crystal structure prediction.
It works by generating an initial set of random configurations and then
evolving them using genetic operators, as described in Ref.~\onlinecite{glass2006uspex}.
An alternative approach is to perform simulated annealing using molecular dynamics.\cite{kirkpatrick1983optimization}
%
\subsection{Reaction Pathways and Transition States}
%
The minimum-energy path (MEP) for changes to the atomic positions
from one stable configuration to another
may be found using the nudged elastic band (NEB) method.\cite{jonsson1998nudged}
The \qatk\ platform implements state-of-the-art NEB,~\cite{Henkelman2000}
including the climbing-image method.\cite{Henkelman2000climbing}
The initial set of images are obtained from linear interpolation
between the NEB end points,
or by using the image-dependent pair potential (IDPP) method.\cite{Smidstrup2014}
The IDPP method aims to avoid unphysical starting guesses,
and leads in general to an initial NEB path that is closer to the (unknown) MEP.
This typically reduces the number of required NEB optimization steps by a factor of 2.

In some implementations, the projected NEB forces for each image are optimized
independently. However, the L-BFGS algorithm is in that case known to behave
poorly.\cite{Sheppard2008}
In \qatk, the NEB forces for each image are combined into a single vector,
$\mathbf{F}_\mathrm{NEB} \in \mathbb{R}^{3mn}$,
where $m$ is the number of images and $n$ the number of atoms.
This combined approach is more efficient when used with L-BFGS,
and has been referred to as the global L-BFGS method.\cite{Sheppard2008}

\subsection{\label{sec:dynamics}Molecular Dynamics}
\begin{figure}
\begin{center}
\includegraphics[width=\columnwidth]{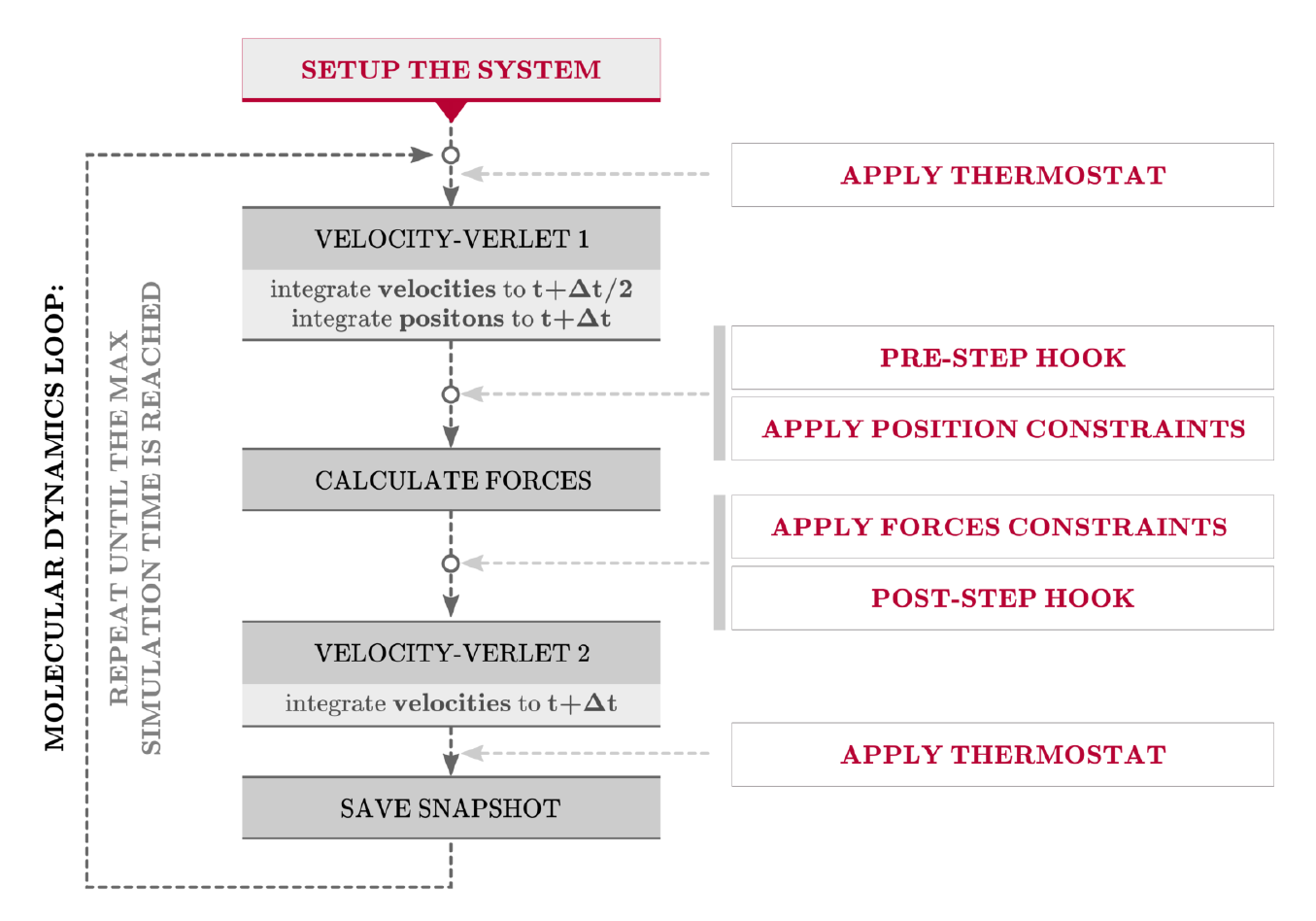}
\caption{Flowchart of a typical \qatk\ MD loop.}
\label{fig:mdloop}
\end{center}
\end{figure}

Molecular dynamics (MD) simulations provide insights into dynamic
atomic-scale processes or sample microscopic ensembles. The essential
functional blocks in a typical \qatk\ MD loop are depicted in
Fig.~\ref{fig:mdloop}. Different thermodynamic ensembles can be simulated.
The basic NVE ensemble uses the well-known velocity-Verlet algorithm.\cite{Swope1982}
Additionally, thermostats or barostats can be applied to
different parts of the system to simulate NVT or NPT ensembles, using for example the
chained Nos\'{e}--Hoover thermostat,\cite{Martyna1992}
an impulsive version of the Langevin thermostat,\cite{Goga2012}
or the barostat proposed by Martyna \textit{et al.} in Ref.~\onlinecite{Martyna1994}
for isotropic and anisotropic pressure coupling.

Figure~\ref{fig:mdloop} also shows that one may apply so-called \textit{pre-step
hooks} and \textit{post-step hooks} during a \qatk\ MD simulation.
These \textit{hook functions} are scripted
in \atkpython, and may vastly increase flexibility with respect to specialized
MD simulation techniques and custom on-the-fly analysis. This makes it easy to
employ predefined or user-defined custom operations during the MD
simulation. The pre-step hook is called before the force calculation, and may modify atomic
positions, cell vectors, velocities, etc. This is often used to implement
custom constraints on atoms or to apply a non-trivial strain to the simulation
cell. The post-step hook is typically used to modify the forces and/or stress. It
may, for example, be used to add external forces and stress contributions, such
as a bias potential, to the regular interaction forces.

\begin{figure}
\begin{center}
\includegraphics[width=\columnwidth]{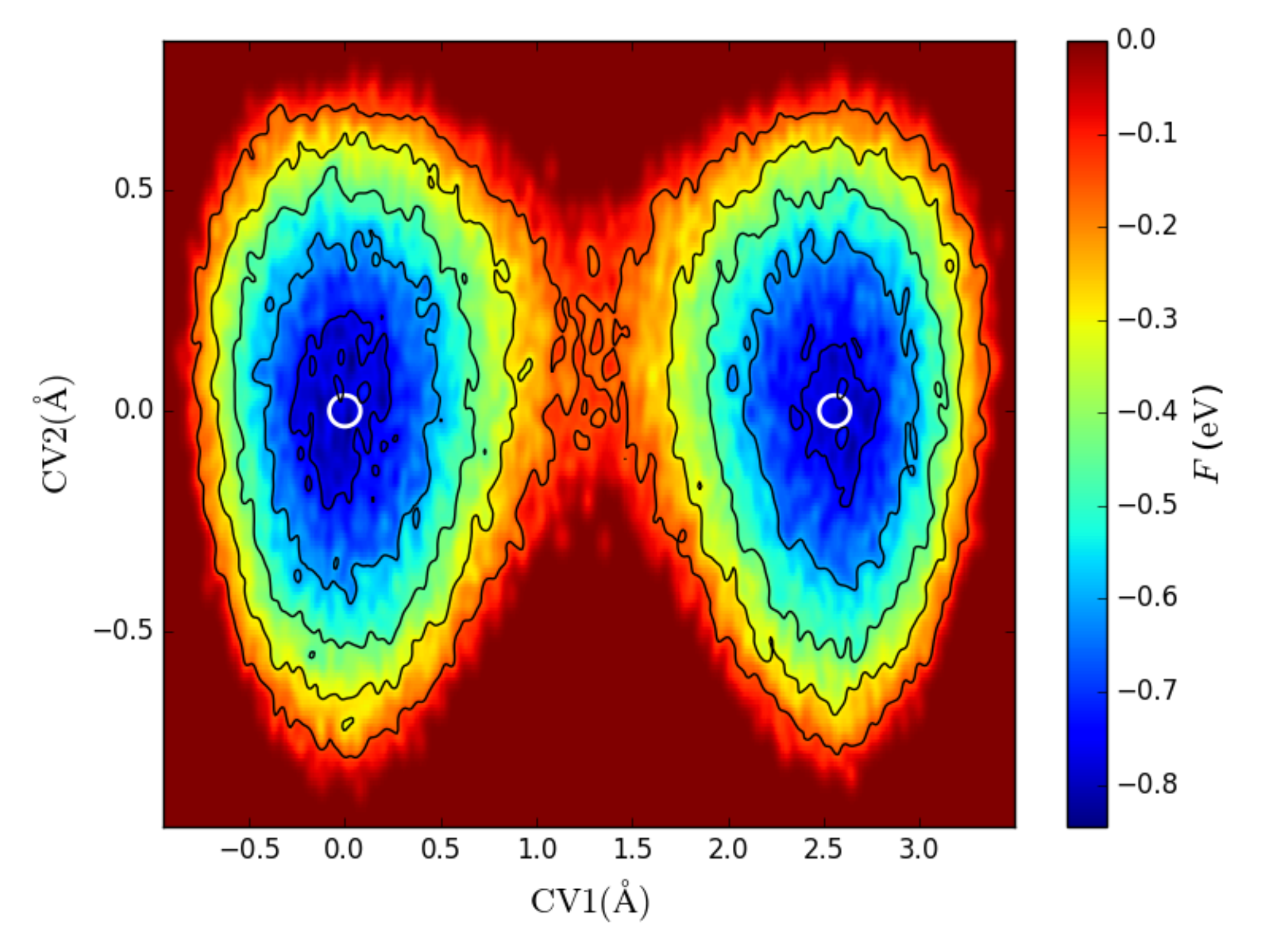}
\caption{Free energy map of a metadynamics simulation of surface vacancy
diffusion on a Cu(111) surface using \qatk.
The collective variables CV1 and CV2 refer to the \textit{x}- and
\textit{y}-position of a surface atom close to the vacancy.
The atom positions of the surface layer of the lattice are
depicted by the white circles.}
\label{fig:plumed}
\end{center}
\end{figure}

\qatk\ is shipped with a number of predefined hook functions, implementing
thermal transport via reverse non-equilibrium molecular dynamics
(RNEMD),\cite{MullerPlathe1997} metadynamics, and other methods.
For metadynamics, \qatk\ integrates with the PLUMED package,\cite{Tribello2014}
so that all methods implemented in PLUMED are available in \qatk\ as well.
Figure~\ref{fig:plumed} illustrates the free-energy map
of surface vacancy diffusion on Cu(111) using the \atkff\ engine
with an EAM potential.\cite{Kondati2017}
%
%
%
\subsection{Adaptive Kinetic Monte Carlo}
%
Adaptive kinetic Monte Carlo (AKMC) is an algorithm for modelling the
long-timescale kinetics of solid-state materials.\cite{Henkelman2001,Xu2008,Chill2014AKMC}
For a given configuration, AKMC involves 3 steps: (1) locate all kinetically relevant
product states; (2) determine the saddle point between the reactant
and product states; (3) select a reaction using kinetic Monte Carlo (KMC).

Step 1 is in \qatk\ performed using high-temperature MD.
At regular intervals, the MD simulation is stopped
and a geometry optimization is performed to check
if the system has left the initial basin.
This procedure is repeated until all relevant reactions
are found within a user-specified confidence.\cite{Chill2014AKMC,Aristoff2016}

In step 2, the saddle-point geometry for each reaction
is determined by performing a NEB optimization for each
reaction, and the reaction rates $k$ are determined via harmonic transition-state
theory (HTST),\cite{Vineyard1957}
\begin{equation}
k_{\rm HTST} = \frac{\prod_i^{3N} \nu_i^\text{min}}{\prod_i^{3N-1} \nu_i^\ddagger}
    \exp \left[ -\left(E^\ddagger - E^\text{min}\right)/k_\text{B} T \right],
    \label{eqn:htst-rate}
\end{equation}
where $N$ is the number of atoms, $\nu^\text{min}_i$ and $\nu^\ddagger_i$ are
the positive (stable) normal-mode frequencies at the minimum and saddle points,
$E^\text{min}$ and $E^\ddagger$ the corresponding energies,
$k_\text{B}$ is the Boltzmann constant, and $T$ the temperature.
The ratio of the products of the vibrational frequencies in \eqref{eqn:htst-rate}
is often called the attempt frequency or the prefactor,
and can be computationally expensive to obtain.
Instead of calculating the prefactor for each reaction mechanism,
a user-given value may therefore be used.

Finally, in step 3, a reaction is selected using KMC, the system evolves to
the corresponding product configuration, and the entire procedure is repeated.
More details of the \qatk\ implementation of AKMC
may be found in Ref.~\onlinecite{Chill2014AKMC}.

\section{Phonons}
\label{sec:phonons}
The ground-state vibrational motion of atoms is of paramount interest in modern
materials science. Within the harmonic approximation, which is valid for small
thermal displacements of atoms around their equilibrium position,
the vibrational frequencies of a configuration are eigenvalues of the dynamical matrix $D$,
\begin{equation}
\label{eq:DynamicalMatrix}
    D_{a, \alpha; b, \beta} =
        \frac{1}{\sqrt{m_a m_b}}\frac{dF_{b, \beta}}{dr_{a, \alpha}} ,
\end{equation}
where $m_a$ ($m_b$) is the atomic mass of atom $a$ ($b$) and $dF_{b, \beta}/dr_{a, \alpha}$ is the
force constant.
Computing and diagonalizing $D$ yields the vibrational modes of the
system (molecular or periodic), and is also used to obtain the phonon density
of states for a periodic crystal.
%
\subsection{Calculating the Dynamical Matrix}
%
\qatk\ calculates the dynamical matrix using a FD method,
where each matrix element in \eqref{eq:DynamicalMatrix}
is computed by displacing atom $a$ along Cartesian direction $\alpha$,
and then calculating the resulting forces on atom $b$ along directions $\beta$.
This approach is sometimes referred to as the frozen-phonon or supercell method,
and applies equally well to isolated (molecular) systems. The method lends
itself to heavy computational parallelization over many computing cores, since
all displacements may be calculated independently. Crystal symmetries are
taken into account in that only symmetrically unique atoms in the unit
cell are displaced, and the forces resulting from displacement of the
equivalent atoms are obtained using the corresponding symmetry
operations.\cite{Alfe2009}
%
\subsection{Wigner--Seitz Method}
%
For crystals with small unit cells, periodic repetition of the cell
is usually needed to accurately account for long-range interactions in $D$.
For larger simulation cells, including cells with defects and amorphous structures,
this is not always necessary, since the cell may already include the entire interaction range.
In order to recover the correct phonon dispersion across periodic boundaries, the
Wigner--Seitz method can be employed. Here, a Wigner--Seitz cell is centered
around the displaced atom and the forces on each atom in the simulation cell
is assigned to its periodic image that is located within the Wigner--Seitz
cell.\cite{Parlinski1997}
%
\subsection{Phonon Band Structure and Density of States}
%
The phonon band structure (or phonon dispersion)
consists of bands with index $\lambda$ of vibrational frequencies $\omega = \omega_{\lambda\bq }$
throughout the Brillouin zone (BZ) of phonon wave vectors $\bq$.
The phonon density of states (phonon DOS) per unit cell, $g(\omega)$, is defined as
\begin{equation}
g(\omega) = \frac{1}{N} \sum_{\bq \lambda} \delta(\omega - \omega_{\lambda\bq }),
\end{equation}
where $N$ is the number of $\bq$-points in the sum. In practice, the phonon DOS is calculated using the tetrahedron method.\cite{Blochl1994}
Additionally, quantities such as vibrational free energy, entropy, and zero-point energy can easily be calculated from the vibrational modes and energies.

\begin{figure}[!htbp]
\centering
{\includegraphics[width=0.99\linewidth]{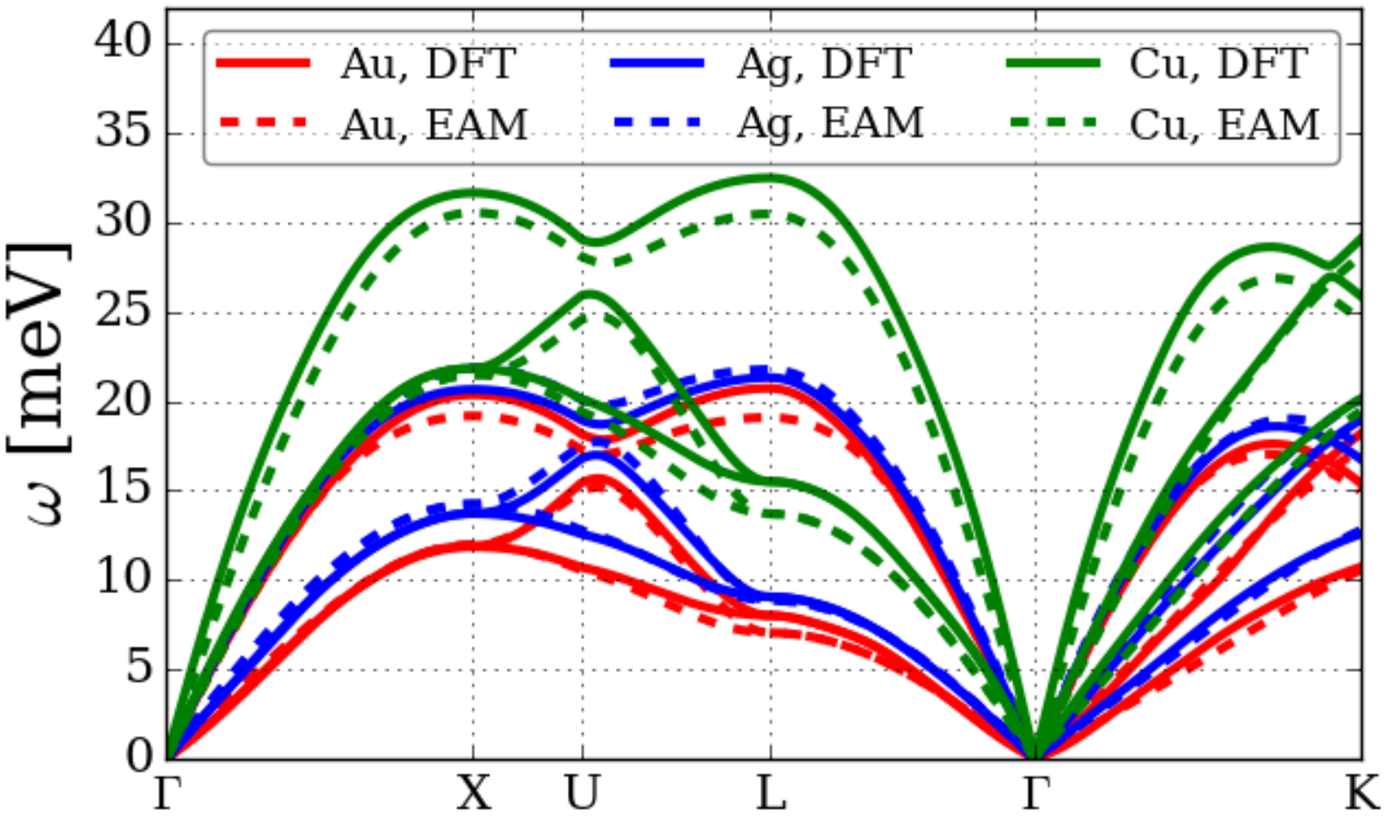}}
\caption{
Phonon dispersions of the three FCC metals Au, Ag and Cu, obtained from supercell calculations using the \atklcao\ engine and the SG15-M LCAO basis set. Supercells were generated from $9\times9\times9$ repetitions of the primitive cells.}
\label{fig:PhononBand}
\end{figure}

Figure~\ref{fig:PhononBand} gives an example of phonon simulations
for different metals using the \atkff\ and \atklcao\ engines.
The \atklcao\ supercell calculation yields accurate vibrational properties, as exemplified by the excellent agreement between the two methods. The dispersions follow the same trends, which is expected, since the three metals have the same FCC crystal symmetry. We note that the higher phonon frequencies in Cu can be understood from the similar bond strength as in Ag and Au, but significantly lower Cu atomic mass.
%
%
\subsection{Electron-Phonon Coupling}
%
The electron-phonon coupling (EPC) is an imporant quantity in modern electronic-structure theory.
It is, for example, used to calculate the transport coefficients in bulk crystals (see Section~\ref{sec:bulktransport})
and inelastic scattering of electrons in two-probe devices (see Section~\ref{sec:inelastic}).

To obtain the EPC, we calculate the derivative of the Hamiltonian matrix with respect to the position of atom $a$, $\br_a$,
\begin{eqnarray}
(\delta \hat{H}_{\br_a})_{ij} = \langle \phi_i | \frac{\partial \hat{H}}{\partial \br_a} | \phi_j \rangle \, ,
\label{eq:dHdR-matrix}
\end{eqnarray}
where $\partial \hat{H} / \partial \br_a$ is calculated using finite differences,
similar to the calculation of the dynamical matrix described above.
A unit cell is repeated to form a supercell
(for a device configuration, only the atoms in the central region are displaced).
The terms that contribute to the Hamiltonian derivative
is the local and non-local PP terms.
The real-space Hamiltonian matrix is expanded in electron eigenstates,
$n \mathbf{k}$, and Fourier transformed using the phonon polarization vectors,
to finally obtain the electron-phonon couplings $g$,
\begin{eqnarray}
g_{\bk \bk' \bq}^{\lambda n n'} = \langle n'\bk' | \delta \hat{H}_{\lambda\bq} | n\bk \rangle ,
\label{eq:M-definition}
\end{eqnarray}
where $\bq$ is the phonon momentum and $\lambda$ the phonon branch index.

Further details of how \qatk\ calculates the EPC
are given in Ref.~\onlinecite{gunst_first-principles_2016}.
%
\subsection{Transport Coefficients}
\label{sec:bulktransport}
%
The electron/hole mobility in a semiconductor material is an important
quantity in device engineering,
and also determines the conductivity of metals.
Electronic transport coefficients for bulk materials,
including the conductivity, Hall conductivity, and thermoelectric response,
may be calculated from the Boltzmann transport equation (BTE)
as linear-response coefficients related to the application
of an electric field, magnetic field, or temperature gradient.
In \qatk, this is done by expanding the current density $\mathbf{j}$ to lowest order
in the electric field $\mathcal{E}$, magnetic field $B$, and temperature gradient $\nabla T$,
\begin{eqnarray}
j_{\alpha} = \sigma_{\alpha \beta}\mathcal{E}_\beta  + \sigma_{\alpha \beta \gamma} \mathcal{E}_\beta B_\gamma + \nu_{\alpha \beta} \nabla_\beta T ,
\label{eqn:LinearResponse}
\end{eqnarray}
where the indices label Cartesian directions and $\sigma_{\alpha \beta}$, $\sigma_{\alpha \beta \gamma}$ and $\nu_{\alpha \beta}$ are the electronic conductivity, Hall conductivity, and thermoelectric response, respectively.
Following Ref.~\onlinecite{madsen_boltztrap_2006},
the band-dependent thermoelectric transport coefficients and Hall coefficients are obtained as
\begin{eqnarray}
\sigma_{\alpha \beta}(n\mathbf{k}) &=& e^2\tau_{n\mathbf{k}}\mathbf{v}_{\alpha}(n\mathbf{k})\mathbf{v}_{\beta}(n\mathbf{k})\,,\nonumber\\
\sigma_{\alpha \beta \gamma}(n\mathbf{k}) &=& e^3\tau^2_{n\mathbf{k}}\epsilon_{\gamma uv}\mathbf{v}_{\alpha}(n\mathbf{k})\mathbf{v}_{v}(n\mathbf{k})\mathbf{M}^{-1}_{\beta u}(n\mathbf{k}) \,, \nonumber \\
\nu_{\alpha \beta}(n\mathbf{k}) &=&   (\varepsilon_{n\mathbf{k}}-\mu) e/T \; \tau_{n\mathbf{k}}\mathbf{v}_{\alpha}(n\mathbf{k})\mathbf{v}_{\beta}(n\mathbf{k}) \, ,
\label{eqn:LinearResponseTensors}
\end{eqnarray}
where $\mu$ is the chemical potential and $\epsilon_{\gamma uv}$ the Levi--Civita symbol.
The band group velocities $\mathbf{v}(n\bk)$ and effective mass tensors $\mathbf{M}(n\bk)$ are obtained from perturbation theory.
Importantly, we may in \eqref{eqn:LinearResponseTensors} include the full scattering rate $\tau_{n\mathbf{k}}$,
and thereby go beyond the constant scattering-rate approximation used in Ref.~\onlinecite{madsen_boltztrap_2006}.
As we will see in Section~\ref{sec:mobility}, this may not only be important in order to obtain quantitatively correct results;
it is also required to reproduce experimental trends in the conductivity of different materials.

The scattering rate is given by
\begin{eqnarray}
\frac{1}{\tau_{n \mathbf{k} }} &=& \sum_{n' \lambda \mathbf{q} } \left[B^{nn'}_{\mathbf{k} (\mathbf{k}+\mathbf{q})} P^{\lambda nn'}_{\mathbf{k}(\mathbf{k}+\mathbf{q}) \mathbf{q}}+B^{nn'}_{\mathbf{k}(\mathbf{k}-\mathbf{q})} \bar{P}^{\lambda nn'}_{\mathbf{k}(\mathbf{k}-\mathbf{q})\mathbf{q}}\right] ,
\label{eqn:tau}
\end{eqnarray}
where $B$ is a temperature-dependent scattering weight,
\begin{eqnarray}
B^{nn'}_{\mathbf{k} \mathbf{k'}}= \frac{1-f_{n'\mathbf{k}'}}{1-f_{n\mathbf{k}}} \left[1-\rm{cos}(\theta_{\mathbf{k}\mathbf{k'}})\right] ,
\label{eqn:RTA_rate}
\end{eqnarray}
where $f$ is the Fermi function, and the scattering angle is defined by
\begin{eqnarray}
\rm{cos}(\theta_{\mathbf{k}\mathbf{k'}}) = \frac{\mathbf{v}(n' \mathbf{k}' )\cdot \mathbf{v}(n\mathbf{k})}{|\mathbf{v}(n' \mathbf{k}')| |\mathbf{v}(n\mathbf{k})|} .
\end{eqnarray}
Furthermore, $P$ ($\bar{P}$) are transition rates due to phonon absorption (emission).
They are obtained from Fermi's golden rule,
\begin{eqnarray}
P^{\lambda n n'}_{\mathbf{k}\mathbf{k}' \mathbf{q}}&=&\frac{2\pi}{\hbar} |g_{\mathbf{k}\mathbf{k}'\mathbf{q}}^{\lambda n n'}|^2
n_{ \lambda \mathbf{ q} }\, \delta \! \left(\varepsilon_{n'\mathbf{k}'}-\varepsilon_{n\mathbf{k}}-\hbar \omega_{\lambda\mathbf{q} } \right), \label{eqn:FGRtransitionRate} \\
\bar{P}^{\lambda n n'}_{\mathbf{k}\mathbf{k}' \mathbf{q}}&=&\frac{2\pi}{\hbar}  |g_{\mathbf{k}\mathbf{k}'-\mathbf{q}}^{\lambda n n'}|^2
(n_{\lambda -\mathbf{q}}+1) \delta \! \left(\varepsilon_{n'\mathbf{k}'}-\varepsilon_{n\mathbf{k}}+\hbar \omega_{\lambda-\mathbf{q} } \right), \nonumber
\end{eqnarray}
where $n_{\lambda \mathbf{q} }$ is the phonon occupation operator,
and $g_{\mathbf{k}\mathbf{k}'\mathbf{q}}^{\lambda n n'}$ the EPC constant from \eqref{eq:M-definition}.

\begin{figure}[!htbp]
\centering
{\includegraphics[width=0.99\linewidth]{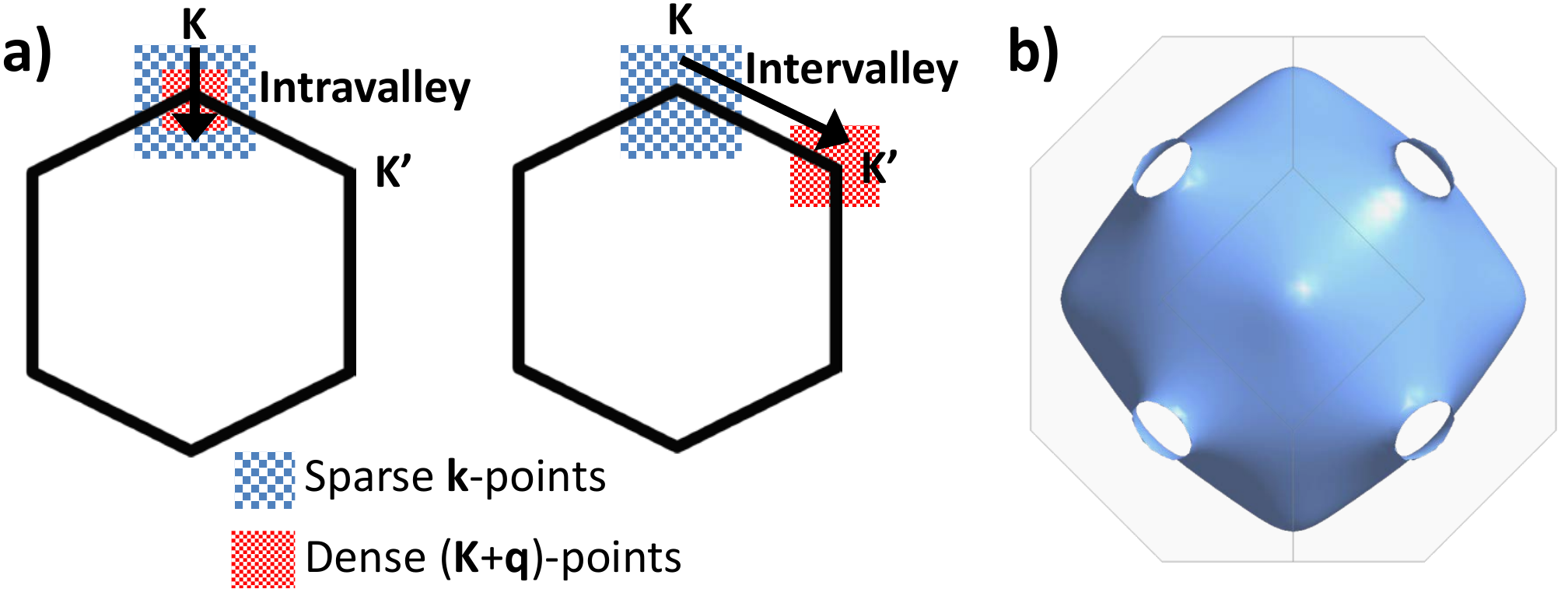}}\\
{\includegraphics[width=0.99\linewidth]{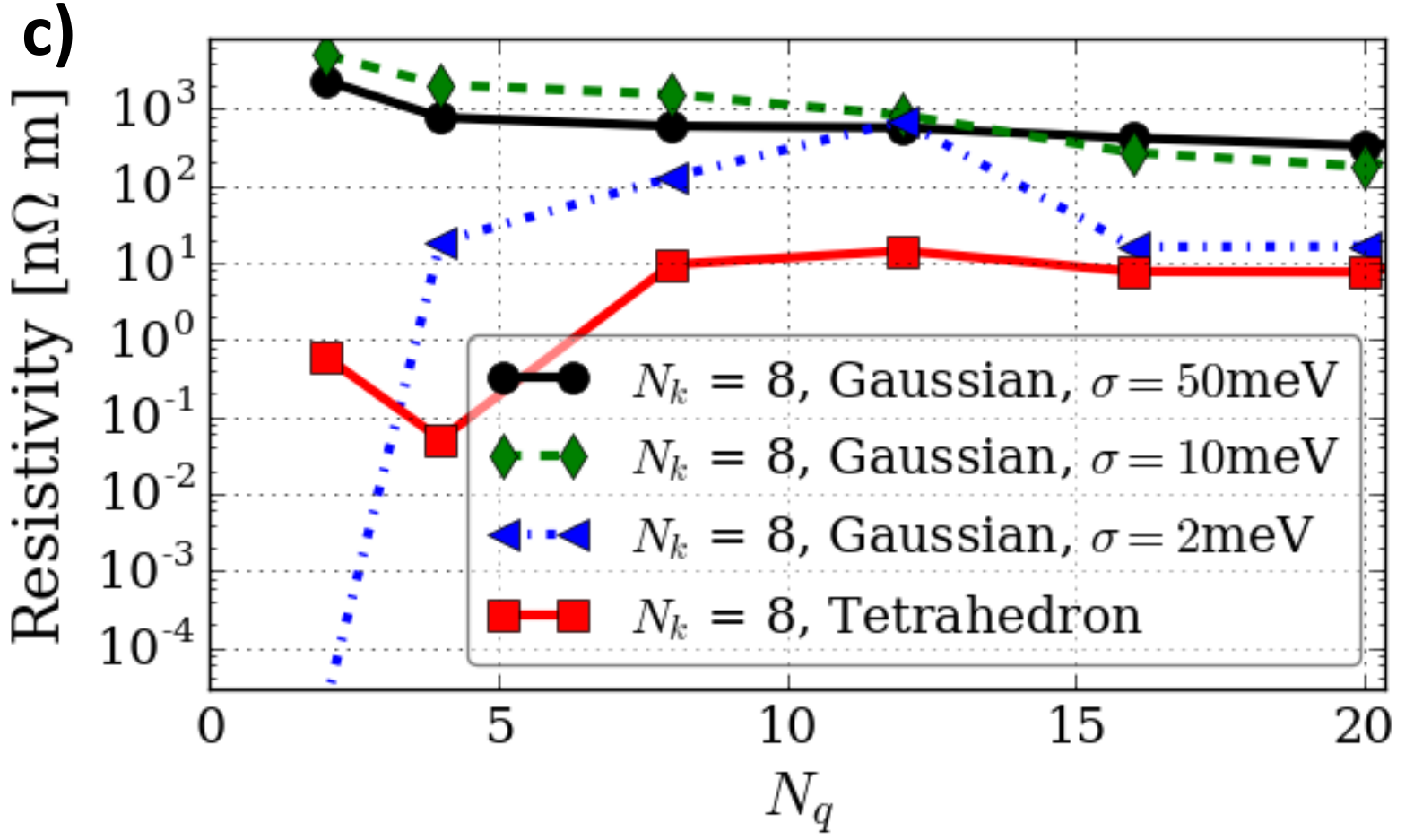}}
\caption{
(a) Illustrative $\bk$- and $\bq$-point selections in the Brillouin zone for the case of a two-dimensional semiconductor with two valleys ($K$ and $K'$). In semiconductors, it is possible to make a clever selection of $\bk$- and $\bq$-points to minimize the computational load while including all relevant scattering processes. Typically, a sparse $\bk$-point sampling is used for the mobility integral, while a denser $\bq$-point sampling is needed to secure a correct scattering rate at each $\bk$-point.
(b) Fermi-surface of bulk Au. In metals, $\bk$-points contributing to the neighborhood of the Fermi surface are not located in a small subset of the Brillouin zone. Therefore, $\bk$- and $\bq$-points are sampled in the full Brillouin zone, and $\bq$-space is integrated using the tetrahedron method to minimize the sampling density.
(c) Resistivity convergence with respect to the number of $\bq$-points for bulk Au with either a direct or tetrahedron integration over the full Brillouin zone. Resistivities were calculated with $N_k\times N_k\times N_k$ $\bk$-points for a sequence of $N_q\times N_q \times N_q$ $\bq$-points.
}
\label{fig:kANDqpoints}
\end{figure}

\qatk\ offers two different methods for performing the $\bq$-integral in \eqref{eqn:tau}.
In the first method, the delta functions in \eqref{eqn:FGRtransitionRate} are represented
by Gaussians with a certain width,
and we perform the discrete sum over $\bq$.
In the second method, we realize that the integral closely resembles
the numerical problem of obtaining a density of states,
and use the tetrahedron method\cite{Blochl1994} for the integration.
In particular for metals, we find the tetrahedron method to be most efficient.
Figure~\ref{fig:kANDqpoints}(c) shows the convergence of the Au resistivity
as the number of $\bq$-points increases, using both Gaussian and tetrahedron integration.
The tetrahedron calculation seems converged for $N_q=20$, that is, a $20\times 20 \times 20$ $\bq$-point sampling.
The result with a finite Gaussian broadening may converge fast if using a rather large broadening,
but the resistivity then appears to converge to a wrong result.
In general, we therefore recommend the tetrahedron integration method for calculation of metallic resistivity.

To further improve the computational performance when calculating transport coefficients,
it is possible to use the energy-dependent isotropic-scattering-rate approximation,
introduced in Ref.~\onlinecite{samsonidze2018thermoelectric}.
A two-step procedure is used for the $\bk$-point sampling,
which significantly reduces simulation time without affecting the
resulting mobilities for many materials
(those that have a fairly isotropic scattering rate in momentum space).
In step one, an initial $\bk$-space with a low sampling density and a well-converged $\bq$-point sampling are used.
The initial $\bk$-point grid is automatically reduced further
by including only $\bk$-points where the band structure has energies in a specific range around the Fermi level.
This limits the simulations to the relevant range of initial states (and relevant carrier densities),
which significantly increases simulation speed and reduces memory usage.
Typically, the variation of the scattering rates from the different directions in momentum space
will be small.
Fom the obtained data, we may therefore generate an isotropic scattering rate that only depends on energy,
\begin{eqnarray}
\frac{1}{\tau(E)} = \frac{1}{n(E)}\sum_{n\mathbf{k}}\frac{1}{\tau_{n\mathbf{k}}} \delta(E_{n\mathbf{k}}-E) ,
\label{eqn:ScatteringRateEnergyDependent}
\end{eqnarray}
where we have integrated over bands $n$ and wave vectors $\mathbf{k}$, and $n(E)$ is the density of states.
In the second step, we then perform a calculation on a fine $\bk$-point grid,
but using the energy-dependent isotropic scattering rate $\tau(E)$. 
Since the scattering rate often varies slowly
on the Fermi surface (for metals),
this is a good approximation.
The second step therefore requires only an evaluation of band velocities and effective masses
on the dense $\bk$-point grid,
while the scattering rate is reused.
This two-step procedure, combined with either direct integration for semiconductors and semimetals,
or tetrahedron integration for metals,
makes \qatk\ an efficient platform for simulating phonon-limited mobilities of materials.

In addition, it is possible to input a predefined scattering rate as a function of energy.
This is relevant for adding extra scattering mechanisms, for example impurity scattering,
on top of the electron-phonon scattering,
or in the case where a scattering-rate expression is known analytically.
One special case of the last situation is the limit of a constant relaxation time,
which is the basis of the popular Boltztrap code.\cite{madsen_boltztrap_2006}
We note that such constant-relaxation-time calculations are easily performed within
the more general \qatk\ framework outlined above.
Moreover, since electron velocities are calculated from perturbation theory,
accuracy is not lost due to band crossings,
which is the case when velocities are obtained from FD methods,
as is done in Ref.~\onlinecite{madsen_boltztrap_2006}.
In some cases, the constant relaxation time approximation
can give a good first estimate of thermoelectric parameters
for a rough screening of materials,
but for quantitative predictions, the more accurate models
of the relaxation time outlined above must be used.

\section{Polarization and Berry Phase}
%
\label{sec:berryphase}
Electronic polarization in materials has significant interest,
for example in ferroelectrics,
where the electric polarization $\mathbf{P}$ can be controlled by application
of an external electric field,
or in piezoelectrics,
where charge accumulates in response to an applied mechanical stress
or strain.\cite{KingSmith1993}

It is common to divide the polarization into ionic and electronic parts, $\mathbf{P}=\mathbf{P}_\mathrm{i} + \mathbf{P}_\mathrm{e}$.
The ionic part can be treated as a classical electrostatic sum of point charges,
\begin{equation}
\mathbf{P}_\mathrm{i} = \frac{|e|}{\Omega}\sum_a Z_a^\mathrm{ion} \br_a ,
\end{equation}
where $Z_a^\mathrm{ion}$ and $\br_a$ are the valence charge
and position vector of atom $a$,
$\Omega$ is the unit-cell volume,
and the sum runs over all ions in the unit cell.

The electronic contribution to the polarization in direction $\alpha$ is obtained as\cite{KingSmith1993}
\begin{eqnarray}
\mathbf{P}_{\mathrm{e},\alpha} = -\frac{|e|}{\Omega}\frac{\Phi_{\alpha}}{2\pi}\mathbf{R}_{\alpha},
\end{eqnarray}
where $\mathbf{R}_{\alpha}$ is the lattice vector in direction $\alpha$, and the Berry phase $\Phi_{\alpha}$ is obtained as
\begin{eqnarray}
\Phi_{\alpha} = \frac{1}{N_\bot} \sum_{\mathbf{k}_{\bot}} \phi_{\alpha}(\mathbf{k}_{\bot}) ,
\end{eqnarray}
where the sum runs over $N_\bot$ $\mathbf{k}_{\bot}$-points in the BZ plane perpendicular to $\mathbf{R}_{\alpha}$, and
\begin{eqnarray}
\phi_{\alpha}(\mathbf{k}_{\bot}) = 2\, \textrm{Im}\left[\textrm{ln} \prod_{j=0}^{J-1} \textrm{det} \mathcal{S}(\mathbf{k}_{j},\mathbf{k}_{j+1}) \right] ,
\end{eqnarray}
with the overlap integrals
\begin{equation}
 \mathcal{S}_{nm}(\mathbf{k}_{j},\mathbf{k}_{j+1})=\braket{\mathbf{u}_{\mathbf{k}_j n}}{\mathbf{u}_{(\mathbf{k}_{j+1})m}} ,
\end{equation}
and with the $J$ $\bk$-points given by $\mathbf{k}_{j}=\mathbf{k}_{\bot}+\mathbf{k}_{\parallel, j}$ lying on a line along the $\mathbf{R}_{\alpha}$ direction.

The polarization depends on the coordinate system chosen since it is related to the real-space charge position, and is determined by the Berry phase, which is only defined modulo $2\pi$. Consequently, the polarization is a periodic function and constitutes a polarization lattice itself. The polarization lattice in direction $\alpha$ is written as
\begin{eqnarray}
\mathbf{P}_{\alpha}^{(n)} = \mathbf{P} + n \mathbf{P}_{Q, \alpha},
\end{eqnarray}
where $n$ is an integer labeling a polarization branch, and the polarization quantum in direction $\alpha$ is $\mathbf{P}_{Q, \alpha} = \frac{|e|}{\Omega}\mathbf{R}_{\alpha}$.
All measurable quantities are related to changes in the polarization, which is a uniquely defined variable, provided that the different polarization values are calculated for the same branch in the polarization lattice.

\qatk\ supports calculation of the polarization itself, as well as the derived quantities piezoelectric tensor,
\begin{eqnarray}
\epsilon_{i\alpha}=\frac{\partial \mathbf{P}_\alpha}{\partial \epsilon_{i}},
\end{eqnarray}
where Voigt notation is used for the strain component, that is, $i \in (xx, yy, zz, yz, xz, xy)$,
and the Born effective charge tensor
\begin{eqnarray}
Z_{a, \alpha\beta}^*=\frac{\partial \mathbf{P}_\alpha}{\partial \br_{a, \beta}},
\end{eqnarray}
where the derivative is with respect to the position of atom $a$ in direction $\beta$.

Table~\ref{tab:born-charges} shows calculated values of the Born effective charges (only the negative components for each structure) and elements of the piezoelectric tensor for III-V wurtzite nitrides and zincblende GaAs. The calculated Born effective charges and piezoelectric tensor components agree well with the reference calculations.

\begin{table}
\caption{\label{tab:born-charges} Born effective charges ($Z^*$)
and piezoelectric tensor components ($\epsilon_{33}$ and $\epsilon_{14}$) for III-V wurtzite nitrides and zincblende GaAs.
Reference vales for the nitrides are from Ref.~\onlinecite{Bernardini1997} and from Ref.~\onlinecite{KingSmith1993} for GaAs.
QuantumATK calculations were performed using the DFT-LCAO engine with the LDA XC functional and a DZP basis set.}
\begin{ruledtabular}
\begin{tabular}{l|cc|cc}
 & \multicolumn{2}{c|}{$Z^*$}  & \multicolumn{2}{c}{$\epsilon_{33}$} \\
 & Reference & QuantumATK & Reference & QuantumATK \\
\hline \cline{1-5} \\ [-2ex]
AlN &   $-$2.70 &    $-$2.67 &  1.46 &  1.65 \\
GaN &   $-$2.72 &    $-$2.75 &  0.73 &  0.86 \\
InN &   $-$3.02 &    $-$2.98 &  0.97 &  1.21 \\
    & \multicolumn{2}{c|}{}  & \multicolumn{2}{c}{$\epsilon_{14}$} \\
GaAs &  $-$1.98 &    $-$2.07 &  $-$0.28 & $-$0.26  \\
\end{tabular}
\end{ruledtabular}
\end{table}

\section{Magnetic Anisotropy Energy}
\label{sec:mae}
%
The magnetic anisotropy energy (MAE) is an important quantity in spintronic magnetic devices. The MAE is defined as the energy difference between two spin orientations, often referred to as in-plane ($\parallel$) and out-of-plane ($\perp$) with respect to a crystal plane of atoms, a surface, or an interface between two materials:
\begin{eqnarray}
\textrm{MAE} = E_\parallel - E_\perp \label{eq:MAE-definition}.
\end{eqnarray}
The MAE can be split into two contributions: A classical dipole-dipole interaction resulting in the so-called shape anisotropy, and a quantum mechanical contribution often refered to as the magnetocrystalline anisotropy, which arises as a consequence of spin-orbit coupling (SOC). In this section we will focus on the magnetocrystalline anisotropy and refer to this as the MAE.

There are at least three different ways of calculating the MAE: (i) Selfconsistent total-energy calculations including SOC with the noncollinear spins constrained in the in-plane and out-of-plane directions, respectively, (ii) using the force theorem (FT) to perform non-selfconsistent calculations (including SOC) of the \textit{band-energy difference} induced by rotating the noncollinear spin from the in-plane to the out-of-plane direction, and (iii) second-order perturbation theory (2PT) using constant values for the SOC. While it has been demonstrated that methods (i) and (ii) give very similar results,\cite{Hafner2009, ArnauArXiv181112100B} the 2PT method can lead to significantly different results.\cite{ArnauArXiv181112100B} In \qatk\ we have implemented an easy-to-use workflow implementing the FT method (ii). Using the FT gives the advantage over method (i) that the calculated MAE can be decomposed into contributions from individual atoms or orbitals, which may give valuable physical and chemical insight.

The \qatk\ workflow for calculating the MAE using the FT method is the following:
\begin{enumerate}
\item Perform a selfconsistent spin-polarized calculation.

\item For each of the considered spin orientations
\begin{enumerate}
\item Perform a non-selfconsistent calculation, in a noncollinear spin representation including SOC, using the effective potential and electron density from the polarized calculation but rotated to the specified spin direction.
\item Calculate the band energies $\epsilon_n$ and projection weights $w_{n,p}$.
\end{enumerate}
\item Calculate the total MAE as
\begin{equation}
\textrm{MAE} = \sum_n f_n^\parallel \epsilon_n^\parallel - \sum_n f_n^\perp \epsilon_n^\perp,
\end{equation}
where $f_n^\parallel$ is the occupation factor for band $n$ (including both band and $\bk$-point index) for the $\parallel$ spin orientation and $\epsilon_n^\parallel$ is the corresponding band energy, and likewise for the $\perp$ spin orientation.
\end{enumerate}

The contribution to the total MAE for a particular projection $p$ (atom or orbital projection) is
\begin{equation}
\textrm{MAE}_p = \sum_n f_n^\parallel \epsilon_n^\parallel w_{n,p}^\parallel - \sum_n f_n^\perp \epsilon_n^\perp w_{n,p}^\perp,
\end{equation}
where the projection weight is
\begin{equation}
w_{n,p} = \langle \psi_n|(\mathbf{S}\mathbf{P} + \mathbf{P}\mathbf{S})/2|\psi_n\rangle,
\end{equation}with $|\psi_n\rangle$ being the eigenstate, $\mathbf{S}$ the overlap matrix, and $\mathbf{P}$ the projection matrix. $\mathbf{P}$ is a diagonal, singular matrix with ones in the indices corresponding to the orbitals we wish to project onto and zeros elsewhere.

Table~\ref{tab:mae-Fe-L10} shows the calculated MAE for a number of Fe-based L$1_0$ alloys. Atomic structures as well as reference values calculated with SIESTA and VASP using the FT method are from Ref.~\onlinecite{ArnauArXiv181112100B}. We first note that the calculated MAEs agree rather well among the four codes, the only exception being FeAu, where the LCAO representations give somewhat smaller values than obtained with PW expansions. In this case it seems that the LCAO basis set has insufficient accuracy, which could be related to the fact that the LCAO basis functions are generated for a scalar-relativistic PP derived from a fully relativistic pseudopotential.

\begin{table}
\caption{\label{tab:mae-Fe-L10} MAE (in units of meV) for various Fe-based L$1_0$ phases. Atomic structures and reference results (SIESTA and VASP) are from Ref.~\onlinecite{ArnauArXiv181112100B}. The \qatk\ selfconsistent and non-selfconsistent calculations were performed with a $17\times\times17\times14$ $\bk$-point grid, while the band energies were sampled on a $40\times40\times34$ $\bk$-point grid. PseudoDojo pseudopotentials were used for both LCAO and PW calculations. The High basis set was used for LCAO.}
\begin{ruledtabular}
\begin{tabular}{lcccc}
Structure & SIESTA$^a$ & VASP$^a$ & QuantumATK & QuantumATK \\
 & LCAO & PW & LCAO & PW \\
\hline \cline{1-5} \\ [-2ex]
FeCo &  0.45 &  0.55 &  0.66 & 0.66 \\
FeCu &  0.42 &  0.45 &  0.45 & 0.45 \\
FePd &  0.20 &  0.13 &  0.12 & 0.15 \\
FePt &  2.93 &  2.78 &  2.43 & 2.57\\
FeAu &  0.36 &  0.62 &  0.22 & 0.56
\end{tabular}
\end{ruledtabular}
\end{table}

\begin{figure}
\begin{center}
\includegraphics[width=\columnwidth]{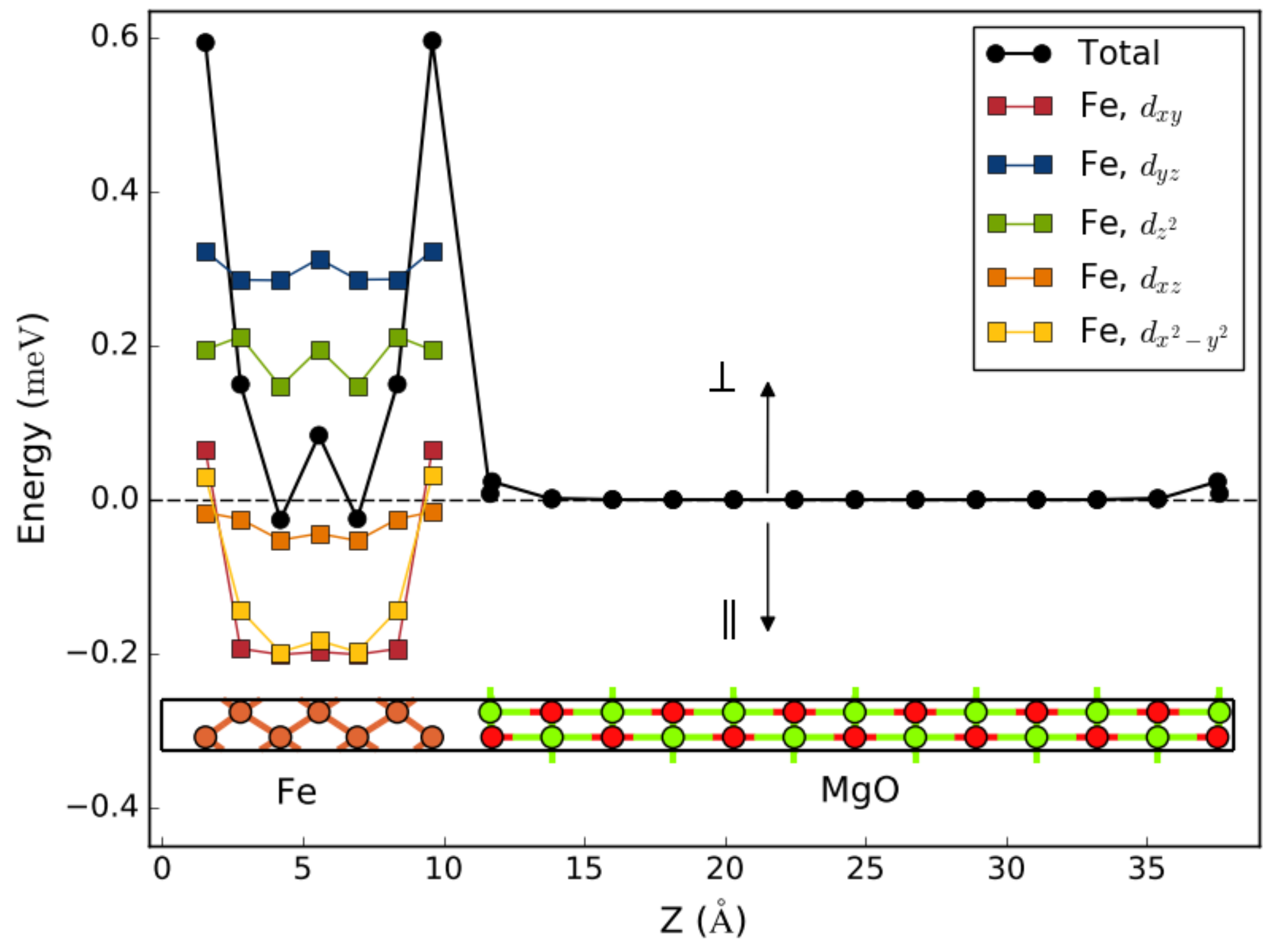}
\caption{MAE for a Fe/MgO interface, calculated using the \qatk\ implementation of the FT method. The total MAE is 1.59~meV, in close agreement with previous results obtained with VASP (1.56~meV).\cite{MasudaPRB2017} The black circles show the atom-projected MAE for all the atoms, while the colored squares show the projection onto the Fe $d$-orbitals, which contribute the most to the total MAE. Positive energies correspond to perpendicular ($\perp$) magnetization, while negative energies correspond to in-plane ($\parallel$) magnetization.}
\label{fig:mae}
\end{center}
\end{figure}

Figure~\ref{fig:mae} shows the atom- and orbital-projected MAE for a Fe/MgO interface. The structure is similar to the one reported in Ref.~\onlinecite{MasudaPRB2017}. We use periodic BCs in the transverse directions. The calculated interfacial anisotropy constant $K_1=\textrm{MAE}/(2A)$, where $A$ is the cross-sectional area, is $K_1=1.41$~mJ/m$^2$, in close agreement with a previous reported value\cite{MasudaPRB2017} of $K_1=1.40$~mJ/m$^2$. From the atom-projected MAE (black circles) it is clear that the interface Fe atoms favor perpendicular MAE (since $\textrm{MAE}>0$), while the atoms in the center of the Fe slab contribute with much smaller values. From the orbital projections it is evident that the MAE peak at the interface is caused primarily by a transition from negative to positive MAE contributaions from the Fe $d_{xy}$ and $d_{x^2-y^2}$ orbitals, which hybridize with the nearby oxygen atom.

\section{\label{sec:negf}Quantum Transport}
%
\begin{figure*}
\begin{center}
    \centering
    \includegraphics[width=\textwidth]{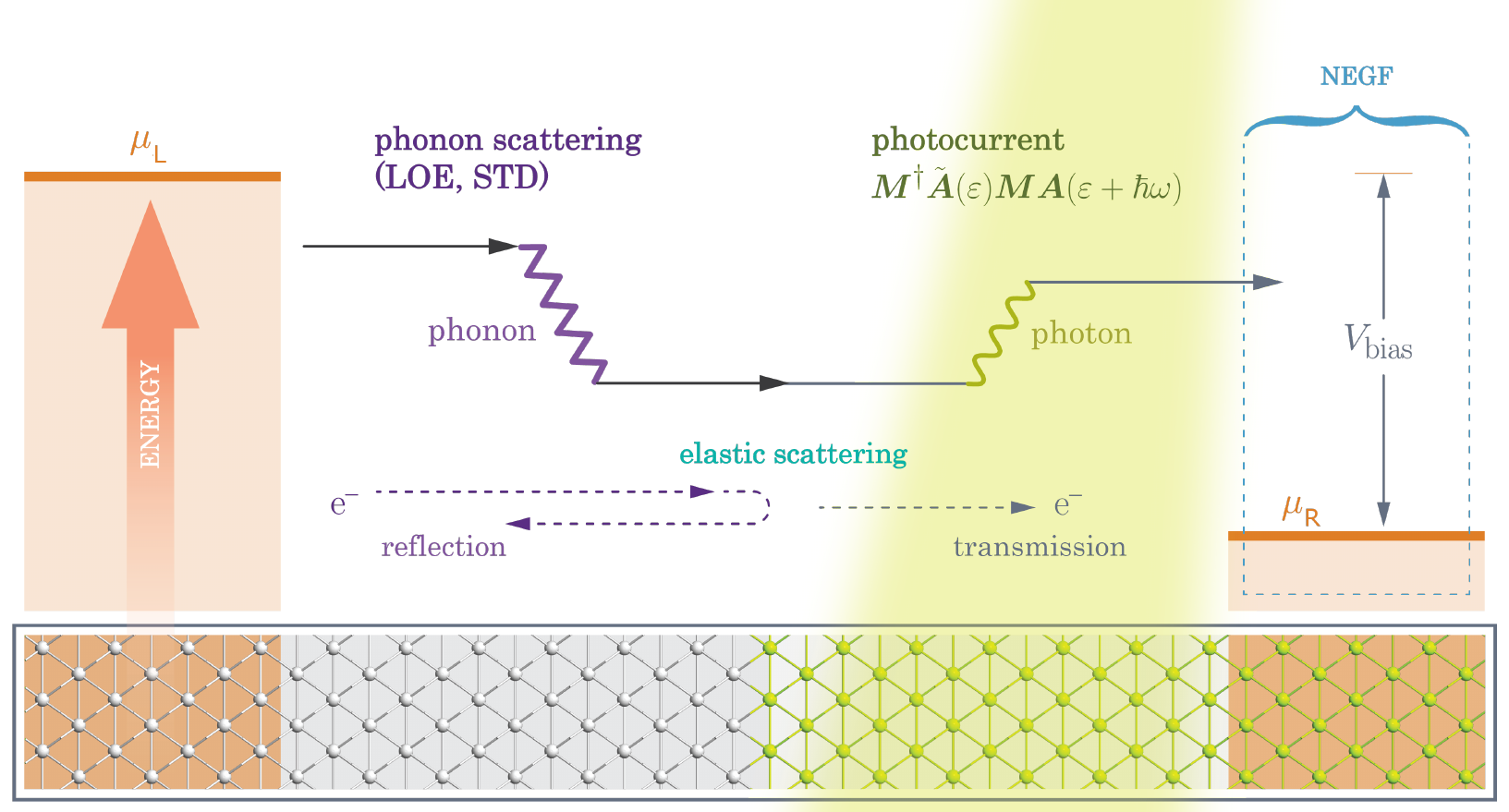}
    \caption{Illustration of the NEGF quantum transport module in QuantumATK. The left and right electrode regions (orange background) have an equilibrium electron distribution with chemical potentials $\mu_\text{L}$ and $\mu_\text{R}$,
  related through the applied sample bias, $\mu_\text{R} - \mu_\text{L} = e V_\textrm{bias}$ . At $T=0$~K, the electrons with energies in the bias
  window, $\mu_\text{L} \leq \varepsilon \leq \mu_\text{R}$, give rise to a steady-state electrical
  current from the right to left electrode. Note that the electron transport direction is from the left to right electrode. For higher temperatures, the electrons above (below) $\mu_\text{L}$ ($\mu_\text{L}$) will also contribute to the current because of the corresponding broadening of the Fermi--Dirac distribution at $T>0$~K. The system is modelled selfconsistently at the DFT or TB level using the NEGF method. It is possible to include the effect of gate potentials in the selfconsistent solution. Inelastic effects due to phonon or photon scattering can be included through perturbation theory.}
 \label{fig:quantumtransport}
\end{center}
\end{figure*}

The signature feature of \qatk\ is simulation of device systems. While most DFT device simulation codes are constructed on top of an electronic structure code designed for simulating bulk systems, \qatk\ is designed from scratch to achieve the highest accuracy and performance for both bulk and device systems.

Figure~\ref{fig:quantumtransport} shows a device (two-probe) geometry. It consists of a left electrode, a central region, and a right electrode.
The three regions have the same BCs in the two lateral directions perpendicular to the left-right electron transport direction, as defined in Fig.~\ref{fig:quantumtransport}.
The left and right electrodes are assumed to have bulk properties, and the first step of the device simulation is to perform a bulk calculation of each electrode with periodic BCs in the transport direction. Using Bloch's theorem, we describe the wave functions in terms of transverse $\bk$-points, and to seamlessly connect the three regions, the same $\bk$-point sampling is used in the transverse directions for all three regions. In the transport direction, the central-region wave functions are described by using scattering BCs, while the electrode wave functions are described by using periodic BCs. To have a seamless connection, it is important that the electrode wave functions very accurately reproduce the infinite-crystal limit in the transport direction. A very dense electrode $\bk$-point grid is therefore needed in the transport direction.

The left and right electrodes are modelled in their ground states with chemical potentials $\mu_\text{L}$ and $\mu_\text{R}$, respectively. This is only a correct model if the electrodes are not affected by the contact with the central region. The central-region electrostatic potential should therefore be sufficiently screened in the regions interfacing with the electrodes (denoted ``electrode extensions''), such that the potential in each electrode extension virtually coincides with that in the electrode.
Furthermore, the approximation is not valid if the finite-bias current density is high; in this case a non-equilibrium electron occupation is needed to accurately model the electrodes. A device with no electron scattering in the central region can therefore not be modelled reliably at finite bias.

The electronic structures of the isolated electrodes are defined with respect to an arbitrary energy reference. When used in a device simulation, they must be properly aligned to a common reference. This is achieved by applying a potential shift to the electronic structure of the right electrode, chosen to fulfill the condition
\begin{equation}
\mu_\text{L}-\mu_\text{R} =-e V_\mathrm{bias},
\end{equation}
where $V_\mathrm{bias}$ is the bias applied on the electrodes. It is clear that $\mu_\text{R} = \mu_\text{L}$ at zero bias.
The electrode electrostatic potentials, including the right-electrode potential shift, sets up the BCs for the central-region electrostatic potential. Thus, the whole system is aligned to a common reference, and device built-in potentials, if any, are properly included.

The electrostatic potential enters the KS equation from which the electron density in the central region is determined. We assume the system is in a steady state, that is, the central-region electron density does not change with time. The density can then be described in terms of extended electronic states from the left and right electrodes, as well as bound states in the central region,
\begin{equation}
 n({\bf r})  = n_\text{L}({\bf r}) + n_\text{R}({\bf r}) + n_\text{B}({\bf r}).
\end{equation}
We now focus on the contribution from the extended states of the left ($n_\text{L}$) and right ($n_\text{R}$) electrodes, and delay the discussion of bound states ($n_\text{B}$) for later. The former may be obtained by calculating the scattering states incoming from the left ($\psi_\alpha^\text{L}$) and right ($\psi_\alpha^\text{R}$) electrodes, which can be obtained by first calculating the Bloch
states in the electrodes, and subsequently solving the KS equation for the
central region using those Bloch states as matching BCs.

The left and right electron densities can then be calculated by summing up the occupied scattering states,
\begin{eqnarray}
   n_\text{L}({\bf r}) & = & \sum_{\alpha}|\psi_\alpha^\text{L}({\bf r})|^2 f\left(\frac{\varepsilon_\alpha-\mu_\text{L}}{k_\text{B} T_\text{L}}\right), \\
   n_\text{R}({\bf r}) & = & \sum_{\alpha}|\psi_\alpha^\text{R}({\bf r})|^2 f\left(\frac{\varepsilon_\alpha-\mu_\text{R}}{k_\text{B} T_\text{R}}\right),
\end{eqnarray}
where $f(x)=(1+\mathrm{e}^x)^{-1}$ is the Fermi--Dirac distribution.
\subsection{NEGF Method}
Instead of using the scattering states to calculate the non-equilibrium electron
density, \qatk\ uses the NEGF method; the two
approaches are formally equivalent and give identical results.\cite{Brandbyge2002}

The electron density is given in terms of the electron density matrix. We split the density matrix into left and right contributions,
\begin{equation}
      D = D^\text{L} + D^\text{R}.
\end{equation}

The left contribution is calculated using the NEGF method as\cite{Brandbyge2002}
\begin{equation}
       D^\text{L} = \int \rho^\text{L}(\varepsilon) f\left(\frac{\varepsilon-\mu_\text{L}}{k_\text{B} T_\text{L}}\right) d\varepsilon ,
\label{eq:DL}
\end{equation}
where
\begin{equation}
      \rho^\text{L}(\varepsilon) \equiv \frac{1}{2\pi} G(\varepsilon) \Gamma^\text{L}(\varepsilon) G^\dagger(\varepsilon)
\end{equation}
is the spectral density matrix, expressed in terms of the retarded Green's function $G$
and the broadening function $\Gamma^\text{L}$ of the left electrode,
\begin{equation}
    \Gamma^\text{L} =  \frac{1}{\mathrm{i}}(\Sigma^\text{L} - (\Sigma^\text{L})^\dagger) ,
\end{equation}
which is given by the left electrode self-energy $\Sigma^\text{L}$.
Note that while there is a non-equilibrium electron
distribution in the central region, the electron distribution in the left electrode is
described by a Fermi--Dirac distribution $f$ with an electron temperature $T_\text{L}$.

Similar equations exist for the right density matrix contribution. The next section
describes the calculation of $G$ and $\Sigma$ in more detail.

We note that the implemented NEGF method supports spintronic device simulations,
using a noncollinear electronic spin representation,
and possibly including spin-orbit coupling.
This enables, for example, studies of spin-transfer torque
driven device physics.\cite{Nikolic2018}

\subsection{Retarded Green's Function}
The NEGF key quantity to calculate is the retarded Green's function matrix for the central region.
It is calculated from the central-region Hamiltonian matrix $H$ and overlap matrix $S$ by adding the electrode self-energies,
\begin{equation}
    G(\varepsilon) =\left[ (\varepsilon+\mathrm{i}
          \delta_+) S  -H  - \Sigma^\text{L}(\varepsilon) -\Sigma^\text{R}(\varepsilon)  \right]^{-1},
\end{equation}
where $\delta_+$ is an infinitesimal positive number.

Calculation of $G$ at a specific energy $\varepsilon$ requires inversion of the central-region Hamiltonian matrix.
The latter is stored in a sparse format, and we only need the density matrix for the same sparsity pattern. This is done by block diagonal inversion,\cite{petersen2008block} which is $\mathcal{O}(N)$ in the number of blocks along the diagonal.

The self-energies describe the effect of the electrode states on the electronic
structure in the central region, and are calculated from the
electrode Hamiltonians. \qatk\ provides a number of different methods,\cite{Sanvito1999, LopezSancho1985, Sorensen2008, Sorensen2009}
where our preferred algorithm use the recursion method of Ref.~\onlinecite{LopezSancho1985},
which in our implementation exploits the sparsity pattern of the electrode.
This can greatly speed up the NEGF calculation as compared to using dense matrices.

\subsection{\label{sec:ComplexContour}Complex Contour Integration}
The integral in \eqref{eq:DL} requires a dense set of energy points due to the rapid variation of the spectral density along the real axis. We therefore follow Ref.~\onlinecite{Brandbyge2002} and divide the integral into an equilibrium part, which can be integrated on a complex contour, and a non-equilibrium part, which needs to be integrated along the real axis, but only for energies within the bias window. We have
\begin{equation}
D = D_{\mathrm{eq}}^\text{L} + \Delta_{\mathrm{neq}}^\text{R},
\label{eq:D2L}
\end{equation}
where
\begin{eqnarray}
\label{eq:Ddelta1}
D_{\mathrm{eq}}^\text{L} & = & \int d\varepsilon (\rho^\text{L}(\varepsilon)+\rho^\text{R}(\varepsilon) + \rho^\text{B}(\varepsilon))  f\left(\frac{\varepsilon-\mu_\text{L}}{k_\text{B} T_\text{L}}\right) , \\
\Delta_{\mathrm{neq}}^\text{R} & = & \int d\varepsilon \rho^\text{R}(\varepsilon ) \left[ f\left(\frac{\varepsilon-\mu_\text{R}}{k_\text{B} T_\text{R}}\right) - f\left(\frac{\varepsilon-\mu_\text{L}}{k_\text{B} T_\text{L}}\right) \right] ,
\label{eq:Ddelta2}
\end{eqnarray}
where $\rho^\text{B}$ is the density of states of any bound states in the central region.
Equivalently, we could write the density matrix as
\begin{equation}
D = D_{\mathrm{eq}}^\text{R} + \Delta_{\mathrm{neq}}^\text{L},
\label{eq:D2R}
\end{equation}
where $\mathrm{L}$ and $\mathrm{R}$ are exchanged in \eqref{eq:Ddelta1} and \eqref{eq:Ddelta2}.

Due to the finite accuracy of the integration along the real axis, \eqref{eq:D2L} and \eqref{eq:D2R} are numerically different. We therefore use a double contour,\cite{Brandbyge2002} where \eqref{eq:D2L} and \eqref{eq:D2R} are weighted such that the main fraction of the integral is obtained from the equilibrium parts, $D_{\mathrm{eq}}^\text{L}$ and $D_{\mathrm{eq}}^\text{R}$, which are usually much more accurate than the non-equilibrium parts, due to the use of high-precision contour integration. We have
\begin{equation}
D_{ij} = W^\text{L}_{ij} \left[ D_{\mathrm{eq}}^\text{L} + \Delta_{\mathrm{neq}}^\text{R}\right]_{ij} + W^\text{R}_{ij}  \left[ D_{\mathrm{eq}}^\text{R} + \Delta_{\mathrm{neq}}^\text{L} \right]_{ij},
\label{eq:D2W}
\end{equation}
where $W^\text{L}$ and $W^\text{R}$ are chosen according to Ref.~\onlinecite{Brandbyge2002}, i.e.\ such that at each site, the equilibrium part of the density matrix gives the largest contribution and $W^\text{L}+W^\text{R}=1$.

\subsection{Bound States}
The non-equlibrium integrals, $\Delta_{\mathrm{neq}}^\text{L}$ and $\Delta_{\mathrm{neq}}^\text{R}$, do not include any density from bound states in the central region. However, the equilibrium part of the density matrix is calculated from a complex contour integral of the retarded Green's function, and this calculation includes bound states with energies below the chemical potential of the contour.

Assume $\mu_\text{L}  < \mu_\text{R}$, then a bound state with energy $\varepsilon_\text{B} < \mu_\text{L}$ will be included in both $D_{\mathrm{eq}}^\text{L}$ and $D_{\mathrm{eq}}^\text{R}$, but a bound state  in the bias window, $\mu_\text{L} < \varepsilon_\text{B} < \mu_\text{R}$, will only be included in $D_{\mathrm{eq}}^\text{R}$. Thus, from \eqref{eq:D2W} we see that the state will be included with weight 1 if $\varepsilon_\text{B} < \mu_\text{L}$ and only with a fractional weight  if $\mu_\text{L} < \varepsilon_\text{B} < \mu_\text{R}$.
The weight will depend on the position of the bound state along the transport direction, that is, if the bound state is in a region that is well connected with the right electrode, the occupation will follow the right electrode and thus be close to 1. If it is in a region that is not well connected with the right electrode, the occupation will follow the left electrode, and thus for the current example the occupation be close to 0.

The \textit{true} occupation of a bound state in the bias window will depend on the physical mechanism responsible for the occupation and de-occupation, for example electron-phonon scattering, defects, etc. However, the matrix element will typically be higher with the electrode that is well connected with the region around the bound state, so we believe that the use of a double contour gives a qualitatively correct description of the occupation of the bound states in the bias window. Furthermore, we find that if we do not use such weighting schemes, bound states in the bias window can cause instabilities in the selfconsistent finite-bias NEGF calculation.

\subsection{Spill-in Terms}
Given the density matrix $D$, the electron density is obtained from the LCAO basis functions $\phi$:
\begin{equation}
      n({\bf r})  = \sum_{ij} D_{ij} \phi_i({\bf r}) \phi_j({\bf r}).
\end{equation}
The Green's function of the central region gives the density matrix of the central region, $D^\text{CC}$.
However, to calculate the density correctly close to the central-region
boundaries towards the electrodes, the terms involving $D^\text{LL}$, $D^\text{LC}$, $D^\text{CR}$, and $D^\text{RR}$ are also needed.
These are denoted spill-in terms.\cite{stradi2016general}

\qatk\ implements an accurate scheme for including all the spill-in terms, both for the electron density
and for the Hamiltonian integrals.\cite{stradi2016general}
This gives additional stability and well-behaved convergence in device simulations.

\subsection{Device Total Energy and Forces}
A two-probe device is an open system where charge can flow in and out of the central region through
the left and right electrode reservoirs. Since the two reservoirs may have different chemical potentials,
and the particle number from a reservoir is not conserved,
it is necessary to use a grand canonical potential to describe the energetics of the system,\cite{todorov2000current}
\begin{equation}
    \Omega[n] = E_{\mathrm{KS}}[n]- N_\text{L} \mu_\text{L} - N_\text{R} \mu_\text{R} ,
\end{equation}
where $N_\text{L/R}$  is the number of electrons contributed to the central region
from the left/right electrode, and $E_{\mathrm{KS}}[n]$ is the KS total energy.

Due to the screening approximation, the central region will be charge neutral, and
therefore $N_\text{L} +  N_\text{R} = N$, where $N$ is the ionic
charge in the central region. At zero bias ($\mu_\text{L} = \mu_\text{R}$),
the particle term is constant, so that $N \mu_\text{L} = N \mu_\text{R} $, and is thus independent of atom displacements in
the central region. However, at finite bias ($\mu_\text{L} \neq \mu_\text{R}$),
the particle terms in $\Omega$ will affect the forces.

If one neglects current-induced forces,\cite{lu2012current,todorov2014current} as done in \qatk\ simulations,
the force acting on atom $a$ at position ${\bf r}_{a}$ in the device central region is given by
\begin{equation}
      {\bf F}_{a}= - \frac{\partial \Omega[n]}{\partial {\bf r}_{a}}.
\end{equation}
It can be shown that the calculation of this force is identical to the calculation of
the equilibrium (zero-bias) force, but in the non-equilibrium (finite-bias) case the density and energy density
matrix must be calculated within the NEGF framework.\cite{Brandbyge2002, todorov2000current, Zhang2011}

\subsection{Transmission Coefficient and Current}
When the selfconsistent non-equilibrium density matrix has been obtained, it is
possible to calculate various transport properties of the system. One of the most notable is
the transmission spectrum from which the current and differential conductance  are obtained.
The transmission coefficient $T$ at electron energy $\varepsilon$ is obtained from the retarded Green's function,\cite{Haug2008}
\begin{equation}
      T(\varepsilon) = \rm{Tr}\left[ G(\varepsilon) \Gamma^\text{L}(\varepsilon) G^\dagger (\varepsilon) \Gamma^\text{R}(\varepsilon) \right] ,
      \label{eq:ElasticTransmission}
\end{equation}
and the electrical current is given by the Landauer formula,
\begin{equation}
      I = \frac{2e}{h} \int_{-\infty}^{\infty} d\varepsilon T(\varepsilon)  \left[ f \left( \frac{\varepsilon- \mu_\text{L}}{k_\text{B} T_\text{L}}\right) - f\left( \frac{\varepsilon- \mu_\text{R}}{k_\text{B} T_\text{R}}\right) \right].
      \label{eq:ElasticCurrent}
\end{equation}

\subsection{Inelastic Transmission and Inelastic Current}
\label{sec:inelastic}
\qatk\ implements the lowest-order expansion (LOE) method\cite{lu2014efficient}
for calculating the inelastic current due to electron-phonon scattering,
which is not included in \eqref{eq:ElasticTransmission} and \eqref{eq:ElasticCurrent}.
The LOE method is based
on perturbation theory in the first Born approximation,
and requires calculation of the  dynamical matrix and the
Hamiltonian derivative with respect to atomic positions in the
central region, $\nabla H(\br)$.
Calculation of these derivatives are described in Section~\ref{sec:phonons}.

First-principles calculation of $\nabla H(\br)$ can be prohibitive for large device systems.
However, if the atomic configuration of the central region can be generated by repeating the
left electrode along the transport direction, then $\nabla H(\br)$ can be obtained to a good approximation by using the
$\nabla H(\br)$ of the left electrode only.\cite{gunst2017new}

From $\nabla H(\br)$ of the central region we get the electron-phonon matrix elements in reciprocal
space,\cite{gunst_first-principles_2016}
\begin{align}
    M_{\lambda,{\bf k},{\bf q}}^{i j} &= \sum_{mn}\mathrm{e}^{\mathrm{i}{\bf k}\cdot({\bf R}_n-{\bf R}_m) - \mathrm{i}{\bf q}\cdot{\bf R}_m} \nonumber \\
&\quad \times \langle \phi_j {\bf R}_m | {\bf v}_{\lambda, {\bf q}}\cdot\nabla H_0(\br)|\phi_i\;{\bf R}_n\rangle,
\end{align}
where the ($mn$)-sum runs over repeated unit cells in the supercell calculation of the Hamiltonian derivatives,\cite{gunst_first-principles_2016} and the subscript 0 indicates that the derivatives are only calculated for atoms in the unit cell with index ‘0’. Moreover, $|\phi_i\;{\bf R}_n\rangle$ ($|\phi_j\;{\bf R}_m\rangle$) denotes the $i$($j$)'th LCAO basis orbital in the unit cell displaced from the reference cell by the
lattice vector ${\bf R}_n$ (${\bf R}_m$), while ${\bf q}$ is the phonon momentum, and ${\bf v}_{\lambda,{\bf q}}$ is the mass-scaled mode vector of phonon mode $\lambda$ with frequency $\omega_{\lambda,{\bf q}}$.

Following Ref. \onlinecite{lu2014efficient}, we obtain the inelastic transmission
functions for a finite transfer of momentum. From these we calculate the total electrical current, including inelastic effects.\cite{vandenberghe2011generalized, gunst2017new}
The complete formulas for the \qatk\ implementation can be found in Ref.~\onlinecite{gunst2017new}.

\subsubsection{Special Thermal Displacement Method}
In Ref.~\onlinecite{markussen_electron-phonon_2017} we
showed that the average transmission from a thermal distribution of
configurations accurately describes the inelastic electron transmission
spectrum due to electron-phonon scattering at this temperature.
In the special thermal displacement (STD) method,
the average is replaced with a single representative configuration,
which may drastically reduce the computational cost
of inelastic transport simulations.\cite{gunst_first-principles_2017}

To obtain the STD configuration, we first calculate the phonon eigenspectrum
using the dynamical matrix of the central region. We consider only $\bq=\textbf{0}$,
since only relative displacements between atoms in the cell will be important,
and to account for finite $\bq$-vectors we will have to increase the cell size.
The phonon modes are labeled by $\lambda$ with frequency
$\omega_{\lambda}$, eigenmode vector ${\bf e}_{\lambda}$, and
characteristic length $l_{\lambda}$.

The STD vector of atomic displacements is given by\cite{gunst_first-principles_2017}
\begin{align}
{\bf u}_\mathrm{STD}(T) = \sum_{\lambda} s_{\lambda} (-1)^{\lambda-1} \sigma_{\lambda}(T) {\bf e}_{\lambda} ,
\label{eqn:OSdisplacement}
\end{align}
where $s_{\lambda}$ denotes the sign of the first non-zero element in ${\bf e}_{\lambda}$, enforcing the same choice of ``gauge'' for the modes. The Gaussian width $\sigma$ is related to the mean
square displacement $\langle {\bf u}_{\lambda}^{2} \rangle =
l^{2}_{\lambda} (2 n_B(\frac{\hbar \omega_\lambda}{k_B T}) +1) = \sigma_{\lambda}^{2}(T)$ at
temperature $T$, where $n_\text{B}$ is the Bose--Einstein distribution.

An essential feature of the STD method is the use of opposite phases for phonons
with similar frequencies; in this way phonon-phonon correlation
functions average to zero and the transmission spectrum of the STD configuration
becomes similar to a thermal average of single phonon excitations.

The final step in the STD method is to calculate the selfconsistent
Hamiltonian of the system displaced by ${\bf u}_\mathrm{STD}$, and use that to calculate the transmission spectrum. Thus, the computational cost of the inelastic transmission calculation is for the STD method similar to that of an ordinary elastic transmission calculation.

Formally, this method becomes accurate for systems where the central region is a large unit cell generated by the repetition of a basic unit cell.
\subsection{Thermoelectric Transport}
%
The thermoelectric figure of merit, $\mathrm{ZT}$,
quantifies how efficiently a temperature difference (heat)
can be converted into a voltage difference in a thermoelectric material,
\begin{equation}
\mathrm{ZT} = \frac{G_\mathrm{e} S^2 T}{\kappa} ,
\label{eq:zt}
\end{equation}
where $G_\mathrm{e}$ is the electronic conductance,
$S$ the Seebeck coefficient,
$T$ the temperature,
and $\kappa = \kappa_\mathrm{e} + \kappa_\mathrm{ph}$ the summed
electron and phonon heat transport coefficients.
Following Ref.~\onlinecite{Markussen2009},
and given a set of electron and phonon transmission spectra
for a device configuration,
\qatk\ uses linear-response theory to compute the above-mentioned
thermoelectric coefficients and the Peltier coefficient, $\Pi$,
\begin{align}
G_\mathrm{e}      &=   \left. \frac{dI}{dV_\mathrm{bias}} \right\rvert_{dT=0} ,  \\
S                 &= - \left. \frac{dV_\mathrm{bias}}{dT} \right\rvert_{I=0}  ,  \\
\kappa_\mathrm{e} &=   \left. \frac{dI_Q}{dT} \right\rvert_{I=0} , \\
\Pi               &=   \left. \frac{I_Q}{I} \right\rvert_{dT=0} = S V_\mathrm{bias} ,
\end{align}
where $I_Q = dQ/dT$ is the electronic contribution to the heat current. It is calculated in a similar way as the electronic current,\cite{SivanPRB1986}
\begin{eqnarray}
I_Q &=& \frac{2e}{h} \int_{-\infty}^{\infty} d\varepsilon T(\varepsilon)\left[\varepsilon - \mu \right] \nonumber \\
&\quad& \times \left[ f \left( \frac{\varepsilon- \mu_\text{L}}{k_\text{B} T_\text{L}}\right) - f\left( \frac{\varepsilon- \mu_\text{R}}{k_\text{B} T_\text{R}}\right) \right],
      \label{eq:ElectronicHeatCurrent}
\end{eqnarray}
where $\mu=(\mu_\mathrm{L} + \mu_\mathrm{R})/2$ is the average chemical potential, and the difference to \eqref{eq:ElasticCurrent} is the inclusion of the factor $(\varepsilon - \mu)$ in the integral.

Note that one may use DFT or a TB model for obtaining
the electron transmission and a force field to calculate the phonon transmission,
constituting a computaionally efficient workflow for investigating thermoelectric materials.
%
\subsection{Photocurrent}
\qatk\ allows for calculating photocurrent  using first-order perturbation theory within the first Born
approximation.\cite{henrickson2002nonequilibrium, chen2012first, zhang2014generation}
In brief, the electron-light interaction is added to the Hamiltonian,
\begin{equation}
  \hat{H} = \hat{H}_0 + \frac{e}{m_0}{\bf A}_\omega \cdot\hat{{\bf p}},
\end{equation}
where $\hat{H}_0$ is the Hamiltonian without the electron-light interaction, $e$ the
electron charge, $m_0$ the free-electron mass, $\hat{{\bf p}}$ the momentum operator,
and ${\bf A}_\omega$ the electromagnetic vector potential from a single-mode monocromatic light source with frequency $\omega$.

The first-order coupling matrix is
\begin{equation}
  M_{ij} = \frac{e}{m_0}\langle i | {\bf A}_\omega \cdot\hat{{\bf p}} | j \rangle ,
\end{equation}
where $|j\rangle$ is an LCAO basis function.

The first Born electron-photon self-energies are
\begin{eqnarray}
  {\bf \Sigma}_\text{ph}^> & = & [N {\bf M}^\dagger {\bf G}_0^>(\varepsilon^+){\bf M} + (N + 1) {\bf M} {\bf G}_0^>(\varepsilon^-){\bf M}^\dagger] , \\
  {\bf \Sigma}_\text{ph}^< & = & [N {\bf M} {\bf G}_0^<(\varepsilon^-){\bf M}^\dagger + (N + 1) {\bf M}^\dagger {\bf G}_0^>(\varepsilon^-){\bf M}] ,
\end{eqnarray}
where $\varepsilon^{\pm} = \varepsilon \pm \hbar\omega$, and $N$ is the number of photons. The Green's function including electron-photon
interactions to first order is then
\begin{equation}
  {\bf G}^{>/<} = {\bf G}_0^r \left({\bf \Sigma}^{>/<}_{L} + {\bf \Sigma}^{>/<}_{R} + {\bf \Sigma}_{ph}^{>/<}\right){\bf G}_0^a ,
\end{equation}
where ${\bf G}_0^{r,>,<}$ denote the non-interacting Green's
functions, and ${\bf \Sigma}^{>/<}_\text{L,R}$ are the lesser and greater self-energies due to
coupling to the electrodes. The current in electrode $\alpha$ (left or right) with spin
$\sigma$ is calculated as
\begin{equation}
  I_{\alpha, \sigma} = \frac{e}{\hbar}\int \frac{d\varepsilon}{2\pi}\sum_{k} T_\alpha(\varepsilon, k, \sigma),
\end{equation}
where the effective transmission coefficients are given by\cite{zhang2014generation}
\begin{equation}
    T_\alpha(\varepsilon, k, \sigma) = {\rm Tr}\left\{\mathrm{i}\Gamma_\alpha(\varepsilon, k) [1 - f_\alpha] G^< + f_\alpha G^>  \right\}_{\sigma\sigma}.
\end{equation}

We note that it is possible to include also the effect of phonons through the STD method,
which is important for a good description of photocurrent in indirect-band-gap materials such as silicon.\cite{palsgaard2018efficient}

\section{\label{sec:parallel}\qatk\ Parallelization}
%
\begin{figure}
  \centering
  \includegraphics[width=0.8\columnwidth]{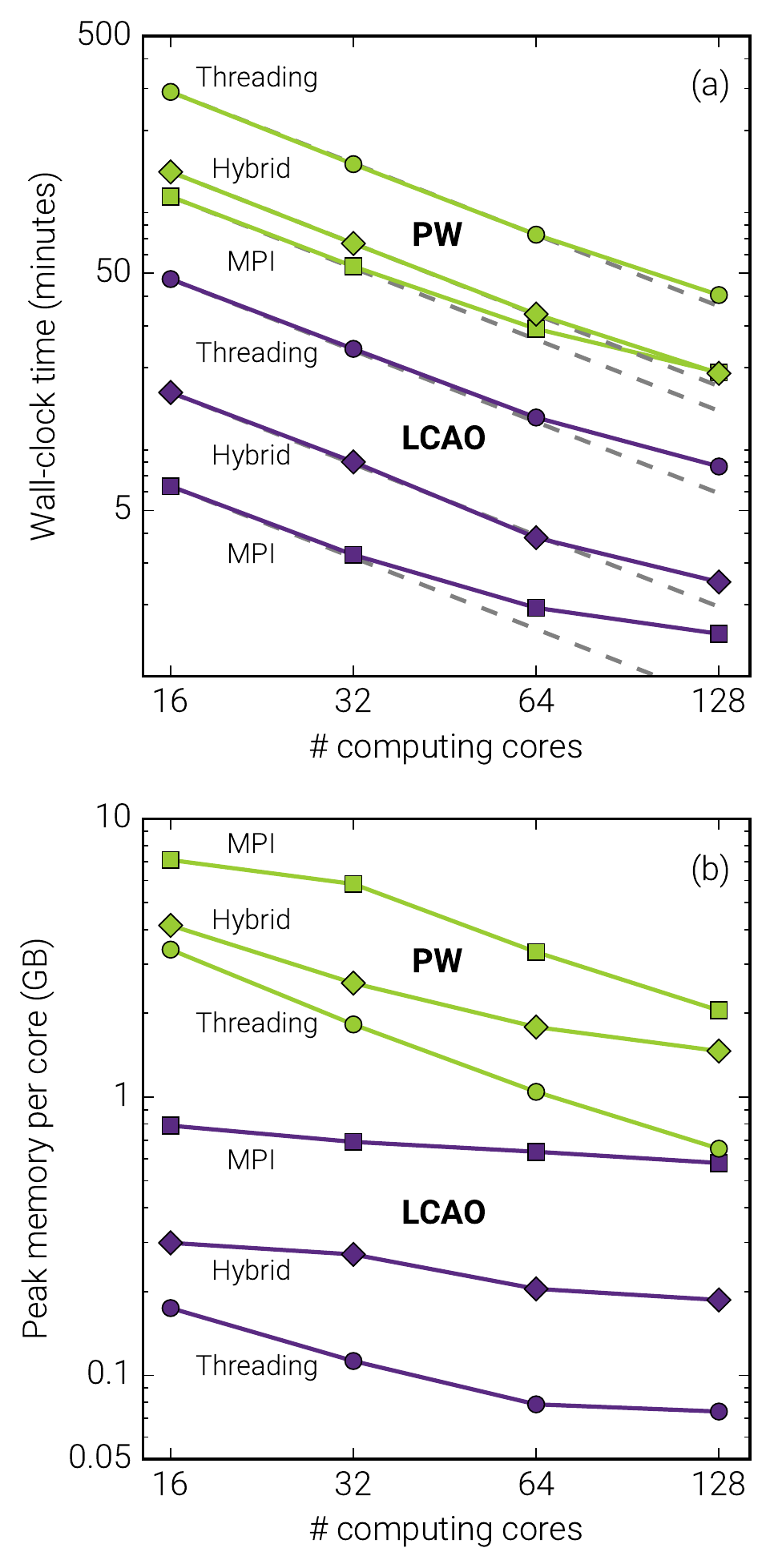}
  \caption{Scaling performance of \qatk\ DFT simulations for a 64-atom
  Si$_{0.5}$Ge$_{0.5}$ random-alloy supercell
  when executed in parallel (using MPI) on 1, 2, 4, and 8
  computing nodes (16 cores per node). a) Total wall-clock times
  for LCAO and PW selfconsistent total-energy calculations,
  and b) the corresponding peak memory requirements per core.
  Grey lines indicate ideal scaling of the wall-clock time.
  PseudoDojo PPs with LCAO-High basis sets were used.
  Note that the Ge PP contains semicore states.
  The supercell has 32 irreducible $\bk$-points,
  corresponding to two computing nodes for full MPI parallelization
  over $\bk$-points. With 4 (8) full nodes, 2 (4) MPI processes
  are assigned to eack $\bk$-point.}
  \label{fig:si64_parallel}
\end{figure}
\begin{figure}
  \centering
  \includegraphics[width=0.8\columnwidth]{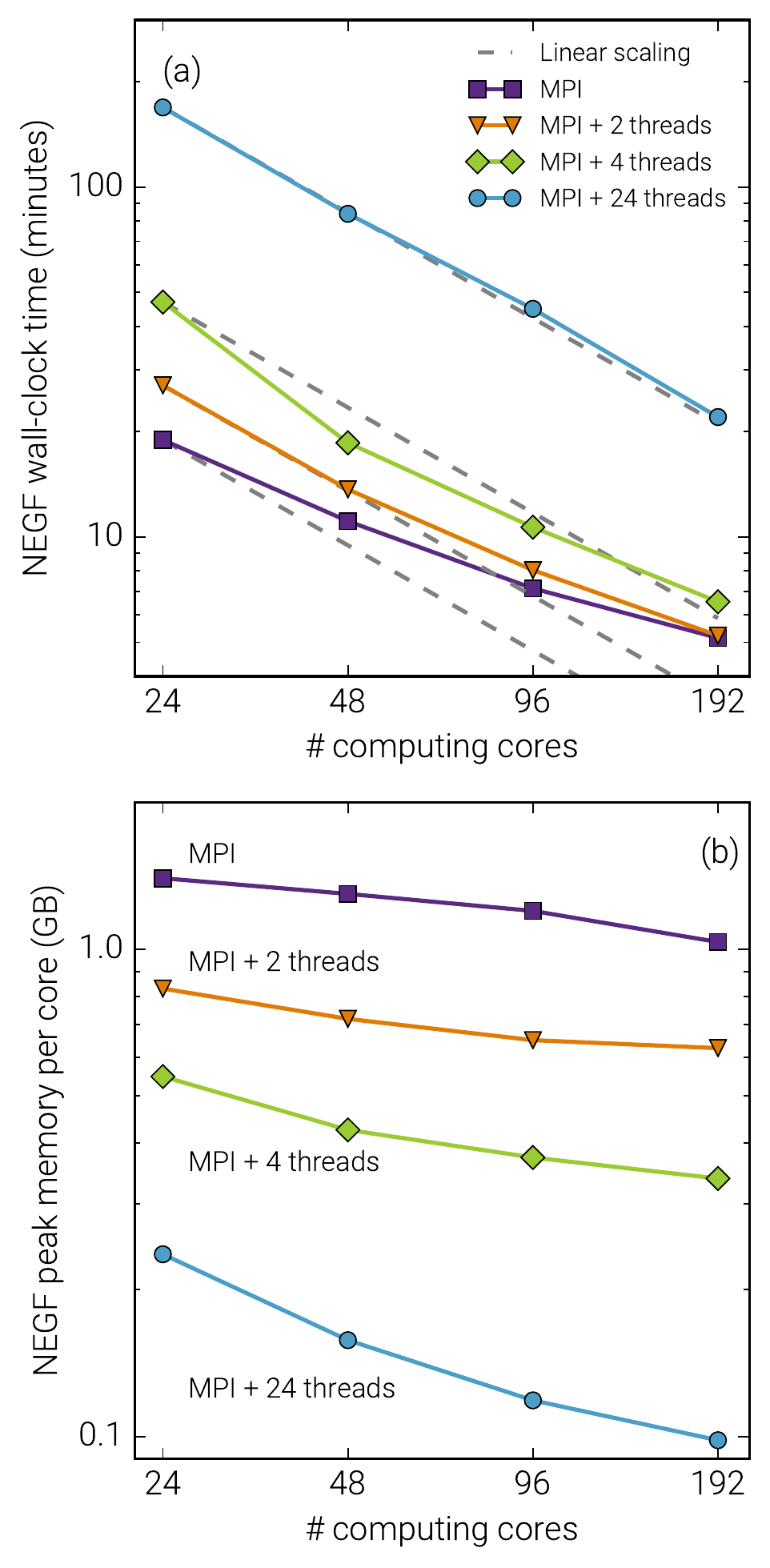}
  \caption{Scaling performance of equilibrium DFT-NEGF simulations
  for a 10~nm long silicon \textit{p-n} junction
  with doping levels of $5 \cdot 10^{20}~\text{cm}^{-3}$.
  The junction cross section is $1.33~\text{nm}^2$,
  corresponding to 684 Si atoms in the device central region.
  The NEGF calculations were done using a PseudoDojo PP
  with the LCAO-Low basis set, and 2 irreducible $\bk$-points
  in the central-region 2D Brillouin zone,
  resulting in 96 generalized contour points.
  The simulations were run on up to eight 24-core Intel Xeon nodes,
  using both MPI (purple) and hybrid parallelization schemes.
  Hybrid parallelization was done using 2 (orange), 4 (green), and 24 (blue) threads
  per MPI process, with processes distributed evenly over the nodes.
  Gray dashed line indicates ideal scaling of the wall-clock time.
  }
  \label{fig:negf_parallel}
\end{figure}
\begin{figure}
  \centering
  \includegraphics[width=0.8\columnwidth]{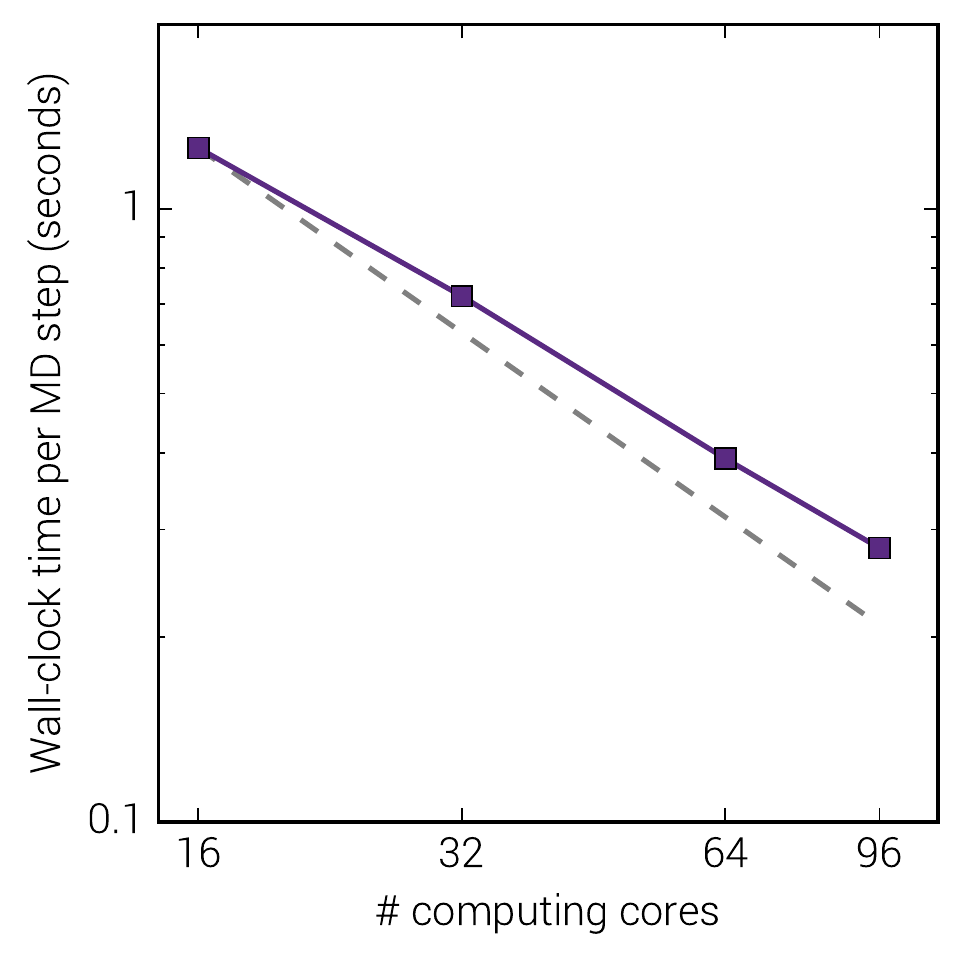}
  \caption{Scaling performance of \atkff\ simulations
  for a SiO$_2$ supercell containing 1 million atoms,
  using a force field from Ref.~\onlinecite{Pedone2006}.
  The simulation used one MPI process per CPU core for parallelization,
  and was run on up to six 16-core Intel Xeon nodes.
  Gray dashed line indicates ideal scaling of the wall-clock time.}
  \label{fig:ff_parallel}
\end{figure}
Atomic-scale simulations for small configurations
(systems with only a few atoms)
may often be executed in serial on a single CPU core,
but most production simulations
require execution in parallel on several cores (often many)
to increase computational speed
and/or to reduce the per-core memory footprint.
The \qatk\ platform offers several parallelization techniques
depending on the type of computational task.
\subsection{Bulk DFT and Semi-Empirical Simulations}
For bulk DFT-LCAO calculations,
the basic unit of computational work to distribute in parallel
is a single $\bk$-point.
\qatk\ uses the message passing interface (MPI) protocol
to distribute such work units as individual computing processes
on individual, or small groups of, CPU cores,
and also allows for assigning multiple processes to each work unit.
Moreover, each MPI process may be further distributed
in a hybrid parallelization scheme
by employing shared-memory threading of each process.

Figure~\ref{fig:si64_parallel} shows an example of how the total wall-clock time
and peak memory requirement for DFT-LCAO and DFT-PW calculations
scale with the number of 16-core computing nodes used with MPI parallelization.
We considered a 64-atom SiGe random-alloy supercell
with $N_\text{k}=32$ $\bk$-points.
In this case, 2 full nodes, $N_\text{n}=2$,
with 32 cores in total ($N_\text{c}=N_\text{n} \times 16 = 32$),
yields full MPI parallelization over $\bk$-points.
The PW calculations were done using a blocked generalized Davidson algorithm\cite{Davidson1975,Morgan1986}
to iteratively diagonalize the Hamiltonian matrix,
which in the \qatk\ implementation parallelizes the computational work over
both $\bk$-points and plane waves.
The LCAO calculations use the LAPACK\cite{lapack}
(when $N_\text{c}/N_\text{k} \leq 1$)
or ELPA\cite{Marek2014}
(when $N_\text{c}/N_\text{k} > 1$)
libraries to distribute Hamiltonian diagonalization over MPI processes.
It is clear from Fig.~\ref{fig:si64_parallel} that
the LCAO engine is both fast and requires less memory than the PW representation
for the 64-atom supercell,
although communication overhead causes
the LCAO computational speed to start breaking off from ideal scaling
when the number of processes (cores)
exceeds the number of $\bk$-points in the DFT calculation
(when $N_\text{c}/N_\text{k} > 1$).
On the contrary, MPI parallelization over both $\bk$-points and plane waves
enables approximately ideal scaling of the PW wall-clock time
up to at least 8 nodes (128 cores), corresponding to 4 MPI processes per $\bk$-point.
\subsection{DFT-NEGF Device Simulations}
As discussed in Section~\ref{sec:ComplexContour},
the NEGF equilibrium density matrix at a single $\bk$-point
is obtained from integrating the spectral density matrix
over $M_\varepsilon$ energy points on a complex contour.
This integral must be performed at all transverse
$\bk$-points in the 2D Brillouin zone of the device central region,
yielding $N_k \times M_\varepsilon$ \textit{generalized contour points}.
Each of these constitute a unit of computational work in
equilibrium NEGF calculations,
equivalent to $\bk$-point parallelization in DFT calculations for periodic bulks.

Since we typically have $M_\varepsilon=48$ contour energies,
an equilibrium NEGF simulation may easily require evaluation of
hundreds of generalized contour points.
MPI parallelization over contour points is therefore a highly efficient strategy.
For devices with relatively large transverse cross sections,
and therefore relatively few contour points (because of small $N_k$),
assignment of several processes to each contour point enables
scaling of NEGF computational speed to numbers of computing cores
well beyond the number of contour points.
This can also be combined with more than one thread per process in a hybrid parallelization scheme,
for a smaller speedup, but with a reduced per-core memory footprint.

Figure~\ref{fig:negf_parallel} shows an example of how the total wall-clock time
and peak memory usage for a DFT-NEGF calculation
scale with the number of computing nodes used with both MPI and hybrid
parallelization schemes.
Calculations for this 10~nm long silicon \textit{p-n} junction
require evaluation of 96 generalized contour points,
in this case corresponding to 4 nodes for full MPI
distribution of computational work.
As expected,
we find that using only MPI parallelization requires most memory per core,
but also results in the smallest wall-clock time for the NEGF calculation,
although communication overhead causes a deviation from ideal scaling for more than 1 node, see Fig.~\ref{fig:negf_parallel}(a).
We also note that the per-core memory consumption is in this case almost constant
in Fig.~\ref{fig:negf_parallel}(b), except for a modest decrease for 8 nodes,
where 2 processes (cores) are assigned to each contour point.
It is furthermore clear from Fig.~\ref{fig:negf_parallel}
that hybrid parallelization enables significant memory reduction,
although at the cost of decreased computational speed.
Taking simulation on 4 nodes as an example,
hybrid parallelization with 4 threads per process (green lines)
requires in this case 50\% more wall-clock time
as compared to the MPI-only simulation (purple lines),
but at a 70\% smaller memory footprint.

Although NEGF computational efficiency and memory consumption
depend significantly on the device length and transverse dimensions,
the general trend is that MPI parallelization over contour points
yields computational speedup, while threading of processes reduce
the NEGF memory footprint at a comparatively smaller computational speedup.
\subsection{FF Simulations}
The \atkff\ engine uses shared-memory threading for parallelization
of relatively small systems,
while additional parallelization by domain decomposition over MPI processes
is available for large systems.
As explained in detail in Ref.~\onlinecite{ATKForceField},
the MPI distribution of \atkff\ workload is implemented via functionality
from the Tremolo-X MD package,\cite{TremoloX}
which is developed by the Fraunhofer Institute for Algorithms
and Scientific Calculations (SCAI).

In Fig.~\ref{fig:ff_parallel}, we show the wall-clock time per MD step for a
simulation of SiO$_2$ with 1 million atoms, using a force field from
Pedone \textit{et al.}\cite{Pedone2006}
This illustrates how the use of domain decomposition over MPI processes results
in a significant speedup when parallelizing over a large number of nodes and cores.

\section{\nanolab\ Simulation Environment}
%
\label{sec:nanolab}
\subsection{Python Scripting}
\label{sec:atkpython}
The \qatk\ software is programmed in the \cpp\ and Python languages. Around 80\% of the code lines are in Python, and only low-level numerically demanding parts are written in \cpp. The use of Python allows for using a large number of high-level physics and mathematics libraries, and this has greatly helped building the rich functionality of \qatk\ in a relatively short time.

The user input file is a Python script and the user has through the script access to the same functionality as a \qatk\ developer. This enables the user to transform input files into advanced simulation scripts, which do not only set up advanced workflows and analysis, but may also alter the functionality of the simulation engines, for example by adding new total-energy terms. \qatk\ supplies a public application programming interface (API) with currently more than 350  classes and functions. These all take a number of arguments with detailed checks of the input parameters to ensure correct usage. For example, if the input argument is a physical quantity, the physical units must be supplied. A wide range of units are supported, e.g., for energy, the user may select units of joule, calorie, electron volt, kilojoule per mole, kilocalories per mole, Hartree, or Rydberg. All physical units are automatically converted to the internal units used by \qatk.
The user also has access to internal quantities such as the Hamiltonian, Green's function, self-energies, etc., through the API.

Through Python scripting it is possible to build advanced workflows that automate complex simulations and analysis. However, some simulations may require a large number of time consuming calculation tasks that are combined into a final result, and scripting such workflows can be impractical. For instance, if the computer crashes during a loop in the script, how to restart the script at the right step in a loop in the middle of the script? Or perhaps some additional tasks are needed after a custom simulation has finished; how to combine the already calculated data with the new data?

To simplify such simulations, \qatk\ has introduced a framework called a \textit{study object}. The study object keeps track of complex simulations that rely on execution and combination of a number of basic tasks. It allows for running the basic tasks in parallel and will be able to resume if the calculation is terminated before completion. A study object also allows for subsequently extending the number of tasks, and will only perform tasks that have not already completed. This framework is currently used for a number of complex simulations, for instance for coupling atomic-scale simulations with continuum-level TCAD tools. Examples include simulation of the formation energy and diffusion paths of charged point defects, scans over source-drain and gate bias for two-terminal devices, relaxation of devices, and calculation of the dynamical matrix and Hamiltonian derivatives by finite differences.

To store data we use the cross-platform HDF5 binary format,\cite{hdf5} which allows for writing and reading data in parallel to/from a single file. This file can also hold many different objects, so the entire output from a \qatk\ simulation can be stored efficently in a single file.
\subsection{NanoLab Graphical User Interface}
While scripting is very efficient for production runs, it requires knowledge of the scripting language, and it takes time to manually build up scripts for setting up the configuration, simulation, and analysis of interest.
The NanoLab GUI eliminates this barrier to productivity
by enabling the user to fully set up the Python input script in a professional GUI environment.
NanoLab is itself programmed in Python,
and each tool in NanoLab can interpret and generate Python scripts, thus, it is
possible to seamlessly shift from using the GUI tools in NanoLab to manually editing the Python scripts. It is the ambition that all NanoLab functions are also available as Python commands, such that any GUI workflow can be documented and reproduced in a Python script.

NanoLab is developed around a plugin concept, which makes it easy to extend it and add new
functionality. Plugins can be downloaded and installed from an add-on server,
and the majority of the plugins are available as source code, making it easy to modify
or extend them with new user-defined functionality.

NanoLab also provides GUI tools for communicating with online databases (``Databases''),
setting up the atomic-scale geometry of configurations (``Builder''),
writing the Python script (``Scripter''),
submitting the script to a remote or local computing unit (``Job Manager''),
and visualizing and analyzing the results (``Viewer'').
It is possible to connect third-party simulation cods with NanoLab by writing plugins that translate the input/output files into the internal NanoLab format.
Such plugins are currently available for the VASP,\cite{VASPKresse} Quantum ESPRESSO,\cite{giannozzi2009quantum} ORCA,\cite{neese2012orca} GPAW,\cite{enkovaara2010electronic} and CASTEP\cite{clark2005first} codes.

The plugin concept also allows for many specialized functions, for example specialized Builder tools like surface builders, interface builders,\cite{stradi2017method} NEB setups,\cite{Smidstrup2014} etc. The Job Manager has plugins that provide support for a wide range of job schedulers on remote computing clusters. Moreover, NanoLab has a large selection of graphical analysis tools, which can be used to visualize and analyze simulations with respect to a wide range of properties, all implemented as plugins. For instance, with the ``MD analyser'' plugin, a MD trajectory can be analyzed with respect to angular and radial distribution functions, or different spatial and time correlation functions. Other examples are interactive band structure analysis with extraction of effective masses, and analysis of transmission in device simulations with on-the-fly inspection of transmission eigenstates at specified points in the transmission spectrum. NanoLab currently ships with more than 100 preinstalled plugins, and additional plugins are available through the add-on server.
\subsection{Documentation}
Keeping an updated documentation system for the large set of \qatk\ classes and functions pose a challenge. To synchronize the documentation with the source code, we have developed an automated documentation system where the information for the \qatk\ reference manual is extracted directly from the Python source code using the Sphinx documentation generator.\cite{sphinx} The reference manual is available from an online platform\cite{qatkdocs} together with tutorials, whitepapers, webinars, etc. Through a search engine it is thus easy to find all available information for a given problem.

\section{\qatk\ applications}
%
\label{sec:applications}
\subsection{Large-scale Simulations of 2D Field-effect Transistors}
\label{sec:2DFET}
\begin{figure*}
\includegraphics[width=\textwidth]{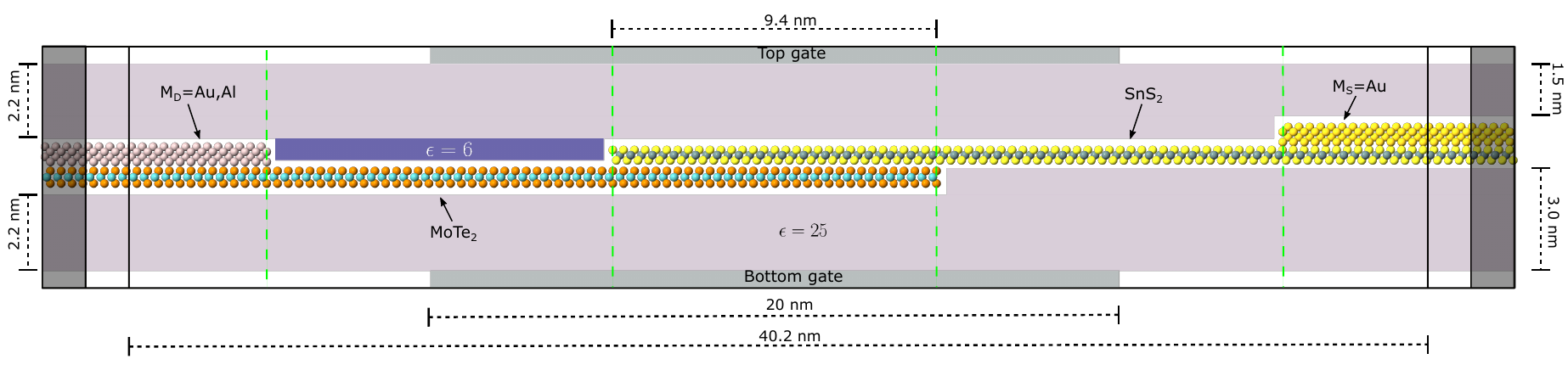}
\caption{Structure of the $\mathrm{M_D}$/MoTe$_2$/SnS$_2$/$\mathrm{M_S}$ device. Mo, Te, Sn and S atoms are shown in cyan, orange, dark green and yellow, respectively. The atoms of the $\mathrm{M_D}$ (Au, Al) and $\mathrm{M_S}$ (Au) regions are shown in pink and yellow, respectively. The metallic gate regions (top and bottom gates) are shown as light grey rectangles. The dielectric regions are shown as dark purple ($\epsilon = 6$) or light purple ($\epsilon = 25$) rectangles. The dashed green lines highlight the boundaries of the different device regions indicated in Fig.~\ref{fig:2Ddevice_pldos}(a,b). Note that the region of 40.2~nm is the 2D device central region without the left and right electrode extensions included, as defined in Sec.~\ref{sec:negf}. A vertical black solid line highlights the boundary between that region and the left (right) electrode extension. The semi-infinite, periodic left (right) electrode is visualized with the corresponding unit cell structure of MoTe$_{2}$ (SnS$_2$), which is highlighted with a dark grey-shaded rectangle adjacent to the left (right) electrode extension region. The Dirichlet BC is imposed on the left (right) boundary plane between the left (right) electrode and its extension. The top (bottom) horizontal black solid line highlights the top (bottom) boundary of the device simulation box. Mixed BCs are imposed on the corresponding boundary planes: Dirichet BCs on the metal gate surfaces, and Neumann BCs on the boundary planes in the vacuum regions (white rectangles). A periodic BC is applied in the lateral direction, which is perpendicular to the transport direction and the MoTe$_2$ (SnS$_2$) sheet.}
\label{fig:2Ddevice_structure}
\end{figure*}
\begin{figure}[!b]
\includegraphics[width=\linewidth]{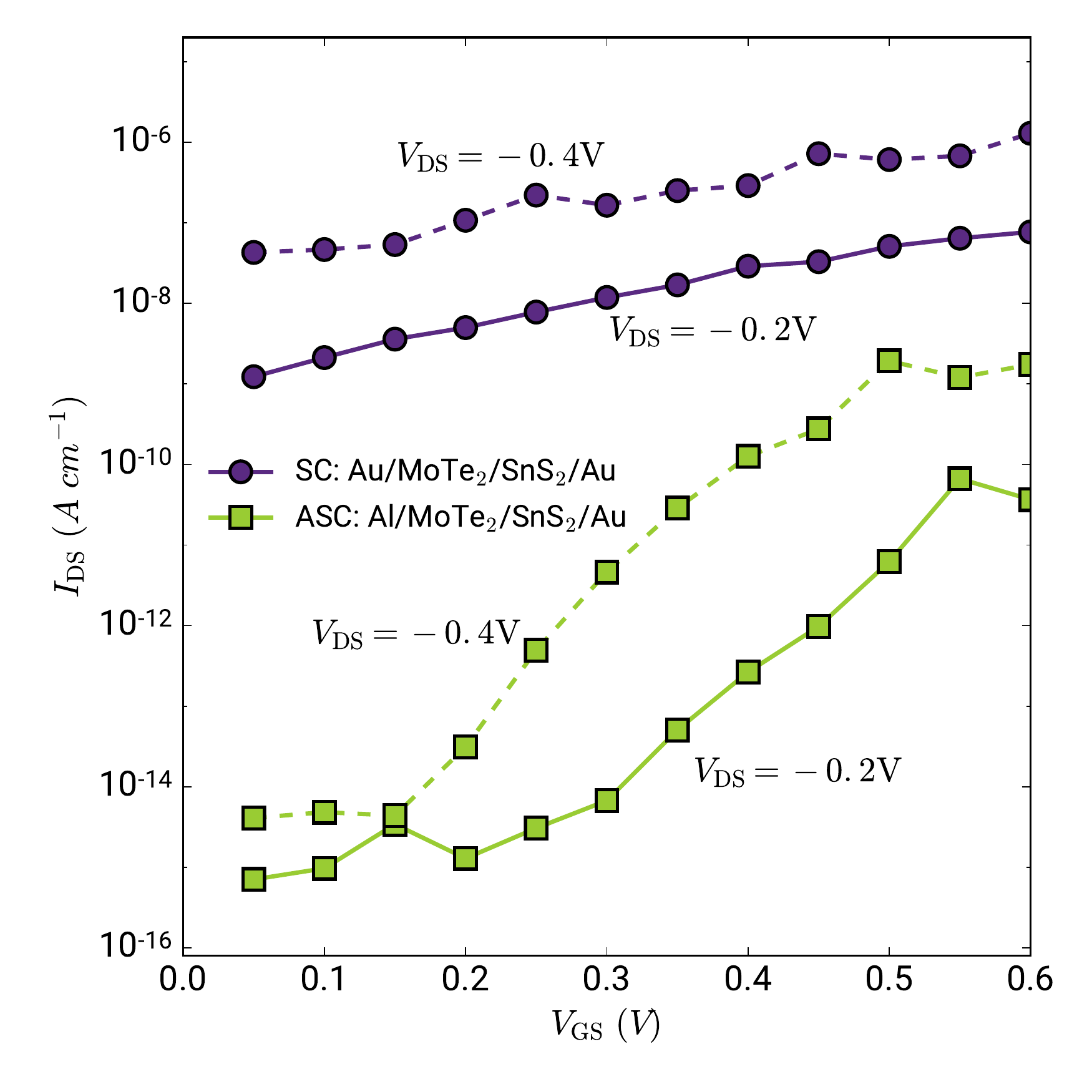}
\caption{(a,c) $I_\mathrm{DS}$-$V_\mathrm{GS}$ transconductance curves calculated for the symmetrically gated Au/MoTe$_2$/SnS$_2$/Au device at drain-source biases of $-0.2$~V (purple circles, solid line) and $-0.4$~V (purple circles, dashed line), and for the asymmetrically gated Al/MoTe$_2$/SnS$_2$/Au device at drain-source biases of $-0.2$~V (green squares circles, solid line) and $-0.4$~V (green circles, dashed line).}
\label{fig:2Ddevice_transconductance}
\end{figure}
\begin{figure*}
\includegraphics[width=1.0\linewidth]{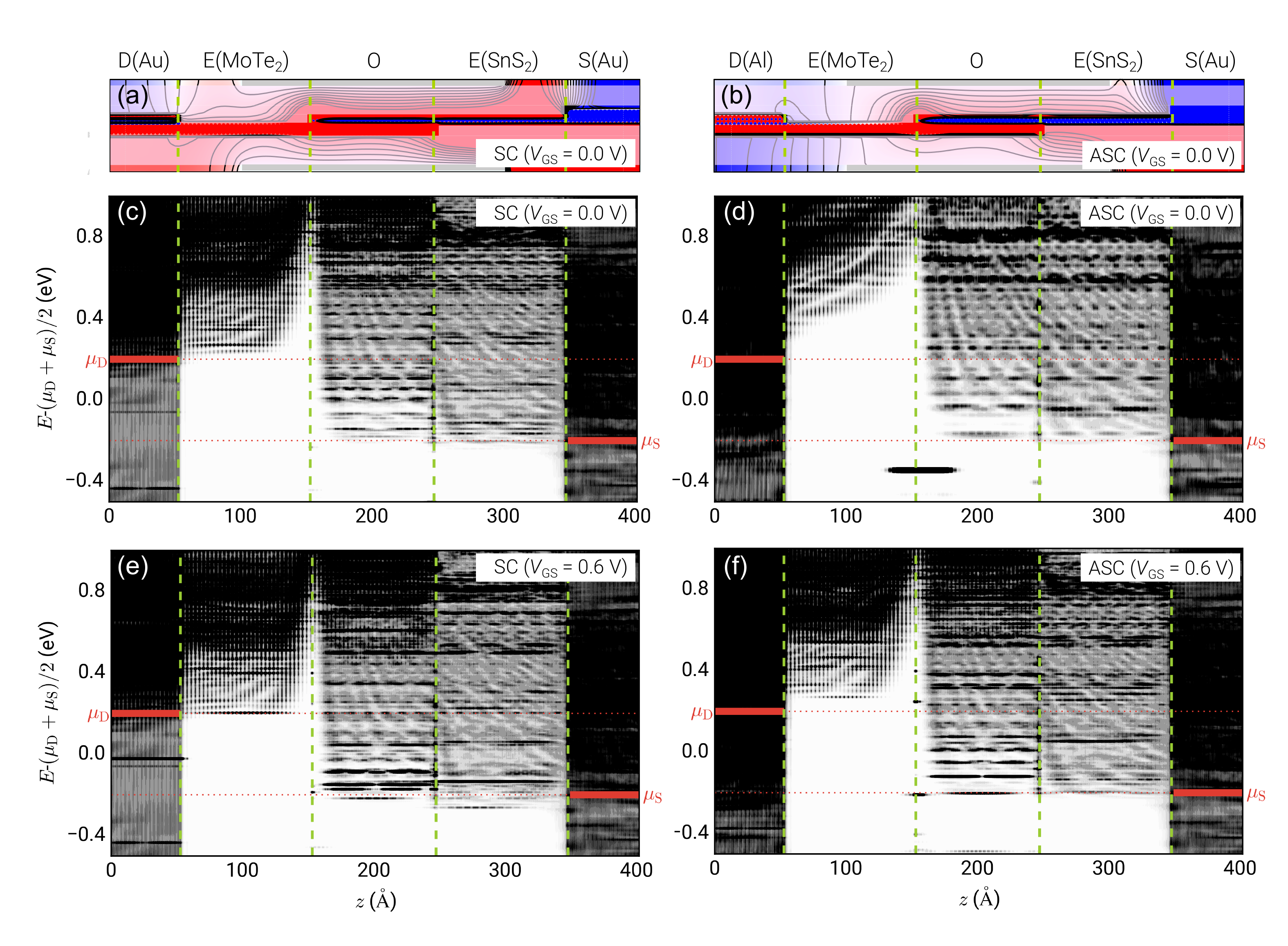}
\caption{(a) Cut-planes of the Hartree difference potential, $\Delta V^\mathrm{H}$, along the transport direction of the symmetrically contacted Au/MoTe$_2$/SnS$_2$/Au device. The potential is plotted in the range $-0.2\ \mathrm{eV} \leq \Delta V^\mathrm{H} \leq 0.2\ \mathrm{eV}$, with equipotential lines shown at every 0.025~meV. Regions of negative, zero, and positive potential are shown in blue, white, and red, respectively. The capital letters indicate the sections of the device corresponding to the drain (D) and source (S) electrodes, the overlap region (O), and the exposed region (E). (c,e) Projected local density of states along the transport direction for the SC device at $V_\mathrm{DS} = -0.2$~V and $V_\mathrm{GS} = 0.0$~V (c) and at $V_\mathrm{GS} = 0.6$~V (e). The red solid lines indicate the position of the left ($\mu_\mathrm{D}$) and right ($\mu_\mathrm{S}$) chemical potentials. The green dashed lines mark the boundaries of the different device regions. (b,d,f) Same as (a,c,e), but for the asymmetrically contacted Al/MoTe$_2$/SnS$_2$/Au device.}
\label{fig:2Ddevice_pldos}
\end{figure*}
As already described in Section~\ref{sec:negf}, the combination of DFT-LCAO with the NEGF method makes it possible to use \qatk\ to simulate the electronic structure and electrical characteristics of devices at the atomistic level. Field-effect transistor (FET) device configurations\cite{Stradi2017,Zhong2016} are simulated by including dielectric regions and electrostatic gates, see Section~\ref{sec:poisson}.

Here, we show how this framework can be used to study the electrical characteristics of a tunnel FET (TFET) device, where the channel is formed by a heterojunction based on two-dimensional semiconductors.\cite{Fiori2014,idrs2017mm} We demonstrate how the characteristics of the device can be tuned by using an asymmetric contact scheme. The latter is similar to that proposed for graphene-based photodetectors,\cite{Mueller2010} where two different metals are used to contact the graphene channel.

Figure~\ref{fig:2Ddevice_structure} shows the 2D-TFET device considered here. The device comprises a semiconducting channel formed by a MoTe$_2$/SnS$_2$ heterojunction.\cite{Szabo2015} We consider two different contact schemes by including atomistic metallic contacts:
In the symmetrically contacted (SC) $\mathrm{M_D}$/MoTe$_2$/SnS$_2$/$\mathrm{M_S}$ device,
Au is used for both the source (M$_\mathrm{S}$) and drain (M$_\mathrm{D}$) metallic contacts,
whereas in the asymmetrically contacted (ASC) device,
we set $\mathrm{M_D = Al}$ and $\mathrm{M_S = Au}$,
in order to have a rather large work function difference ($\Delta \Phi$) between $\mathrm{M_D}$ and $\mathrm{M_S}$.\cite{Singh2009}
In both devices, the metallic contacts to MoTe$_2$ and SnS$_2$ are represented by $\langle 110 \rangle$-oriented 4-layer slabs.

The device configurations were constructed from the optimized structures of the Au(110)/MoTe$_2$ and Au(110)/SnS$_2$ electrodes,
and the interlayer distance in the overlap region was set to 3.1~\AA.
Following Ref.~\onlinecite{Szabo2015}, the devices were encapsulated in a high-$\kappa$ dielectric region
(HfO$_2$, $\kappa = 25.0$), and a thin low-$\kappa$ dielectric region (h-BN, $\kappa = 6.0$)
was placed above the ``exposed'' MoTe$_2$ region that is not contacted or forms part of the overlap region, hereafter denoted $\mathrm{E(MoTe_2)}$.
Electrostatic top and bottom gates were defined outside the high-$\kappa$ dielectric region,
covering the overlap and half of the $\mathrm{E(MoTe_2)}$ and $\mathrm{E(SnS_2)}$ regions.
The ASC device was constructed by replacing the Au atoms in the left electrode with Al atoms,
with no further structural optimization.\cite{CheckElectrode2DFET}
Additional computational details are given in Appendix~\ref{appendix}.

To study the impact of the contact asymmetry on the device characteristics,
the reverse-bias $I_\mathrm{DS}$-$V_\mathrm{GS}$ curves (the transconductance)
were simulated for both devices and for two values of the drain-source voltage,
$V_\mathrm{DS} = -0.2$~V and $V_\mathrm{DS} = -0.4$~V,
by grounding the top gate and by sweeping the bottom gate.
The same physical picture emerges for both values of $V_\mathrm{DS}$,
and we discuss here only the results obtained for $V_\mathrm{DS} = -0.4$~V.
The $I_\mathrm{DS}$-$V_\mathrm{GS}$ curves in Fig.~\ref{fig:2Ddevice_transconductance}
show that the drain-source current is higher in the SC device than in the ASC device
across the entire range of gate-source voltages.
However, in the SC device, $I_\mathrm{DS}$ increases only by a factor of $\sim$10,
from $I_\mathrm{DS} (V_\mathrm{GS} = 0.05\ \mathrm{V}) = 4.28 \times 10^{-8}\ \mathrm{A\ cm^{-1}}$
to $I_\mathrm{DS} (V_\mathrm{GS} = 0.6\ \mathrm{V}) = 1.29 \times 10^{-6}\ \mathrm{A\ cm^{-1}}$.
Conversely, in the ASC device, $I_\mathrm{DS}$ increases by about six orders of magnitude
in the same $V_\mathrm{GS}$ range,
from $I_\mathrm{DS} (V_\mathrm{GS} = 0.05\ \mathrm{V}) = 4.09 \times 10^{-15}\ \mathrm{A\ cm^{-1}}$
to $I_\mathrm{DS} (V_\mathrm{GS} = 0.6\ \mathrm{V}) = 1.74 \times 10^{-9}\ \mathrm{A\ cm^{-1}}$.

Understanding these trends requires considering that the asymmetric contact scheme
has a two-fold effect on the electronic structure of the device.
On the one hand, the use of two metals with different work functions
leads to an additional built-in electric field in the channel region,
when the chemical potentials of the drain and source electrodes,
$\mu_\mathrm{D}$ and $\mu_\mathrm{S}$,
are aligned on a common energy scale.
On the other hand, the interaction between the metallic contact and $\mathrm{MoTe_2}$
is expected to depend also on the chemical nature of the metal.

The presence of an additional built-in electric field,
and its effect on the device electrostatics,
are evident by comparing the Hartree difference potential ($\Delta V^\mathrm{H}$)
in the two devices at $V_\mathrm{GS} = 0$~V along the channel,
as shown in Fig.~\ref{fig:2Ddevice_pldos}(a,b).
While in the SC device the potential changes smoothly along the channel region,
a sudden increase in the potential is observed in the ASC device around the $\mathrm{E(MoTe_2)}$ region.
Here, the potential lines run parallel to the transport direction,
indicating the presence of a left-pointing local electric field.
The sign of this field is consistent with that generated by an asymmetric contact scheme
with $\Phi^{M_\mathrm{S}} > \Phi^{M_\mathrm{D}}$, that is, the same as that of the ASC device.

The projected local density of states (PLDOS) along the devices reveal that the different electrostatics
also affect their electronic structure. For both contact schemes, the DOS within the bias window,
$[\mu_\mathrm{D}-\mu_\mathrm{S}]\pm k_\mathrm{B}T = \Delta\mu\pm k_\mathrm{B}T$,
is strongly inhomogeneous along the channel,
as the conduction bands (CBs) of MoTe$_2$ and SnS$_2$ are pinned
to $\mu_\mathrm{D}$ and $\mu_\mathrm{S}$, respectively (see Fig.~\ref{fig:2Ddevice_pldos}(c,d)).
This results in a vanishing DOS in the $\mathrm{E(MoTe_2)}$ region within the bias window.
Here, the DOS is even smaller in the ASC device,
due to (i) the weaker pinning of the CBs to $\mu_\mathrm{D}$,
and (ii) the effect of the local electric field, which bends and depletes even more the CBs,
moving them further away from the bias window.
In the SC device, the field is much weaker, and the CBs are bent only in the proximity of the overlap region.

The transconductance behavior can be understood from the combined analysis of $\Delta V^\text{H}$ and of the PLDOS.
The DOS within $\Delta\mu \pm k_\mathrm{B} T$ in the $\mathrm{E(MoTe_2)}$ region,
described in terms of an effective barrier $\phi_\mathrm{MoTe_2}$,
ultimately determines the reverse-bias current in the channel.
In the SC device, $\phi_\mathrm{MoTe_2}$ is lower for the case $V_\mathrm{GS} = 0$~V,
and depends only weakly on $V_\mathrm{GS}$, as shown in Fig.~\ref{fig:2Ddevice_pldos}(e).
This results in a higher absolute value of $I_\mathrm{DS}$,
and in a lower variation of $I_\mathrm{DS}$ with $V_\mathrm{GS}$.
Conversely, in the ASC device, $\phi_\mathrm{MoTe_2}$ is higher at comparable values of $V_\mathrm{GS}$,
and varies appreciably when $V_\mathrm{GS}$ is increased, see Fig.~\ref{fig:2Ddevice_pldos}(f).
This explains the lower values of the drain-source current,
and its higher variation with the gate-source voltage.
These trends are consistent with those of the transconductance curves shown in Fig.~\ref{fig:2Ddevice_transconductance}.

In summary, DFT-NEGF simulations for
$\mathrm{M_D}$/MoTe$_2$/SnS$_2$/$\mathrm{M_S}$ ultra-scaled 2D-TFET devices
show that the transconductance can be engineered by an appropriate choice
of the metallic electrodes, and highlight the importance of atomistic device simulations
for optimization of the electrical characteristics of devices based on non-conventional semiconductors.

\subsection{Phonon-limited Mobility of Metals}
\label{sec:mobility}
\begin{figure}
\centering
{\includegraphics[width=0.9\linewidth]{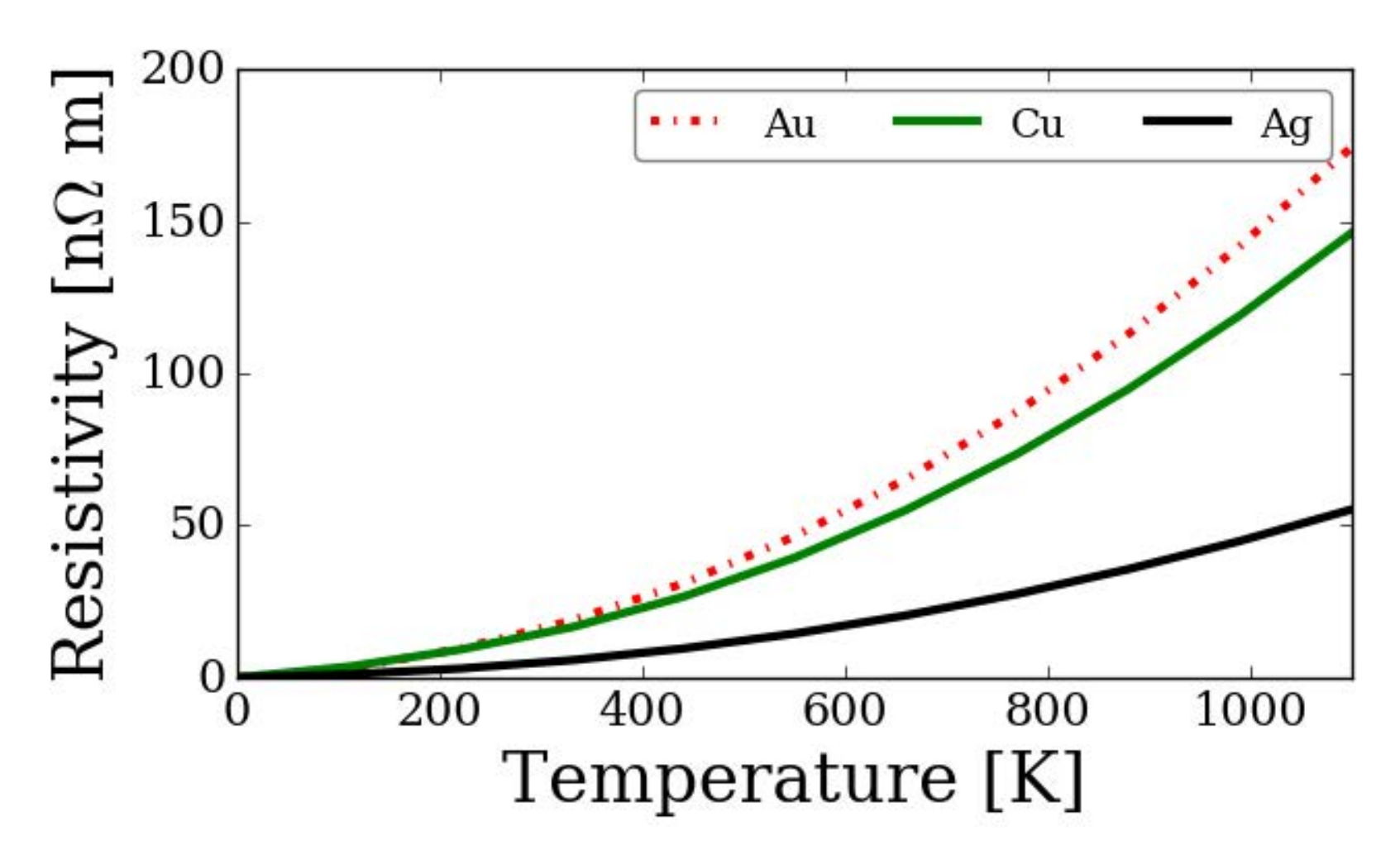}}
\caption{
Temperature-dependent phonon-limited resistivity of the three metals Au, Ag and Cu
evaluated from first-principles simulations using the \atklcao\ engine.
}
\label{fig:ResistivityVsT}
\end{figure}
\begin{table}
\begin{ruledtabular}
\begin{tabular}{lcc}
           & DFT  & Experiment  \\
\hline \cline{1-3} \\ [-2ex]
  Au, Bulk & 15.9 &  20.5 \\ 
  Au, NW ($d\approx1$~nm) & 56.0 & - \\ 
  Ag, Bulk & 4.9 &  14.7\\ 
  Ag, NW ($d\approx1$~nm) & 28.7 & - \\ 
  Cu, Bulk & 14.2  & 15.4 \\ 
  Cu, NW ($d\approx1$~nm) & 98.3 & - \\ 
\end{tabular}
\caption{First-principles phonon-limited resistivities at 300~K
(in units of $\textrm{n}\Omega\cdot \textrm{m}$),
compared with experimental values from Ref.~\onlinecite{kasap_electrical_2017}.
Au nanowire results from Ref.~\onlinecite{markussen_electron-phonon_2017}.}
\label{Tbl:Resistivities}
\end{ruledtabular}
\end{table}

The continued downscaling of nanoelectronics makes the metal interconnects an increasingly critical part of transistor designs.\cite{josell_size-dependent_2009}
Present-day transistors use Cu as an interconnect material, and a good understanding of the origin of resistance increase with downscaling of interconnects will be important for the design and performance of future nanoscale devices.

We here present first-principles calculations of the phonon-limited resistivity of three FCC metals; Cu, Ag, and Au.
We solve the Boltzmann transport equation for the mobility,
using first-principles EPC constants, as described in section~\ref{sec:bulktransport}.
Such DFT calculation of the resistivity of metals is computationally demanding,
as one needs to integrate the EPC over both electron and phonon wave vectors ($\bk$- and $\bq$-space),
and we know of only few studies of the EPC in metals that includes a full integration.\cite{bauer_electron-phonon_1998,gall_electron_2016,markussen_electron-phonon_2017} We here show that the tetrahedron integration method
enables computationally efficient mobility calculations.
The method may therefore be used for computational screening of materials,
and first-principles simulations become accessible for identifying promising replacement materials for future interconnects.

To calculate the scattering rate related to EPC, the phonon modes and derivatives of the Hamiltonian with displacements are needed.
The supercell method for calculation of phonons and EPC from first principles was described in section~\ref{sec:phonons},
and Fig.~\ref{fig:PhononBand} showed the phonon band structures of Cu, Ag and Au,
calculated using the \atklcao\ simulation engine.
For the integration of the scattering rate in \eqref{eqn:tau} we use a sampling of
$20 \times 20 \times 20$ $\bq$-points and tetrahedron integration.
In addition, we apply the two-step procedure, where a $\bk$-space isotropic but energy-dependent scattering rate
is used to efficiently evaluate the resistivity.

Figure~\ref{fig:ResistivityVsT} shows the DFT results
for the temperature-dependent phonon-limited resistivity of bulk Cu, Ag, and Au
(Debye temperatures of 347, 227 and 162~K,\cite{stewart_measurement_1983} respectively).
The resistivity increases with temperature as the phonon occupation increases,
and becomes linearly dependent on temperature above the Debye temperature.

Table~\ref{Tbl:Resistivities} presents the calculated room-temperature bulk resistivities,
and compares them to experiments and to calculated values for metal nanowires (NWs) with diameters $d=1$~nm.
In agreement with experiments, we find that Au has the largest resistivity,
and that Ag is more conductive than Cu.
In addition, the resistivity increases significantly when forming nanowires of the elements.
Despite the fact that the phonon dispersions of bulk Au and Ag are very similar,
the resistivity is quite different.
In the minimal free-electron model of metals,
the conductivity is given by $1/\rho(T) = \frac{1}{3}e^2v_\text{F}^2 \tau(T) n(\varepsilon_\text{F})$.
In the three FCC metals considered here, the Fermi velocity, $v_\text{F}$, and the DOS, $n(\varepsilon_\text{F})$,
(and resulting carrier density) are almost identical,
and the difference in the resistivity is traced back to the variation in the scattering rate.
This shows how full first-principles Boltzmann transport simulations of the scattering rate
is needed to capture the origin of the resistivities of different metals.
While the resistivity of bulk Ag is slightly underestimated by the simulations,
we find good agreement with experiments for bulk Au and Cu,
as well as the correct ranking of the individual metals.
This illustrates the predictive power of the method.
In general, we find that the resistivity of $d=1$~nm nanowires
is increased by a factor of three for Au and even more for Ag and Cu,
as compared to bulk, due to the increased EPC in nanowires.

\subsection{Multi-model Dynamics with an Applied Electric Field}
\label{sec:mdefield}
\begin{figure}[!t]
\centering
\includegraphics[width=\linewidth]{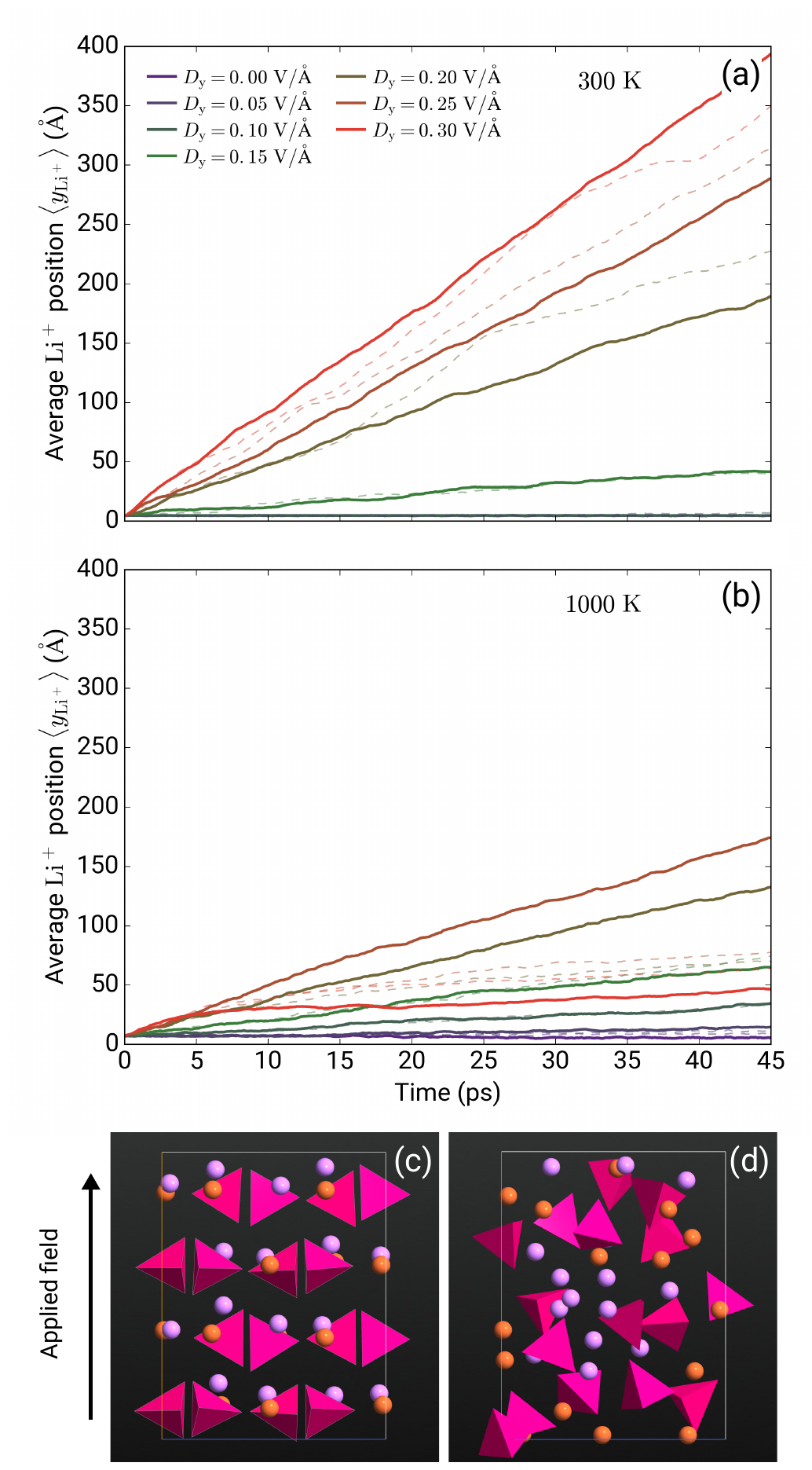}
\caption{Average position of the $\mathrm{Li^+}$ ions in $\mathrm{LiFePO_4}$ along the $y$ Cartesian direction (along the [010] channel), as a function of time, calculated at temperatures 300~K (a) and 1000~K (b), and for electric field strengths from $D_y = 0.0$ (purple lines) to $D_y = 0.3$ (red lines) V\AA. The data obtained from multi-model and FF simulations are shown as solid and dashed lines, respectively. (c,d) Snapshots obtained from the simulations after 40~ps of simulation for 300~K (c) and 1000~K (d). The lithium and iron ions are shown in pink and orange, respectively, while the magenta tetrahedra represent the phosphate groups. The black arrow indicates the direction of the applied field.}
\label{fig:multimodel1}
\end{figure}
The tight integration of different atomic-scale simulation engines within the same software framework
allows for straight-forward combination of multiple atomistic models into one single simulation workflow.
This enables elaborate computational workflows and extend the functionality of \qatk\
beyond that of methods based on a single atomistic model.
We here show how such a \textit{multi-model approach}
can be used to implement a hybrid method that combines classical FF MD simulations
with a DFT description of time-dependent fluctuations of the atomic charges as the MD simulation progresses.

We study here LiFePO$_4$, a promising cathode material of the olivine family for Li-ion batteries.\cite{Islam2014,Zhang2011}
In this class of materials, the olivine scaffold provides natural diffusion channels
for the $\mathrm{Li^+}$ ions, which have been shown to diffuse via
a hopping mechanism, preferentially through [010]-oriented channels.\cite{Boulfelfel2011,Hu2015}
MD simulations aimed at understanding the diffusion process
have focused mainly on its temperature ($T$) dependence.
In this case, relatively high temperatures,
usually in the range 500 to 2000~K,
are required to reach a sufficiently high hopping probability
within a reasonable MD simulation time,
and allow for calculation of the associated diffusion constants.
These simulations have demonstrated that the diffusion increases with $T$,
as a natural consequence of the increased hopping probability favored by Brownian motion.

However, in an electrochemical cell under operating conditions,
the motion of the $\mathrm{Li^+}$ ions may also
have a non-negligible drift component,
due to the displacement field resulting from the voltage difference
applied between the anode and the cathode.
This potentially rather important effect is rarely taken into account in atomistic simulations.\cite{English2015,Rungger2010}

Another significant issue in the simulation of Li-ion batteries
is related to the inclusion of electronic effects.
In order to reach reasonably long simulation times,
to describe atom diffusion at temperatures close to 300~K,
most low-$T$ MD simulations are based on FFs,
which by construction neglect any time-dependent fluctuations
of the electronic density during the MD run.
A number of models have tried to address this issue
by either including approximate models to account for the charge fluctuation,\cite{Kim2011}
or by running semi-classical dynamics on precalculated potential-energy surfaces based on DFT.\cite{Kahle2018}

A \qatk\ multi-model approach can be used to address these issues
by including first-principles charge fluctuations in the FF MD.
The applied displacement field should add a force term
$\mathbf{F}^\prime_a = Q_a \, \mathbf{D}$ on the $a$-th ion
with formal charge $Q_a$ and $\mathbf{D}$ being the field vector.
However, in a FF, $Q_a$ is a time-independent parameter,
so the field-induced force will also be time-independent.
In a multi-model approach, we instead use DFT simulations
to determine the instantaneous charge $Q_a$
at regular intervals during the MD.
Time-dependency in the field-induced force term is then included
by use of a MD hook function (see Section~\ref{sec:dynamics})
by defining the time-dependent formal charge $Q^\prime_a(t)$ as
\begin{equation}
Q^\prime_a = \mathrm{Q}_a^\mathrm{FF} + \Delta Q_a(t),
\label{multimodel1}
\end{equation}
where $\Delta Q_a$ describes the time-dependent fluctuation.
In principle, $\Delta Q_a$ can be defined arbitrarily,
provided that charge neutrality is maintained in the system.
In the present case, we chose a simple definition,
\begin{equation}
\Delta Q_a(t) = \mathrm{Q}_a^\mathrm{DFT}(t) - \mathrm{Q}_a^\mathrm{DFT,ref} ,
\label{multimodel2}
\end{equation}
where $\mathrm{Q}_a^\mathrm{DFT}(t)$ and $\mathrm{Q}_a^\mathrm{DFT,ref}$
are the time-dependent charge of the $i$-th atom obtained from a DFT calculation
for the MD configuration at time $t$,
and a time-independent charge obtained for a reference configuration at $T = 0$~K, respectively.
We note that, in the present case, the lack of consistency between the methods used to calculate
$\mathrm{Q}_a^\mathrm{FF}$ and $\Delta Q_a(t)$ does not constitute an issue,
since the charge fluctuations during the dynamics are of the order $\Delta Q_a \sim 0.1\ e^-$.

We have applied this multi-model approach to investigate the interplay between Brownian
and drift components of the diffusion of $\mathrm{Li^+}$ ions along the [010] channels in LiFePO$_4$
in the presence of an applied displacement field.
The system was described by a $1 \times 2 \times 1$ LiFePO$_4$ 112-atom supercell,
that is, 2 times the conventional unit cell (16 formula units).
For the classical part of the multi-model simulations,
we used a FF potential by Pedone \textit{et al.},\cite{Pedone2006}
which has been shown to describe qualitatively correctly the geometry and transport
properties of olivine materials.\cite{Kutteh2014}
The \atklcao\ engine was used for the DFT part.
MD simulations were performed at temperatures 300~K
and 1000~K for a displacement field $\mathbf{D} = [0, D_y, 0]$,
with $0.0\ \text{V/\AA} \leq D_y \leq 0.3\ \text{V/\AA}$.
For each temperature, a 5~ps equilibration run using a NPT ensemble
was performed, starting from the structure optimized at 0~K,
using a Maxwell--Boltzmann distribution of initial velocities,
followed by a 45 ps production run using a NVT ensemble.
The MD time step was 1.0~fs,
and $\Delta Q_a(t)$ was recalculated every 100 MD steps, see \eqref{multimodel2},
with $\mathrm{Q}_a^\mathrm{DFT}(t)$ and $\mathrm{Q}_a^\mathrm{DFT,ref}$
obtained from Mulliken population analysis.
Further computational details are given in Appendix~\ref{appendix}.

Figure~\ref{fig:multimodel1}(a,c) shows the average displacement
$\langle y_\mathrm{Li^+} \rangle$ of the $\mathrm{Li^+}$ ions along the $y$ Cartesian direction,
that is, along the [010] channels of the FePO$_4$ scaffold,
calculated for temperatures 300 and 1000~K
and for increasingly higher values of the applied field,
using either FFs only or the FF+DFT multi-model approach.
In the absence of an applied field and at 300~K,
the average $\mathrm{Li^+}$-ion displacement remains constant
at $\langle y_\mathrm{Li^+} \rangle = 4.67\pm0.11$~\AA\
during the entire simulation, indicating the absence of hopping events.
At 1000~K, the situation is rather similar,
as $\langle y_\mathrm{Li^+} \rangle$ increases only slightly from an initial value of
$7.12\pm0.19$~\AA\ (obtained from an average of the snapshots collected
during the first picosecond of the FF-only MD)
to a final value of $9.16\pm0.13$~\AA\
(obtained from an average of the snapshots collected during the last picosecond).
For the multi-model simulation, we observe instead a small decrease of $\langle y_\mathrm{Li^+} \rangle$ over time.
This indicates that, at both temperatures, $\mathrm{Li^+}$ hopping due to Brownian motion is a rare event.

Applying an increasingly stronger displacement field
leads to a progressive increase in the $\mathrm{Li^+}$ hopping probability.
At 300~K, the average $\mathrm{Li^+}$-ion displacement increases steadily
from the beginning of the MD run for $D_y \geq 0.20$~V/\AA,
indicating that, for these values of $D_y$,
$\mathrm{Li^+}$ hopping is primarily due to field-induced drift.
The $\mathrm{Li^+}$ ions accelerate until they reach a constant velocity,
as shown by the tendency of the $\langle y_\mathrm{Li^+} \rangle$ vs.\ time curves
to continually decrease their slope, corresponding to a straight line on a linear plot.

In the absence of an applied field,
increasing the temperature should increase the probability of $\mathrm{Li^+}$ ion diffusion
due to increased Brownian motion.\cite{Kutteh2014,Hu2015}
However, in the present case we find that the $\mathrm{Li^+}$ ions move less at
1000~K than at 300~K. For comparable values of $D_y$,
the $\langle y_\mathrm{Li^+} \rangle$ vs.\ time curve
has a smaller slope at 1000~K than those calculated at 300~K.
The reason is that collision events of the $\mathrm{Li^+}$ ions
with the LiFePO$_4$ lattice,
where phonons are considerably more excited at higher temperatures than at room temperature,
limits the effective velocity of the $\mathrm{Li}^+$ ions.

This is evident by comparing the LiFePO$_4$ structures at the two temperatures.
Figure~\ref{fig:multimodel1}(c,d) shows two snapshots extracted at the end of the MD runs
at $D_y = 0.3$~V/\AA\ and at temperatures 300 and 1000~K, respectively.
At 300~K, the LiFePO$_4$ structure is relatively unperturbed.
Consequently, the $\mathrm{Li^+}$ ions are able to travel
through the [010] channels with relatively few scattering events
with the LiFePO$_4$ lattice.
Conversely, at 1000~K, the LiFePO$_4$ structure is significantlty perturbed,
leading to a high probability of collisions between the $\mathrm{Li^+}$ ions and the olivine lattice.

In summary, we have studied the diffusion of $\mathrm{Li^+}$ in olivine LiFePO$_4$,
using a multi-model computational approach
that combines a classical FF with DFT,
the latter to include the effect of the field and of time-dependent charge fluctuations.
Our analysis highlights the importance of considering the combined effect of both Brownian
and drift contributions to the $\mathrm{Li^+}$ hopping to describe the overall process,
which strongly depends on not only the temperature itself,
but also on the probability of collision events between the diffusing ions
and the FePO$_4$ lattice.
\subsection{\label{sec:SiGe_alloys}Electronic Structure of Binary Alloys}
\begin{figure}[!h]
\includegraphics[width=\columnwidth]{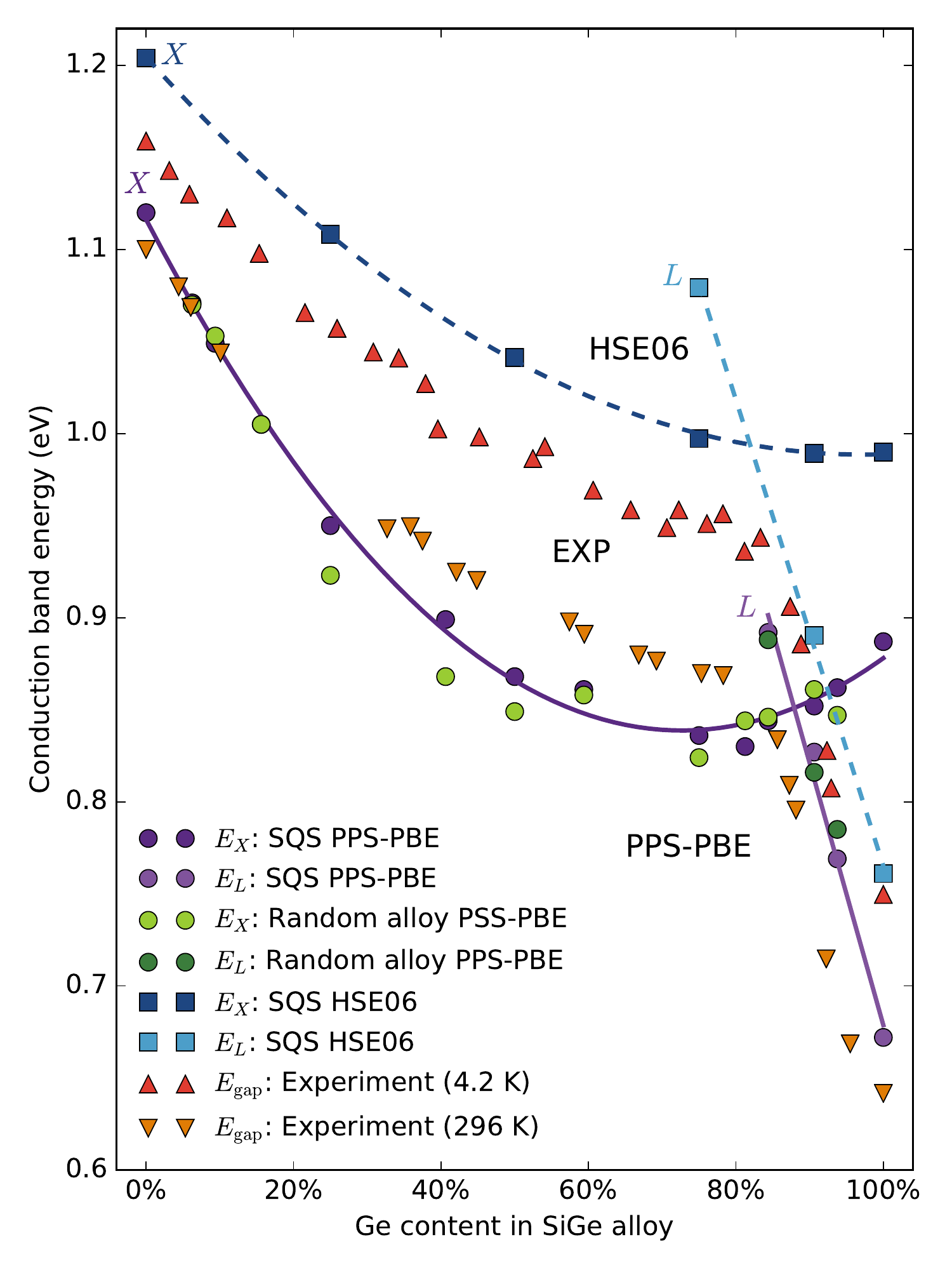}
\caption{Conduction band energies ($E_\text{X}$ and $E_\text{L}$) of Si$_{1-x}$Ge$_{x}$ alloy as a function of Ge content, $x$, calculated using PPS-PBE and HSE06 functionals in combination with the LCAO basis set and the PW basis set, respectively. The $E_\text{X}$ and $E_\text{L}$ energies are defined with respect to the top of the valence band ($E_\text{val}$) at the $\Gamma$-point. Details on the definition of band energies at special $\bk$-points for disordered alloys can be found elsewhere.\cite{khomyakov2015alloys} Reference experimental data (open markers) on the band-gap compositional dependence, $E_\mathrm{gap}(x)$, are given for low (4.2~K) and room (296~K) temperatures.\cite{braunstein1958exp} The dashed (solid) lines correspond to linear (quadratic) interpolation of the DFT-calculated band energies, $E_\text{L}$ ($E_\text{X}$), given with filled markers; the interpolation formulas are given in Table~\ref{tab:sige}.}
\label{fig:SiGe_bandgap}
\end{figure}
\begin{figure*}
\begin{tabular}{cc}
\includegraphics[width=\columnwidth]{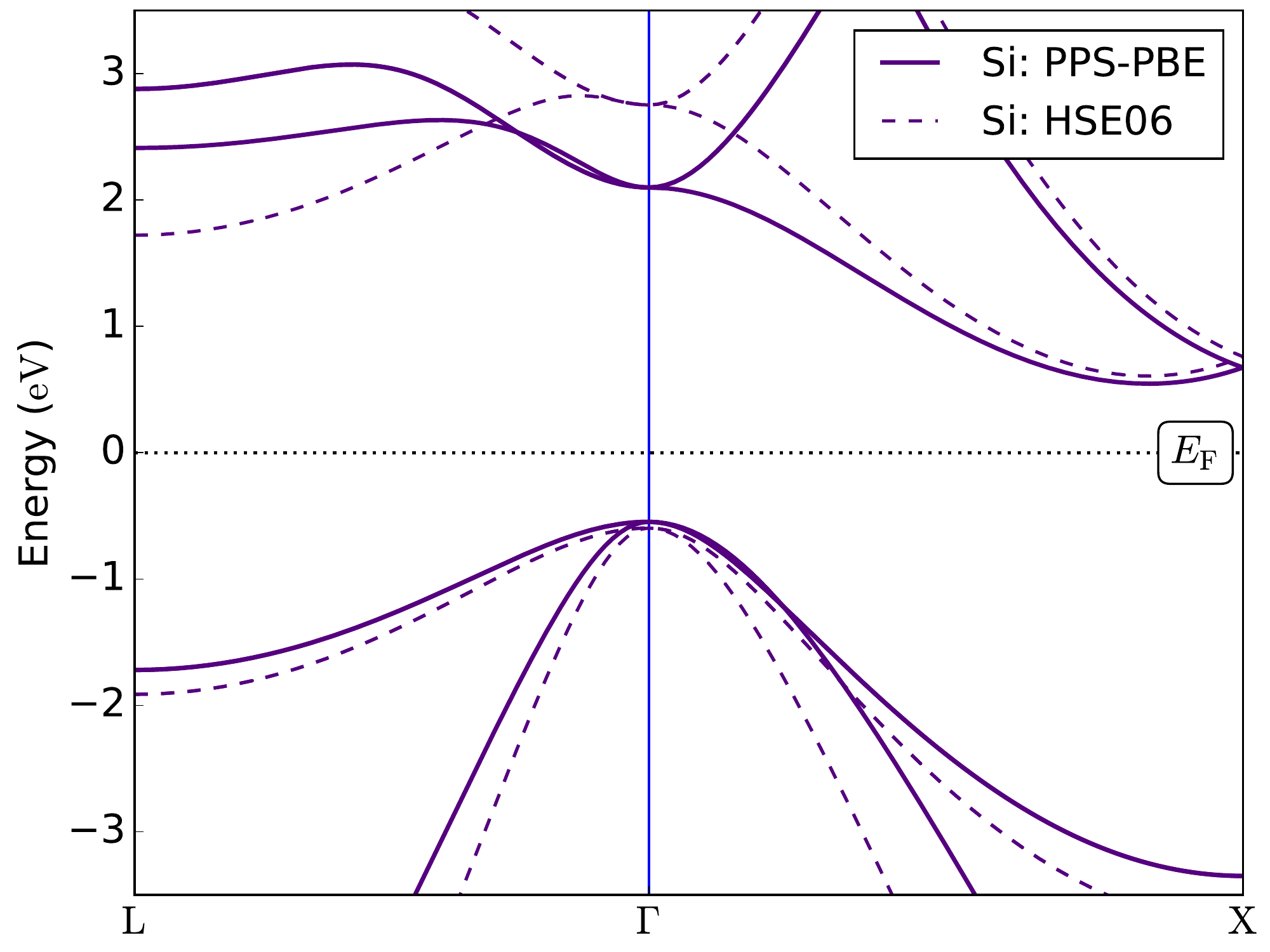}&
\includegraphics[width=\columnwidth]{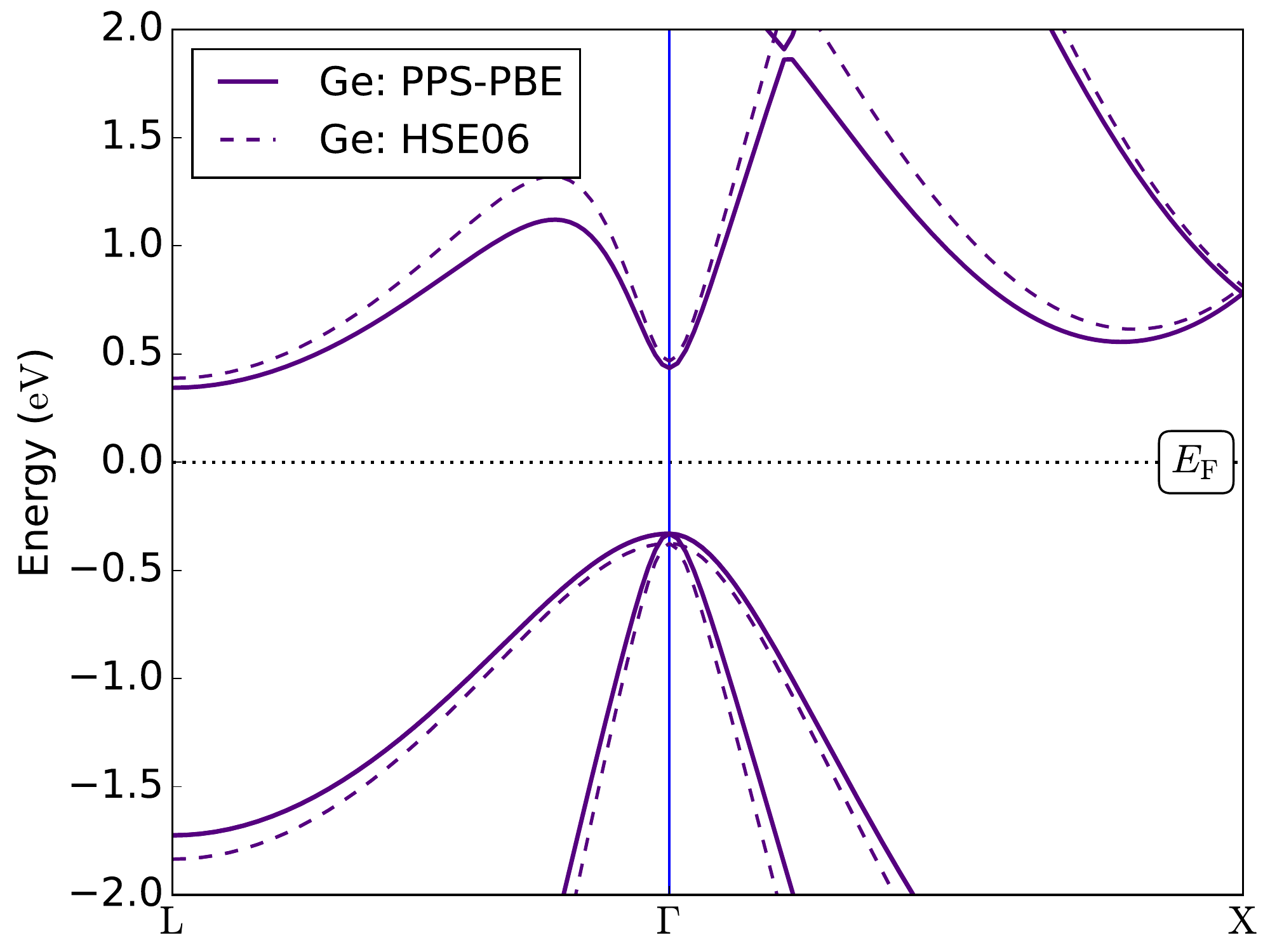}\\
\end{tabular}
\caption{Band structure of bulk Si (left panel) and Ge (right panel) obtained using the PPS-PBE (solid line) and HSE06 (dashed line) methods. The calculations used a $\Gamma$-centered $12\times 12\times 12$ $\bk$-point grid to sample the Brillouin zone of the 2-atom primitive cells.}
\label{fig:Si_Ge_band_structure}
\end{figure*}
Understanding the physical properties of semiconductor alloys,
such as silicon-germanium binary compounds,
is highly relevant, since such alloys are commonly used in
microelectronics as a semiconductor material for, e.g.,
heterojunction bipolar transistors or as a strained semiconductor layer in CMOS transistors.\cite{taur1997sige}
Moreover, device-level TCAD simulations,
frequently used in industrial semiconductor research and development,
usually require material-dependent input parameters such as band gap,
effective masses, deformation potentials, and many others.\cite{li2012tcad}
Atomic-scale simulations may be used to calculate such parameters from first principles
if experimental values are not available,
including composition dependence.\cite{khomyakov2015alloys}
However, simulating randomly disordered alloys may be computationally challenging
since the traditional approach to random-alloy (RA) simulations
use stastical sampling of multiple large supercells
with random atomic arrangements (configurational averaging)
to take into account the effect of disorder
on the physical properties of alloys.

We here adopt the special quasi-random structure (SQS) approach\cite{zunger1990sqs}
for DFT modelling of SiGe random alloys,
which significantly reduces the computational cost.
Unlike in the RA approach, in the SQS method the configurational averaging
of band energies is captured by a single supercell structure.
We study 64-atom Si$_{1-x}$Ge$_{x}$ supercells in the full range of compositions,
$0\leq x\leq 1$, by calculating composition-dependent lattice constants
and band energies.
The PPS-PBE method,~\cite{Smidstrup2017} discussed in Section~\ref{sec:xc},
was used with the \atklcao\ simulation engine,
and we compare the band energies to those obtained with the HSE06 hybrid functional
using the \atkpw\ engine.
We also compare SQS band energies to those calculated using the traditional RA approach,
obtained by averaging over 5 randomly generated RA configurations.
In both the SQS and RA cases, the band energies were computed by averaging the energies
of the conduction (valence) band states split by alloy disorder.\cite{khomyakov2015alloys}

We used a NanoLab SQS module to generate the SQS configurations.
The module uses a genetic algorithm to optimize finite-size alloy configurations
to reproduce selected correlation functions of the infinite alloy system.
The genetic optimization algorithm is very efficient,
and systems with many hundred atoms are easily handled.
In this case, the SQS structures were generated by fitting all pair,
triplet, and quadruplet correlation functions
with figure sizes up to 7.0, 5.0 and 4.0~\AA, respectively,
such as to match those correlation functions for the truly random alloy,
as detailed in Refs.~\onlinecite{zunger1990sqs,vandewalle2009sqs}.
Generation of a single 64-atom SiGe SQS alloy takes about 4 minutes on a modern 4-core processor.
The alloy configurations were then relaxed using PPS-PBE
followed by band structure analysis.
HSE06-level band structures were calculated
without further relaxation.
More computational details are given in Appendix~\ref{appendix}.
\begin{table}
\caption{\label{tab:sige}
First-principles interpolation formulas for the Si$_{1-x}$Ge$_{x}$ composition-dependent
band gap and lattice constant.
The variables $b_\text{w}$, $b_\text{w}^{\prime}$, and $b_\text{w}^{\prime\prime}$ are bowing parameters.}
\begin{ruledtabular}
\begin{tabular}{lc}
        & Band gap interpolation formula (eV) \\
[0.5ex] \hline \cline{1-2} \\ [-2ex]
\multirow{3}{*}{PPS-PBE} & $E_\text{X} = 1.116 - 0.764 x + b_\text{w}^{\prime} x^{2}$ \\
        & $b_\text{w}^{\prime}=0.526$~eV \\
        & $E_\text{L} = 2.104 - 1.425 x$ \\
\hline \\ [-2ex]
\multirow{3}{*}{HSE06} & $E_\text{X} = 1.204 - 0.444 x + b_\text{w}^{\prime\prime} x^{2}$ \\
        & $b_\text{w}^{\prime\prime}=0.228$~eV \\
        & $E_\text{L} = 2.032 - 1.267 x$ \\
\hline \\
        & Lattice constant interpolation formula (\AA) \\
[0.5ex] \hline \cline{1-2} \\ [-2ex]
\multirow{2}{*}{PPS-PBE} & $a(x) = 5.431 + 0.257 x + b_\text{w} x^{2}$ \\
        & $b_\text{w}=0.034$~\AA
\end{tabular}
\end{ruledtabular}
\end{table}

Figure~\ref{fig:SiGe_bandgap} shows the Si$_{1-x}$Ge$_{x}$ composition dependent
conduction band minima (CBM), referenced to the valence band maximum,
for both the $X$- and $L$-valley in the SiGe BZ.
We first note that SQS band energies
are very similar to the those calculated using the more expensive RA approach.
It is well known that the Si$_{1-x}$Ge$_{x}$ fundamental band gap
changes character at $x\sim 0.85$.
The PPS-PBE and HSE06 predictions of the transition point
are $x\sim 0.88$ and 0.82, respectively.
As expected, the calculated $X$-valley conduction-band energies exhibit bowing, i.e.,
nonlinear behavior of these quantities with respect to Ge content, $x$.
The best-fit interpolation formulas, shown as lines in Fig.~\ref{fig:SiGe_bandgap},
are listed in Table~\ref{tab:sige},
including the band-gap compositional bowing parameters.
The PPS-PBE band gaps
are in good agreement with room-temperature experiments
(within $\sim$50~meV for the entire range of Ge content),
while the HSE06 band gaps are in better agreement with low-temperature experiments.
Moreover, the HSE06-based approach appears to more accurately describe the
band-gap bowing parameter,
while PPS-PBE tends to overestimate it.
Finally, the calculated SiGe lattice constant also exhibits compositional bowing,
as indicated by the interpolation formula in Table~\ref{tab:sige}.
The bowing parameter of 0.034~\AA\ is
overestimated by $\sim$26\% as compared to experiments (0.027~\AA).

To benchmark the empirical PPS-PBE method against the parameter-free HSE06 approach,
we also calculated the band structure of bulk Si and Ge
using both methods, as shown in Fig.~\ref{fig:Si_Ge_band_structure}.
The PPS-PBE conduction and valence bands around the Fermi energy
are in good agreement with the HSE06 band structure.
This is consistent with the fact that the PPS-PBE method was fitted to experimental data,
and that the HSE06 hybrid functional accurately simulates the band structure of bulk semiconductors.

In summary, we find that the SQS approach is well suited to describe
the compositional bowing of the band energies in Si$_{1-x}$Ge$_{x}$ random alloys,
suggesting that SQS provides an accurate and efficient approach to random-alloy simulations.
The HSE06 hybrid functional accurately describes
the conduction-band energies of SiGe alloys and their compositional bowing,
while the PPS-PBE method offers a computationally efficient alternative
if only bands around the Fermi level are important.

\section{Summary}
\label{sec:Summary}
%
In this paper we have presented the \qatk\ platform
and details of its atomic-scale simulation engines,
which are ATK-LCAO, ATK-PlaneWave, ATK-SE, and ATK-ForceField.
We have compared the accuracy and performance of the different engines,
and illustrated the application range of each.
The platform includes a wide range of modules for application of the different
simulation engines in solid-state and device physics,
including electron transport, phonon scattering, photocurrent,
phonon-limited mobility, optical properties, static polarizations, molecular dynamics, etc.

The simulation engines are complimentary and through the seamless Python integration
in the \qatk\ platform, it is easy to shift between different levels of theory
or integrate different engines into complex computational workflows.
This has been illustrated in several application examples,
where we for example showed how ATK-LCAO and ATK-ForceField can be combined
to study $\mathrm{Li^+}$-ion drift in a battery cathode material.
We also presented applications of \qatk\ for simulating electron transport in 2D materials,
phonon-limited resistivity of metals, and electronic-structure simulations of SiGe random alloys.

While several of the simulation engines and methods have been described independently
before,\cite{stokbro2010semiempirical, ATKForceField, stradi2016general,gunst_first-principles_2016, gunst_first-principles_2017, Smidstrup2017, palsgaard2018efficient, petersen2008block, Smidstrup2014,stradi2017method}
we have here provided an overview of the entire platform, including implementation details not previously published.
We expect that this paper can become a general reference for documenting
the \qatk\ platform, and is a reference to its applications for atomic-scale modelling
in semiconductor physics, catalysis, polymer materials, battery materials, and other fields.

\begin{acknowledgments}
Authors acknowledge funding from the European Union's Horizon 2020 research
and innovation programme under grant agreements No 713481 (SPICE),
No 766726 (COSMICS), and No 723867 (EMMC-CSA),
as well as funding from the Quantum Innovation Center (QUBIZ)
and the Lundbeck Foundation (R95-A10510).
CNG is sponsored by the Danish National Research Foundation (DNRF103).
\end{acknowledgments}

\appendix
\vspace{1cm}
\section{Computational Details}
\label{appendix}
In the simulations presented in Fig.~\ref{fig:ComparingModels},
we have considered non–crystalline a-$\mathrm{Al_2O_3}$ structures
with a constant density of 2.81 g/cm$^3$.
The system sizes considered were formed by 5, 30, 60, 120, 240, 480,
960, 1920, 3840, 7680, 15360, and 30720 atoms, respectively,
each system with the appropriate unit-cell volume.
The amorphous phases were generated by randomizing the structure at 5000~K
and then quenching to 0~K to avoid extremely small bond distances.
The MD simulations were then performed at 300~K
using a random Boltzmann distribution of initial velocities and a Langevin thermostat.
The FF simulations used a Pedone potential,\cite{Pedone2006}
while the TB simulations used a Slater--Koster parametrization.
For the DFT-LCAO and DFT-PW simulations,
we used normconserving PseudoDojo pseudopotentials
with a Medium basis set and a kinetic-energy cutoff energy of 1360~eV (50~Ha), respectively.
For TB, DFT-LCAO, and DFT-PW simulations,
the Brillouin zone was sampled using a
Monkhorst--Pack\cite{MonkhorstPack1976} (MP)
$\bk$-point density of 3--4~\AA.
For systems with sizes between 240 and 960 atoms,
2 processes/$\bk$-point was used,
whereas for the 1920-atom system, 16 processes/$\bk$-point was used.

For the DFT-NEGF device simulations presented in Section~\ref{sec:2DFET},
we used the PBE density functional with SG15-Medium (FHI-DZP)
combinations of pseudopotentials and basis sets
for MoTe$_2$ and SnS$_2$ (Au and Al).
The real-space cutoff energy was 2721~eV (100~Ha),
and MP $\bk$-point grids of $12 \times 1 \times 100$ and $12 \times 1$
were used to sample the BZ of the electrode and of the device, respectively.

In the study of multi-model dynamics presented in Section.~\ref{sec:mdefield},
we used the \atklcao\ engine with a DZP basis set
and a real-space cutoff energy of 2180~eV (80~Ha).
Exchange-correlation effects were described by the PBE functional,
and the FePO$_4$ BZ was sampled using a $3 \times 3 \times 2$ MP $\bk$-point grid.

For the electronic-structure calculations for SiGe random alloys presented in Section~\ref{sec:SiGe_alloys},
we used a $3\times 3\times 3$ MP $\bk$-point grid
and an electron temperature of 0.025~eV for the Fermi--Dirac occupation function.
SG15 (FHI) pseudopotentials were used for the PSS-PBE (HSE06) simulations.
The LCAO mesh density cutoff was 2721~eV (100~Ha),
and the PW kinetic-energy cutoff was 544~eV (20~Ha).
The LCAO simulations used Medium (High) bais sets for silicon (germanium).
Relaxation of unit-cell volume and ion positions
was done using the PPS-PBE method
with total energy, forces and stress converged to $10^{-5}$~eV,
0.01~eV/\AA, and 0.05~GPa, respectively.
%
%
%
%

%

\begin{thebibliography}{193}
\bibitem [{Che()}]{CheckElectrode2DFET}%
  \BibitemOpen
  \href@noop {} {}\bibinfo {note} {We checked that the electronic structures of
  the relaxed and unrelaxed Al(110)/MoTe$_2$ electrodes are very
  similar.}\BibitemShut {Stop}%
\end{thebibliography}
\end{document}